\documentclass{winnower}

\usepackage{hyperref}
\hypersetup{
    colorlinks = true,
	citecolor=blue,
	linkcolor=blue
}

\usepackage[subrefformat=parens]{subcaption}

\def\qed{\hfill $\Box$}
\newcommand{\bs}{\boldsymbol}

\newcommand{\argmin}{\mathop{\arg\min}}

\newcommand{\best}[1]{\textbf{#1}}
\newcommand{\second}[1]{\underline{#1}}

\usepackage{amsmath}
\usepackage{pdflscape}
\usepackage{multirow}
\usepackage{boldline}
\usepackage{amsthm}
\usepackage{enumerate}
\theoremstyle{definition}
\newtheorem{theo}{Theorem}
\newtheorem{prop}{Proposition}
\newtheorem{ex}{Example}

\usepackage{algorithmic,algorithm}
 \usepackage{setspace}
\let\Algorithm\algorithm
\renewcommand\algorithm[1][]{\Algorithm[#1]\setstretch{1.6}}

\begin{document}

\title{Hyperlink Regression via Bregman Divergence}

\author[1]{Akifumi Okuno\thanks{oknakfm@gmail.com}}
\author[1,2]{Hidetoshi Shimodaira\thanks{shimo@i.kyoto-u.ac.jp}}
\affil[1]{RIKEN Center for Advanced Intelligence Project}
\affil[2]{Graduate School of Informatics, Kyoto University}


\date{}

\maketitle

\begin{abstract}
A collection of $U \: (\in \mathbb{N})$ data vectors is called a \emph{$U$-tuple}, and the association strength among the vectors of a tuple is termed as the \emph{hyperlink weight}, that is assumed to be symmetric with respect to permutation of the entries in the index. 
We herein propose Bregman hyperlink regression~(BHLR), which learns a user-specified symmetric similarity function such that it predicts the tuple's hyperlink weight from data vectors stored in the $U$-tuple. 
BHLR is a simple and general framework for hyper-relational learning, that minimizes Bregman-divergence~(BD) between the hyperlink weights and estimated similarities defined for the corresponding tuples; 
BHLR encompasses various existing methods, such as logistic regression~($U=1$), Poisson regression~($U=1$), link prediction~($U=2$), and those for representation learning, such as graph embedding~($U=2$), matrix factorization~($U=2$), tensor factorization~($U \geq 2$), and their variants equipped with arbitrary BD. 
Nonlinear functions~(e.g., neural networks), can be employed for the similarity functions. 
However, there are theoretical challenges such that some of different tuples of BHLR may share data vectors therein, unlike the i.i.d.\ setting of classical regression.
We address these theoretical issues, and proved that BHLR equipped with arbitrary BD and $U \in \mathbb{N}$ is 
(P-1) statistically consistent, that is, it asymptotically recovers the underlying true conditional expectation of hyperlink weights given data vectors, and 
(P-2) computationally tractable, that is, it is efficiently computed by stochastic optimization algorithms using a novel generalized minibatch sampling procedure for hyper-relational data. 
Consequently, theoretical guarantees for BHLR including several existing methods, that have been examined experimentally, are provided in a unified manner. 
\end{abstract}

\section{Introduction}
\label{sec:introduction}
Many real-world datasets are in the form of undirected graphs comprising nodes and their links, where nodes may have attributes called \emph{data vectors} and the links are specified by \emph{link weights} representing the strength of association between the corresponding data vectors. 
 A friend network is an example whose data vectors and binary link weights represent properties of people and their friendships, respectively. 

Although such a graph-structured dataset contains rich information, a large number of underlying link weights may be missing in practice~\citep{clauset2008hierarchical,lu2011link}. 
Such missing link weights may be inferred by considering the observed link weights; for instance, two nodes that are connected to the same types of nodes in common are supposed to have high link weights~\citep{lu2011link,liben2007link}. 
However, such an inference deteriorates easily when no or only a few positive link weights to the target nodes are observed.

Even in a severe situation, missing link weights can be inferred by additionally utilizing node data vectors, as their similarities imply the link weights. 
Thus, various methods inferring link weights through data vectors, which are often implemented with neural networks these days, have been developed. We generalize these methods as \emph{link regression}.

A simple implementation of link regression is similarity learning, where a user-specified similarity function defined for pairs of data vectors is trained to predict link weights. 
Although arbitrary similarity functions can be employed, many existing studies leverage the Mahalanobis distance~\citep{de2000mahalanobis} and Mahalanobis inner product~\citep{kung2014kernel}. 
Using these Mahalanobis similarities is mathematically equivalent to using the Euclidean distance or inner product between low-dimensional linearly transformed data vectors~\citep{goldberger2005neighbourhood}, 
implying that Mahalanobis similarity learning implicitly obtains the optimal low-dimensional linear transformation of data vectors. 


Obtaining such an optimal transformation is also known as graph embedding~(GE). 
GE is a method for representation learning; it computes feature vectors such that their inner products predict link weights, and the obtained feature vectors can be used for a variety of downstream tasks in machine learning and statistics. 
For computing the feature vectors, neural networks~(NN) have been  incorporated recently~\citep{tang2015line} to enhance its expressive power. 
Graph embedding with NNs demonstrates promising performance experimentally with some theoretical justification; 
\citet{okuno2018probabilistic} theoretically proved that the inner product similarity~(IPS) between NN-based transformation of data vectors can approximate arbitrary positive-definite~(PD) similarities. 
Furthermore, \citet{okuno2019graph} proposed a shifted IPS by introducing NN-based bias terms to approximate a larger class of similarities called conditionally PD similarities that includes PD similarities and some other non-PD similarities as special cases; an example is the recently popular negative Poincar{\'e} distance \citep{nickel2017poincare,nickel2018learning} for embedding in a Hyperbolic space.
Furthermore, \citet{kim2019representation} proposed a weighted IPS for approximating general similarities. 
Therefore, GE equipped with these similarities can be regarded as a theoretically guaranteed and highly expressive link regression.

Along with the development of highly expressive GEs, replacing loss functions for learning GE has shown progress. 
Whereas many GEs minimize logistic loss~\citep{tang2015line} or the Kullback--Leibler~(KL) divergence~\citep{okuno2018probabilistic} between the observed link weights and those predicted from data vectors, 
\citet{okuno2019robust} recently proposed $\beta$-GE that instead minimizes $\beta$-divergence~\citep{basu1998robust}, which reduces to KL divergence when $\beta=0$.
In addition to the robustness of $\beta$-GE against noisy link weights, \citet{okuno2019robust} proved that $\beta$-GE exhibited the following two desirable properties: 
\textbf{(P-1) statistical consistency}, that is, it asymptotically recovers the underlying true conditional expectation of link weights given data vectors, and
\textbf{(P-2) computational tractability}, that is, it can be computed efficiently by stochastic algorithms using a minibatch sampling for relational data.

Although the existing GEs above achieved success from both theoretical and application perspectives, several challenges still remain.

The first challenge is that the existing GEs are limited to considering the link weight defined between only two nodes, despite the fact that link weights can be similarly defined for a set of three or more nodes. 
We call the weight defined for three or more nodes as \emph{hyperlink weight}. 
A hyperlink weight appears in many practical situations; 
in a friend network, the existence of a group to which all the selected $U(\geq 2)$ people belong should be expressed as a binary hyperlink weight. Similarly, the number of co-authored papers written by all the selected $U (\geq 2)$ people in a co-authorship network should be represented as hyperlink weights assuming values in non-negative integers. 
The existing link regression, including metric learning and GE, cannot address such complicated hyperlink weights.

The second challenge is that, it is unclear whether the properties (P-1) and (P-2) above only hold for the $\beta$-divergence function class, or if they hold for some larger function classes. 
Because only the $\beta$-GE is theoretically proven to  exhibit such favorable properties, the present circumstance may limit the choice of loss function and may result in a missed opportunity to improve the GE's performance.

For simultaneously solving these two challenges, 
we propose the Bregman hyperlink regression (BHLR) by 
(i) extending link regression to hyperlink regression~(HLR) such that it predicts the hyperlink weight defined for a collection of $U(\in \mathbb{N})$ vectors called \text{$U$-tuple}, and
(ii) employing the Bregman divergence~(BD) that includes many loss functions such as logistic loss, KL divergence, and $\beta$-divergence as special cases. 
BHLR is a general framework for hyper-relational learning, that encompasses various existing methods; 
BHLR is in general demonstrated to possess
the two desirable properties~(P-1) statistical consistency and (P-2) computational tractability. 

\subsection{Contribution}
The contribution of this study is summarized as follows.

\begin{enumerate}
    \item In Section~\ref{subsec:HLR}, we propose BHLR, that is a simple and general framework for hyper-relational learning. 
    BHLR predicts hyperlink weight $w_{i_1,i_2,\ldots,i_U} \in \mathcal{S} \: (\subset \mathbb{R})$ from the corresponding tuple of data vectors $\bs x_{i_1},\bs x_{i_2},\ldots,\bs x_{i_U} \in \mathcal{X} \: (\subset \mathbb{R}^p)$ through a user-specified symmetric similarity function $\mu_{\bs \theta}(\bs x_{i_1},\bs x_{i_2},\ldots,\bs x_{i_U})$; 
    highly expressive nonlinear functions, e.g., neural networks, can be employed for the similarity function.
    \item In Section~\ref{sec:BHLR_family}, we demonstrate that BHLR encompasses various existing methods, such as logistic regression~($U=1$), Poisson regression~($U=1$), and link prediction~($U=2$). 
    Furthermore, BHLR also includes methods for representation learning, such as graph embedding~($U=2$), matrix factorization~($U=2$), tensor factorization~($U \geq 2$), and their variants equipped with arbitrary BD; obtained feature vectors through the representation learning methods can be used for a variety of downstream tasks~(e.g., clustering and visualization) besides just predicting hyperlink weights. 
    \item In Section~\ref{sec:properties_of_bhlr}, we generally prove the following properties (P-1) and (P-2) for BHLR equipped with arbitrary BD and $U \in \mathbb{N}$:
    \begin{enumerate}[{(P-1)}]
    \item \textbf{Statistical consistency.} 
    Some tuples in hyper-relational learning may share some data vectors therein. 
    For instance, two different tuples $\bs X_{(1,2,3)}=(\bs x_1,\bs x_2,\bs x_3),\bs X_{(1,3,4)}=(\bs x_1,\bs x_3,\bs x_4)$ share two data vectors $\bs x_1,\bs x_3$. 
    This interesting data structure results in the difference between underlying theories for BHLR and classical regression; 
    Proposition~\ref{prop:bregman_converge} proves that the convergence rate of the loss function used in BHLR is $O(1/\sqrt{n})$ even if $O(n^U)$ tuples are leveraged; the convergence rate is similar to $U$-statistic, and is different from the rate $O(1/\sqrt{n^U})$ of classical regression using $O(n^U)$ i.i.d. data vectors.
    Also, Theorem~\ref{theo:consistency} generally proves that the similarity $\mu_{\hat{\bs \theta}_{\varphi,n}}(\bs x_{i_1},\bs x_{i_2},\ldots,\bs x_{i_U})$ estimated via BHLR asymptotically recovers the underlying true conditional expectation of the tuple's hyperlink weight $\mu_*(\bs x_{i_1},\bs x_{i_2},\ldots,\bs x_{i_U}):=\mathbb{E}(w_{i_1,i_2,\ldots,i_U} \mid \bs x_{i_1},\bs x_{i_2},\ldots,\bs x_{i_U})$, i.e., 
    $\|\mu_{\hat{\bs \theta}_{\varphi,n}}-\mu_*\| \overset{p}{\to} 0$ as the number of data vectors $n$ goes to infinity. 
    Theorem~\ref{theo:consistency} assumes that the similarity function $\mu_{\bs \theta}$ is correctly specified, i.e., $\exists \bs \theta \in \bs \Theta$ such that $\mu_{\bs \theta_*}=\mu_*$, but it is free from specifying the probability distribution of $w_{i_1,i_2,\ldots,i_U}$. 

    \item \textbf{Computationally tractability.} 
    Due to the non-negligible significant computational complexity for dealing with $O(n^U)$ hyperlink weights appeared in hyper-relational learning, 
    we employ stochastic optimization algorithms using a novel generalized mini-batch sampling procedure for hyper-relations. 
    The proposed procedure is a hyper-relational extension ($U \ge 2$) of negative-sampling~\citep{mikolov2013distributed}, that is often used for graph embedding $(U=2)$. 
    Our numerical experiments empirically demonstrate that BHLR is efficiently computed by the stochastic optimization, 
    and our Theorem~\ref{theo:sgd_convergence} also provides a theoretical guarantee for the entire optimization procedure, in the sense that the full-batch gradient of a loss function, evaluated at each step in the stochastic optimization using mini-batch, approaches $\bs 0$ in probability as the number of iterations goes to infinity. 
    \end{enumerate}
    Consequently, BHLR including several existing methods, that have been examined experimentally, is theoretically justified in a unified manner. 
    \item In Section~\ref{sec:experiments}, we perform BHLR on real-world datasets. 
\end{enumerate}

\subsection{Organization}
The remainder of this paper is organized as follows. 
In Section~\ref{sec:bregman_divergence}, we first introduce the Bregman divergence. 
In Section~\ref{sec:hyperlink_regression}, we formally formulate the hyperlink regression and propose the BHLR. 
In Section~\ref{sec:BHLR_family}, we explain the BHLR family members and related works. 
In Section~\ref{sec:properties_of_bhlr}, we show the two favorable properties (P-1) statistical consistency and (P-2) computational tractability for BHLR. 
In Section~\ref{sec:experiments}, we describe the numerical experiments conducted for performing BHLR. 
In Section~\ref{sec:conclusion}, we present our conclusions and future works.

\section{Bregman Divergence}
\label{sec:bregman_divergence}
In this section, we introduce Bregman divergence~(BD) for formulating the Bregman hyperlink regression later in Section~\ref{sec:hyperlink_regression}. 

Here, we consider an index set $\mathcal{I}$, which is specifically defined as the set of tuple indices in our problem setting explained in Section~\ref{subsec:problem_setting}. 
With a continuously differentiable and strictly convex \emph{generating function} $\varphi:\text{dom}(\varphi) \to \mathbb{R}$ whose domain is a set $\text{dom}(\varphi) \subset \mathbb{R}$, 
the BD~\citep{bregman1967relaxation,censor1997parallel} between 
$\bs a:=\{a_{\bs i} \in \text{dom}(\varphi) \mid \bs i \in \mathcal{I}\}$ and $\bs b:=\{b_{\bs i} \in \text{dom}(\varphi) \mid \bs i \in \mathcal{I}\}$ is defined by 
\begin{align}
D_{\varphi}(\bs a , \bs b)
&:=
\frac{1}{|\mathcal{I}|}
    \sum_{\bs i \in \mathcal{I}}
d_{\varphi}(a_{\bs i},b_{\bs i}), 
\label{eq:bregman_divergence}
\end{align}
where $d_{\varphi}:\text{dom}(\varphi)^2 \to \mathbb{R}$ indicates the difference between $\varphi(a)$ and the first-order Taylor approximation of $\varphi(a)$ around $b \in \text{dom}(\varphi)$ as
\begin{align*}
d_{\varphi}(a,b)
&:=
\varphi(a)-(\varphi(b)+\varphi'(b)(a-b)),
\quad (a,b \in \text{dom}(\varphi)).
\end{align*}
Because $\varphi$ is strictly convex, $d_{\varphi}(a,b)$ is always non-negative, and attains the minimum value $0$ at $b=a$ for any fixed $a \in \text{dom}(\varphi)$. 
Similarly, $D_{\varphi}(\bs a,\bs b) \geq 0 \: (\forall \bs a,\bs b \in \text{dom}(\varphi)^{|\mathcal{I}|})$, and the equality holds if and only if $\bs a=\bs b$~(basic property~2 in \citep{cichocki2009nonnegative} p.101). 
Thus, for any fixed $\bs a \in \text{dom}(\varphi)^{|\mathcal{I}|}$, minimizing $D_{\varphi}(\bs a,\bs b)$ with respect to $\bs b \in \text{dom}(\varphi)^{|\mathcal{I}|}$ is expected to cause $\bs b$ to be closer to $\bs a \in \text{dom}(\varphi)^{|\mathcal{I}|}$. 
In our proposed BHLR, $\bs a,\bs b$ are specifically defined as observed hyperlink weights and their predicted weights, respectively, as explained in Section~\ref{subsec:HLR}; the predicted weights are expected to be closer to the observed weights, due to the BD's property. 


Some of the BD family members such as the KL divergence are originally defined for measuring the difference between two probability distributions. That is, they assume that $\bs a,\bs b$ satisfy (1) $a_{\bs i},b_{\bs i} \geq 0 \: (\forall \bs i \in \mathcal{I})$, and (2) $\sum_{\bs i \in \mathcal{I}}a_{\bs i}=\sum_{\bs i \in \mathcal{I}}b_{\bs i}=1$. However, assumptions (1) and (2) are in fact not required for the BD to hold the favorable property above. 
Thus, we do not assume (1) and (2) hereinafter, similarly to some existing studies~\citep{cichocki2009nonnegative,banerjee2005clustering,sra2006generalized}.

The BD includes a variety of loss functions such as the KL divergence, $\beta$-divergence, quadratic loss, and logistic loss, as shown in Table~\ref{table:bregman}. 

\begin{table}[htbp]
\centering
\begin{tabular}{l|l|l|l}
\hlineB{2.5}
$\varphi(x)$ & $\text{dom}(\varphi)$ & $d_{\varphi}(a,b)$ & Name of $D_{\varphi}(\bs a,\bs b)$\\
\hline
$x \log x+(1-x)\log(1-x)$ & $[0,1]$ & 
$\substack{-a \log b - (1-a)\log(1-b) \\ \qquad +a \log a + (1-a) \log (1-a)}$  & Logistic loss$^{\dagger}$~\citep{banerjee2005clustering}\\ 
$x\log x-x$ & $\mathbb{R}_{\geq 0}$ & $a \log \frac{a}{b}-(a-b)$ & Kullback--Leibler div.~\citep{cichocki2009nonnegative} \\
$\frac{x^{1+\beta}}{\beta(1+\beta)} - \frac{x}{\beta} $ & $\mathbb{R}_{\geq 0}$ & $\frac{a^{1+\beta}}{\beta(1+\beta)}-\frac{ab^{\beta}}{\beta}+\frac{b^{1+\beta}}{1+\beta}$ & $\beta$-div.$^{\ddagger}$~\citep{basu1998robust}\\
$-\log x$ & $\mathbb{R}_{> 0}$ & $\frac{a}{b}-\log\frac{a}{b}-1$ & Itakura-Saito div.~\citep{cichocki2009nonnegative} \\
$\frac{1}{x}$ & $\mathbb{R}_{> 0}$ & $\frac{(a-b)^2}{ab^2}$ & Inverse div.~\citep{cichocki2009nonnegative} \\
$\frac{x^2-x}{2}$ & $\mathbb{R}$ & $\frac{1}{2}(a-b)^2$ & Quadratic loss~\citep{cichocki2009nonnegative} \\
$\exp(x)$ & $\mathbb{R}$ & $\exp(a)-(a-b+1)\exp(b)$ & Exponential div.~\citep{cichocki2009nonnegative} \\
$\log(1+\exp(x))$ & $\mathbb{R}$ & $\log \frac{1+\exp(a)}{1+\exp(b)}-(a-b)\frac{\exp(b)}{1+\exp(b)}$ & Dual logistic loss~\citep{boissonnat2010bregman} \\
\hlineB{2.5}
\end{tabular}

$^{\dagger}$By specifying $a \in \{0,1\}$ and $0 \cdot \log 0=0$, the logistic loss reduces to $-a\log b-(1-a)\log (1-b)$. 

$^{\ddagger} \beta>0$ is a user-specified parameter. $\beta$-div. generalizes the Kullback--Leibler div. ($\beta \downarrow 0$) and quadratic loss ($\beta=1$).

\caption{Bregman divergence family. See, e.g. \citet{cichocki2009nonnegative} Section 2.4 and \citet{banerjee2005clustering} Table~1 for details.}
\label{table:bregman}

\end{table}

By removing the strict convexity assumption on $\varphi$ and additionally assuming $a \in \{0,1\}$, 
the BD includes margin-based loss functions. 
For instance, $\varphi(x)=\max\{-x,x-1\}$ results in the misclassification loss $d_{\varphi}(a,b)=I(a \neq I(b>1/2))$, where $I(\cdot)$ represents the indicator function; other examples can be found in \citet{zhang2009new} Section 6.2.

\section{Bregman Hyperlink Regression (BHLR)}
\label{sec:hyperlink_regression}

In this section, we first describe the problem setting in Section~\ref{subsec:problem_setting}; subsequently, we formally define the conditional distribution of hyperlink weights in Section~\ref{subsec:conditional_dist_hyperlink}. 
We compare two different approaches to HLR in Section~\ref{subsec:comparison_approaches_HLR}, and propose BHLR in Section~\ref{subsec:HLR}. 
In Section~\ref{subsec:relation_bhlr_exponential}, we demonstrate that the BHLR can be interpreted as a maximum likelihood estimation using some exponential family model.

\subsection{Problem Setting}
\label{subsec:problem_setting}

For fixed $p,n,U \in \mathbb{N}$ and non-empty sets $\mathcal{X} \subset \mathbb{R}^p,\mathcal{S} \subset \mathbb{R}$, 
our dataset comprises $p$-dimensional data vectors $\{\bs x_i\}_{i=1}^{n} \subset \mathcal{X}$ and symmetric hyperlink weights $\{w_{\bs i}\}_{\bs i \in \mathcal{I}_n^{(U)}} \subset  \mathcal{S}$, where 
$\bs i=(i_1,i_2,\ldots,i_U)$ is an index in a set 
$\mathcal{I}_n^{(U)} \subset [n]^U$, and $[n]$ represents the set $\{1,2,\ldots,n\}$. 
Formal descriptions for tuple of data vectors, hyperlink weights and the index set are provided in the following.

\begin{itemize}
\item \textbf{$U$-tuple} $\bs X = (\bs x, \bs x', \bs x'', \ldots ) \in \mathcal{X}^U$ is an array of $U$ vectors, where $\bs x, \bs x', \bs x'',\ldots \in \mathcal{X}  \: (\subset \mathbb{R}^p)$ are $p$-dimensional vectors. 
For an index $\bs i=(i_1,i_2,\ldots,i_U) \in \mathcal{I}_n^{(U)} \: ( \subset [n]^U)$, a collection of $U$ data vectors
$\bs x_{i_1},\bs x_{i_2},\ldots,\bs x_{i_U} \in \mathcal{X}$ constitute $U$-tuple  $\bs X_{\bs i}= (\bs x_{i_1},\bs x_{i_2},\ldots,\bs x_{i_U})$ indexed by $\bs i$.
Although the order of the vectors is provided, it is in effect ignored in the proposed method, by considering only the symmetric function for the tuple. 
Note that two different tuples may share same data vectors. For instance, $\bs X_{(1,2,3)}=(\bs x_{1},\bs x_{2},\bs x_3)$ and $\bs X_{(1,3,4)}=(\bs x_1,\bs x_3,\bs x_4)$ share two data vectors $\bs x_1,\bs x_3$; we use the multiple index $\bs i \in \mathcal{I}_n^{(U)}$ for dealing with the duplicate data vectors that appear in several different tuples.

\item \textbf{Hyperlink weight} $w_{\bs i} \in \mathcal{S} \: (\subset \mathbb{R})$ represents the strength of association defined for the $U$-tuple $\bs X_{\bs i}$. 
Hyperlink is also called hyperedge in hypergraph theory, and is assumed to be symmetric with respect to permutation of the entries $i_1,i_2,\ldots,i_U$ in the index $\bs i$. 
Although we practically consider non-negative hyperlink weights in many cases, i.e.,  $\mathcal{S}:=\mathbb{R}_{\geq 0}$ such that the weight taking value $0$ represents no association among the tuple, $\mathcal{S}$ is not restricted to be non-negative; $\mathcal{S}$ can be arbitrary specified depending on the setting. 

\item \textbf{Index set} $\mathcal{I}_n^{(U)} \subset [n]^U$ is typically defined as $\mathcal{I}_n^{(U)} = [n]^U$, or $\mathcal{I}_n^{(U)} = \{\bs i \in [n]^U \mid u \neq u' \Rightarrow i_u \neq i_{u'}\}$ such that any tuple do not contain any duplicate vectors in itself, though different tuples may share some data vectors. 
A particular set $\mathcal{I}_n^{(U)}=\mathcal{J}_n^{(U)}:=\{\bs i \in [n]^U \mid 1 \leq i_1 <i_2< \cdots<i_U\}$ is employed later in Section~\ref{subsec:expectation_consistency}, for showing asymptotic properties of the proposed method. 
Although the examples of $\mathcal{I}_n^{(U)} $ mentioned above basically cover all the combinations of indices under some constraints, we can think of even a subset of them
for $\mathcal{I}_n^{(U)} $ in order to allow the practical situation that a limited number of hyperlink weights are actually observed.
\end{itemize}



Such hyperlink weights defined for $U$-tuples appear in many practical situations. 
Two different types of hyperlink weights for $\mathcal{S}:=\mathbb{R}_{\ge 0}$ are shown in the following Examples~\ref{ex:friend_network} and \ref{ex:co-authorship_network}. 
They are also referred to as a hypernetwork~\citep{jeffrey2013hypernetworks}.

\begin{ex}[Friend network]
\label{ex:friend_network}
Data vector $\bs x_i$ represents the property of person $i \in [n]$, e.g., age, gender, education, etc., and the hyperlink weight $w_{\bs i} \in \{0,1,2,\ldots\}(\subset \mathcal{S})$ represents the number of social groups to which all the $U$ people indexed by $\bs i=(i_1,i_2,\ldots,i_U)$ belong. 
\end{ex}

\begin{ex}[Co-authorship network]
\label{ex:co-authorship_network}
Data vector $\bs x_i$ represents the attributes of researcher $i \in [n]$ such as number of publications in each journal, and the hyperlink weight $w_{\bs i} \in \{0,1,2,\ldots\}(\subset \mathcal{S})$ represents the number of co-authored papers written by all the $U$ researchers indexed by $\bs i=(i_1,i_2,\ldots,i_U)$. 
\end{ex}

Here, we consider a user-specified parametric model of \emph{similarity function} $\mu_{\bs \theta}:\mathcal{X}^U \to \mathcal{S}$ with parameter vector $\bs \theta \in \bs \Theta \subset \mathbb{R}^q$.
For $U$-tuple $\bs X=(\bs x, \bs x', \bs x'',\ldots) \in \mathcal{X}^U$, we consider a random variable $w\in \mathcal{S}$ with conditional expectation $\mu_*(\bs X) := 
\mathbb{E}(w \mid \bs X)$. 
$w$ and $\bs X$ are linked by a conditional probability mass (or density) function $q$, as will be formally described in the following Section~\ref{subsec:conditional_dist_hyperlink}. 
Then, learning the similarity function $\mu_{\bs \theta}$ so that 
\[
\mu_{\bs \theta}(\bs X) \approx \mu_*(\bs X),\quad \bs X \in \mathcal{X}^U
\]
is called hyperlink regression~(HLR);
this is analogous to the ordinary regression analysis, where $w$ and $\bs X$ correspond to the response and explanatory variables, respectively. 
For illustrating the HLR, two simple instances are provided in the following Examples~\ref{ex:linear_regression} and \ref{ex:graph_embedding}.

\begin{ex}[Linear regression] 
\label{ex:linear_regression}
As will be explained in Section~\ref{subsec:u1}, linear regression~(LR) is the simplest case of HLR~($U=1$); ``LS reg.'' in Table~\ref{table:bhlr_family_members}.
Given data vectors $\bs x_1,\bs x_2,\ldots,\bs x_n \in \mathcal{X}$ and the corresponding response variables $w_1,w_2,\ldots,w_n \in \mathbb{R}$, LR considers a probabilistic model $w_i=\langle \bs \theta_*,\bs x_i \rangle+\varepsilon_i$, where $\langle \cdot,\cdot \rangle$ represents the inner product and $\bs \theta_* \in \mathbb{R}^p$ is an underlying true parameter. 
Assuming that $\mathbb{E}(\varepsilon_i \mid \bs x_i)=0$, the conditional expectation is $\mu_*(\bs x_i)=\mathbb{E}(w_i \mid \bs x_i)=\langle \bs \theta_*,\bs x_i \rangle$; 
linear regression aims at learning the function $\mu_{\bs \theta}(\bs x):=\langle \bs \theta,\bs x \rangle$, so that it satisfies $\mu_{\bs \theta}(\bs x) \approx  \mu_*(\bs x)$ for all $\bs x \in \mathcal{X}$. 
\end{ex}

\begin{ex}[Graph embedding] 
\label{ex:graph_embedding}
As will be explained in Section~\ref{subsec:u2}, 
graph embedding is a special case of HLR~($U=2$). 
Let $\bs x_1,\bs x_2,\ldots,\bs x_n \in \mathcal{X}$ be data vectors, and $\{w_{i_1 i_2}\}_{1 \le i_1,i_2 \le n}$ be the corresponding weights, where $w_{i_1 i_2} $ represents the strength of association between a pair of two vectors $\bs X_{i_1,i_2}=(\bs x_{i_1},\bs x_{i_2})$.
We consider that $\{\bs x_i \}_{i=1}^n$ are nodes of a graph, and $(w_{i_1 i_2})_{1 \le i_1, i_2 \le n} \in \mathbb{R}^{n\times n}$ represents the adjacency matrix of the graph.  
In addition to the
conditional expectation $\mu_*(\bs X_{i_1, i_2}):=\mathbb{E}(w_{i_1 i_2} \mid \bs X_{i_1, i_2})$,
we may also specify a conditional distribution $q(w \mid \bs X)$ of $w$ given $\bs X=(\bs x, \bs x')$;
typically, Bernoulli distribution $q(w \mid \bs X)=\mu_*(\bs X)^{w}(1-\mu_*(\bs X))^{1-w}$
is considered for binary $w\in\{0,1\}$.
Furthermore, we assume that the data vectors $\bs x_1,\bs x_2,\ldots,\bs x_n$ are i.i.d. generated from a pdf $q_X$
the link weights $\{w_{i_1i_2}\}_{1 \le i_1 < i_2 \le n}$ and data vectors $\{\bs x_i\}_{i=1}^{n}$ follow a joint distribution
\begin{align}
    \prod_{1 \le i_1 < i_2 \le n}
    \underbrace{
        \mu_*(\bs X_{i_1,i_2})^{w_{i_1i_2}}
        (1-\mu_*(\bs X_{i_1,i_2}))^{1-w_{i_1i_2}}
    }_{=q(w_{i_1i_2} \mid \bs X_{i_1,i_2})}
    \prod_{i=1}^{n} q_{X}(\bs x_i).
    \label{eq:generative_model_ge}
\end{align}
The remaining link weights are specified by $w_{i_2i_1}=w_{i_1i_2}$ for $1 \le i_1 < i_2 \le n$ and $w_{ii}=0$. 
See Figure~\ref{fig:generative_model} for the generative model~(\ref{eq:generative_model_ge}); it is straightforwardly generalized to arbitrary $U \in \mathbb{N}$ and arbitrary $q(w \mid \bs X)$, in the following Section~\ref{subsec:conditional_dist_hyperlink}. 
For fully describing the generative model,
we also define a similarity function
$\mu_{\bs \theta}(\bs X_{i_1,i_2}):=\sigma(\langle \bs f_{\bs \theta}(\bs x_i),\bs f_{\bs \theta}(\bs x_j) \rangle)$, where 
$\sigma(z)=(1+\exp(-z))^{-1}$ represents the sigmoid function, and 
$\bs f_{\bs \theta}:\mathcal{X} \to \mathbb{R}^K \: (K \in \mathbb{N})$ is an user-specified parametric function such as neural networks. 
Then, graph embedding learns the function  $\bs f_{\bs \theta}$ so that the similarity function $\mu_{\bs \theta}$ satisfies $\mu_{\bs \theta}(\bs X) \approx \mu_*(\bs X)$ for any $\bs X=(\bs x,\bs x') \in \mathcal{X}^2$. A better feature vector $\bs y_i=\bs f_{\bs \theta}(\bs x_i) \in \mathbb{R}^K$ can be obtained by applying a trained $\bs f_{\bs \theta}$ to the data vector $\bs x_i \in \mathcal{X}$, which is often used for several tasks including ``link prediction''  by looking at the value of $\sigma(\langle \bs y_i,\bs f_{\bs \theta}(\bs x) \rangle)$, $i=1,\ldots, n$, for a newly obtained vector $\bs x \in \mathcal{X}$.
\begin{figure}[htbp]
\centering
\includegraphics[width=0.45\textwidth]{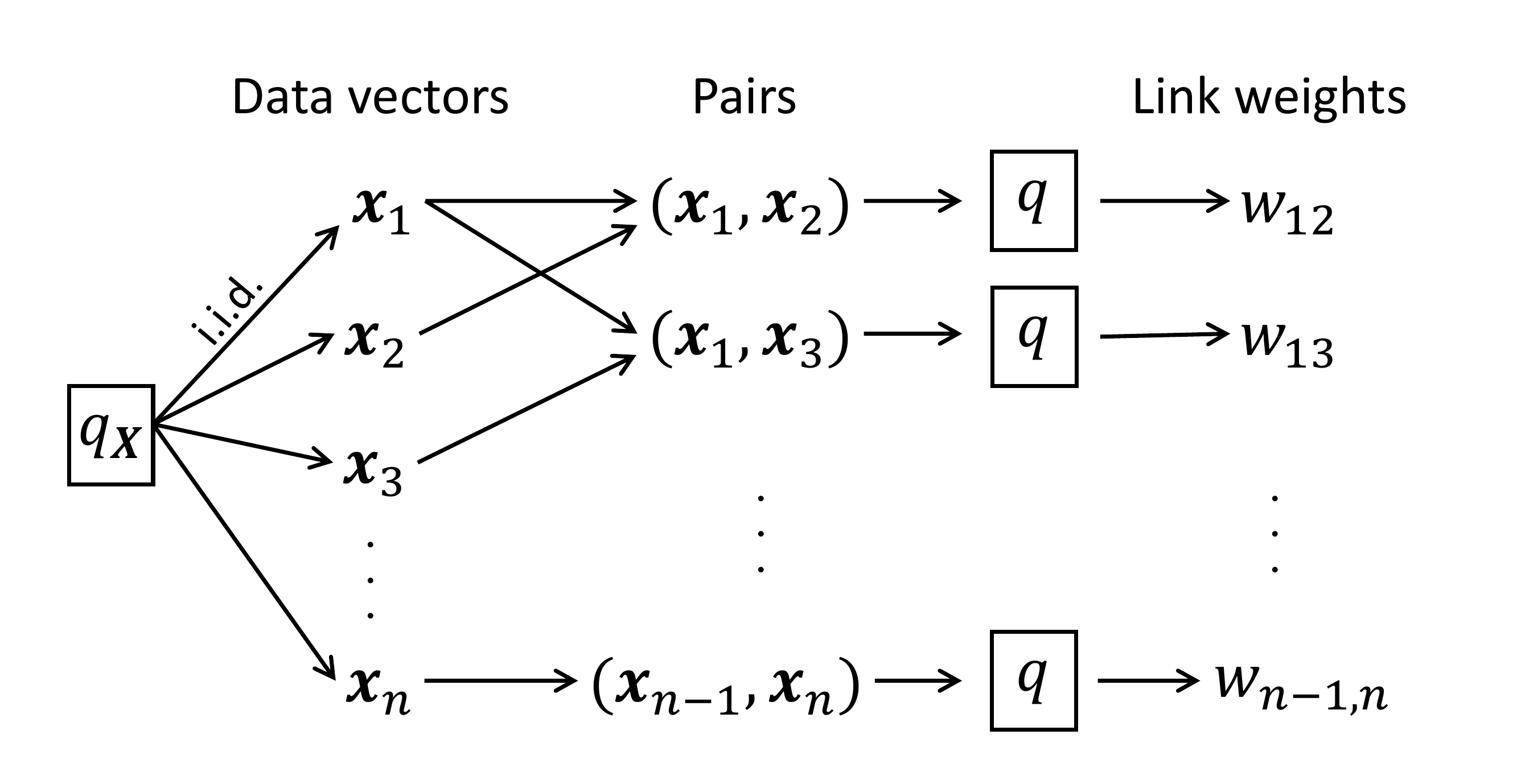}
\caption{Generative model for the graph embedding (Example~\ref{ex:graph_embedding}; $U=2$): $\bs x_i$ i.i.d. follows a distribution $q_X(\bs x)$, and each of link weights $w_{i_1,i_2}$ follows a conditional distribution $q(w \mid (\bs x, \bs x'))$
given the pair $\bs X_{i_1,i_2}=(\bs x_{i_1},\bs x_{i_2})$. The joint distribution of all the hyperlink weights $\{w_{\bs i}\}_{\bs i \in \mathcal{I}_n^{(U)}}$ and all the data vectors $\{\bs x_i\}_{i=1}^{n}$ is defined by (\ref{eq:generative_model_ge}) when considering Bernoulli distribution.}
\label{fig:generative_model}
\end{figure}
\end{ex}

Given our dataset consists of data vectors $\{\bs x_i\}_{i=1}^{n}$ and hyperlink weights $\{w_{\bs i}\}_{\bs i \in \mathcal{I}_n^{(U)}}$,
the parameter vector $\bs \theta$ is optimized by minimizing an empirical loss function so that
\[
\mu_{\bs \theta}(\bs X_{\bs i}) \approx w_{\bs i},\quad
\bs i \in \mathcal{I}_n^{(U)}
\]
hold.
This paper aims at providing a general framework for HLR, named BHLR, such that it encompasses a variety of existing methods. 
This paper also intends to provide theoretical guarantees for general BHLR; several existing methods, that have been examined experimentally, are also theoretically justified in a unified manner.

\subsection{Probability Distributions of Hyperlink Weights and Tuples}
\label{subsec:conditional_dist_hyperlink}
In order to obtain the conditional expectation $\mu_*(\bs X_{\bs i})=\mathbb{E}(w_{\bs i} \mid \bs X_{\bs i})$, 
we first formally define the conditional distribution of hyperlink weights given data vectors by straightforwardly generalizing the probabilistic model for $U=2$ shown in Example~\ref{ex:graph_embedding} and Figure~\ref{fig:generative_model}.

Here, we explain why the extra attention is required for defining the conditional distribution of hyperlink weights given data vectors. 
For any $\bs i'$ obtained by permutating the elements of $\bs i$, tuples $\bs X_{\bs i},\bs X_{\bs i'}$ consist of the same vectors $\bs x_{i_1},\bs x_{i_2},\ldots,\bs x_{i_U}$, and it holds that $w_{\bs i}=w_{\bs i'}$ since the hyperlink weights are assumed to be symmetric. 
In the case of $U=2$, this symmetry coincides with considering undirected links; link weights should satisfy $w_{i_1 i_2}=w_{i_2 i_1}$ for all $i_1$ and $i_2$, implying the constraints on the distributions for $w_{i_1,i_2}$ and $w_{i_2 i_1}$. 

For specifying the distribution appropriately, we employ a simple idea. 
We first specify the conditional probability density function (cpdf) or conditional probability mass function (cpmf) $\tilde{q}$ only for $w_{i_1 i_2} \mid \bs X_{i_1 i_2}$ whose index is in non-decreasing order $i_1 \leq i_2$. 
Then, the cpdf or cpmf $q$ of $w_{i_2 i_1} \mid \bs X_{i_2 i_1}$ whose index is in reverse order, can be defined as that of $w_{i_1 i_2} \mid \bs X_{i_1 i_2}$, since the weights satisfy the symmetry $w_{i_1 i_2}=w_{i_2 i_1}$ and both tuples $\bs X_{i_1,i_2},\bs X_{i_2 i_1}$ consist of the same vectors $\bs x_{i_1},\bs x_{i_2}$. 
This idea of symmetry is readily generalized to $U \in \mathbb{N}$; we specify the cpdf or cpmf $\tilde{q}$ of $w_{\bs i'} \mid \bs X_{\bs i'}$ only for non-decreasing order index $\bs i' \in [n]^U$ such that $i'_1 \le i'_2 \le \cdots \le i'_U$, and consider a mapping $r:\bs i \mapsto \bs i'$ such that $\bs i'=r(\bs i)$ is obtained by sorting the elements of $\bs i$ in non-decreasing order.
Then cpdf or cpmf $q$ of $w_{\bs i} \mid \bs X_{\bs i}$ is defined as
\begin{align}
    q(w_{\bs i} \mid \bs X_{\bs i})
    :=
    \tilde{q}(w_{r(\bs i)} \mid \bs X_{r(\bs i)}), 
    \quad (\bs i \in \mathcal{I}_n^{(U)}).
    \label{eq:def_cpdf}
\end{align}
Therefore, we have well-defined conditional distribution for hyperlink weights. 
Then, the cpdf (or cpmf) of all the hyperlink weights $\{w_{\bs i}\}_{\bs i \in \mathcal{I}_n^{(U)}}$ given data vectors $\bs x_1,\bs x_2,\ldots,\bs x_n \in \mathcal{X}^n$ is
\begin{align}
    \prod_{\bs i \in \mathcal{I}_n^{(U)}}
    q(w_{\bs i} \mid \bs X_{\bs i}),
    \label{eq:all_conditional}
\end{align}
meaning that hyperlink weight $w_{\bs i}$ is conditionally independently generated by following the probabilistic model (\ref{eq:all_conditional}). 
When considering the case that $U=2$, $q(w \mid \bs X)=\mu_*(\bs X)^w(1-\mu_*(\bs X))^{1-w}$ represents the cpmf of Bernoulli distribution whose expectation is $\mu_*(\bs X):=\mathbb{E}(w \mid \bs X)$, and $\bs X=(\bs x,\bs x')$ is a pair of latent variables, 
the probabilistic model~(\ref{eq:all_conditional}) is also known as latent position random graph~(LPRG) model with kernel $\mu_*$. 
LPRG model is considered in \citet{tang2013universally} and \citet{athreya2018statistical} Definition~6, and it is originated from the random dot product graph model~\citep{young2007random}, that corresponds to a case $\mu_*(\bs X):=\langle \bs x,\bs x' \rangle$ for $\bs X=(\bs x,\bs x') \in \mathcal{X}^2$. Our probabilistic model~(\ref{eq:all_conditional}) generalizes the LPRG model to arbitrary probability distribution with arbitrary $U \in \mathbb{N}$, though the previous studies focus on the spectral analyses on the matrix $\bs W=(w_{ij})$ of Bernoulli link weights with $U=2$, and they assume that $\bs x_1,\bs x_2,\ldots,\bs x_n$ are latent variables.

Hereinafter, we note the probability distribution of the tuple $\bs X_{\bs i}$. 
We will simply assume that the data vectors $\{\bs x_{i}\}_{i=1}^{n}$ are i.i.d. randomly generated from a distribution $q_{X}$ in Section~\ref{sec:properties_of_bhlr} for showing statistical consistency of BHLR. 
Then, the joint distribution over all the hyperlink weights and data vectors is specified as $\prod_{\bs i \in \mathcal{I}_n^{(U)}}  q(w_{\bs i} \mid \bs X_{\bs i})\prod_{i=1}^{n}q_X(\bs x_i)$.
Note that the marginal distribution for $\bs Z_{\bs i}:=(w_{\bs i},\bs X_{\bs i})$ does not depend on the index $\bs i$, thus $\bs Z_{\bs i}  ,  \bs i \in   \mathcal{I}_n^{(U)} $ are identically distributed.
However, even if data vectors $\{\bs x_{i}\}_{i=1}^{n}$ are i.i.d. generated, 
two different $\bs Z_{\bs i},\bs Z_{\bs i'}$ can be dependent, as their tuples $\bs X_{\bs i},\bs X_{\bs i'}$ may share same data vectors in common. 
For instance, $\bs X_{(1,2,3)}=(\bs x_{1},\bs x_2,\bs x_3)$ and $\bs X_{(1,3,4)}=(\bs x_1,\bs x_3,\bs x_4)$ share two data vectors $\bs x_1$ and $\bs x_3$. 
Therefore, $\bs Z_{\bs i}  ,  \bs i \in   \mathcal{I}_n^{(U)} $ are NOT independently distributed.
This property for $U\ge 2$ makes our setting interesting and needs a special care in the asymptotic theory. 
In this regard, theories for HLR, that predicts hyperlink weights from the constrained tuples, can be different from those of classical regression, that typically predicts response variables from i.i.d. data vectors. We consider such constrained tuples, and the statistical consistency for BHLR is proved later in Section~\ref{sec:properties_of_bhlr}.

\subsection{Two Different Approaches to HLR} 
\label{subsec:comparison_approaches_HLR}
In this section, we show two different approaches to HLR with $\mathcal{S}:=\mathbb{R}_{\geq 0}$,
and explain why we employ the second approach.
Although the case of $U=1$ is illustrated here, it can be easily generalized to arbitrary $U \in \mathbb{N}$. 

\begin{figure}[htbp]
\centering
 \begin{minipage}[!b]{0.47\hsize}
 \centering
  \includegraphics[width=0.6\textwidth]{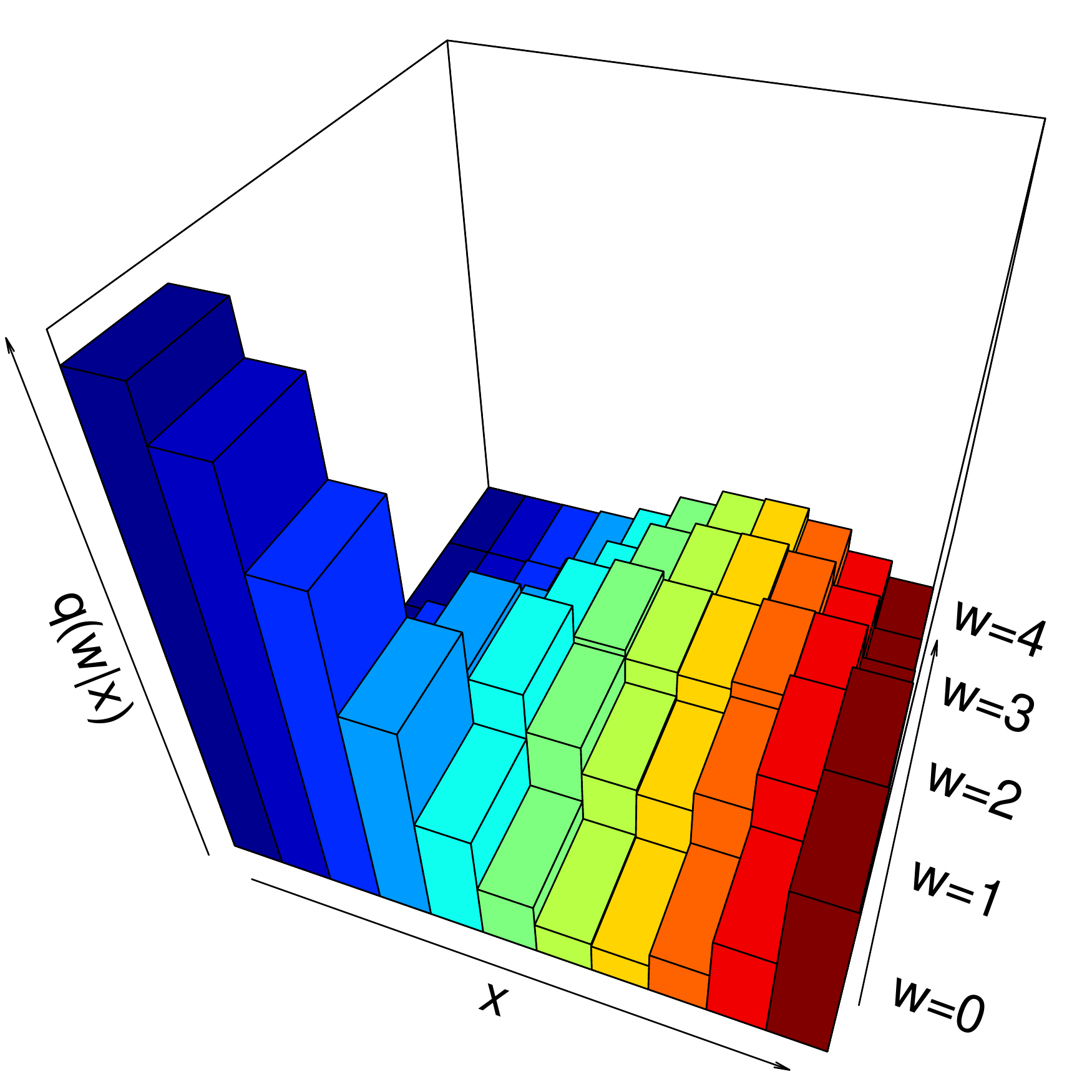}
  \caption*{(a) $q(w \mid \bs x)$}
 \end{minipage}
 \hspace{0.5em}
 \begin{minipage}[!b]{0.47\hsize}
 \centering
  \includegraphics[width=0.5\textwidth]{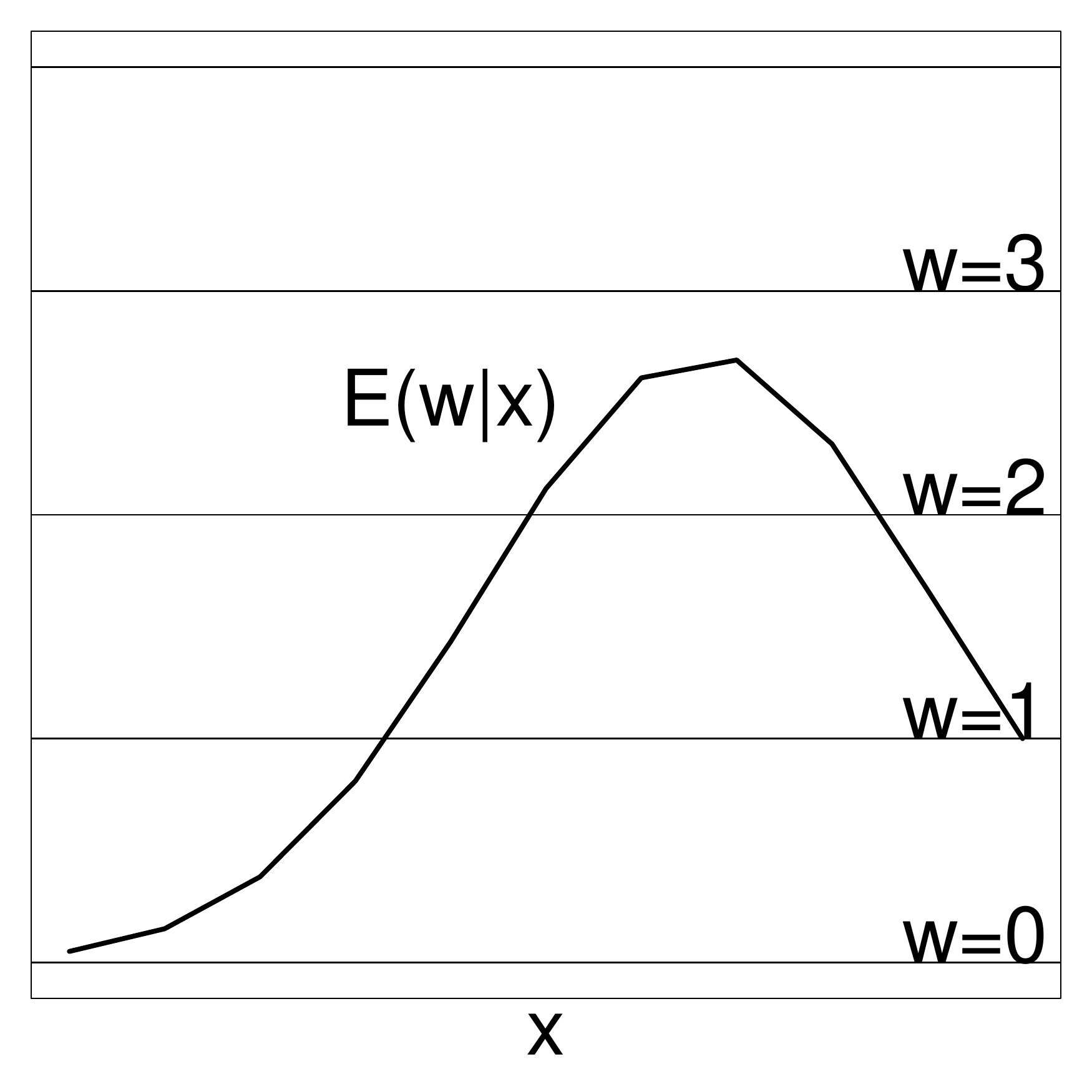}
  \caption*{(b) $E(w \mid \bs x)$}
 \end{minipage}
 \caption{Examples of (a) underlying conditional probability mass function $q(w \mid \bs x)$ whose conditional expectation is $E(w \mid \bs x)=\sum_{w \in \mathbb{N}_0}w q(w \mid \bs x)$, and (b) the conditional expectation function $\mu_*(\bs x)=E(w \mid \bs x)$.}
 \label{fig:underlying_conditional_pmf}
\end{figure}

Considering a weight $w_i$ taking a value in the set $\{0,1,2,\ldots\} \subset \mathcal{S}$ and a data vector $\bs x_i \in \mathbb{R}^p$ $(i=1,2,\ldots,n)$, HLR predicts the weight $w_i \in \mathcal{S}$ through the function $\mu_{\bs \theta}(\bs x_i) \in \mathcal{S}$. 
However, there are two different approaches to this problem. 
The first approach is based on matching conditional probability mass function~(pmf) $q(w_i \mid \bs x_i)$ shown in Fig.~\ref{fig:underlying_conditional_pmf} (a) and the parametric generative model $p_{\bs \theta}(w_i \mid \bs x_i)$ whose expectation is $\mu_{\bs \theta}(\bs x_i)=\sum_{w \in \mathbb{N}_0} w p_{\bs \theta}(w \mid \bs x_i)$. 
Although this approach naturally extends the maximum likelihood regression, there remain several challenges explained below. 
For  solving these challenges, we also consider the second approach, that instead matches only the conditional expectation function $\mu_*(\bs x_i):=E(w_i \mid \bs x_i)$ shown in Fig.~\ref{fig:underlying_conditional_pmf} (b) and the model $\mu_{\bs \theta}(\bs x_i)$. 
Consequently, we employ and generalize the second approach, and propose \emph{Bregman-HLR~(BHLR)} in Section~\ref{subsec:HLR}. 

Hereinafter, we describe the details of the two approaches to HLR.

The first approach is, matching the underlying conditional pmf $q(w_i \mid \bs x_i)$ and the parametric generative model $p_{\bs \theta}(w_i \mid \bs x_i)$. 
Let $q_{iw} = q(w \mid \bs x_i)$ and $p_{\bs \theta,iw} = p_{\bs \theta}(w \mid \bs x_i)$ for $w\in  \mathbb{N}_0$, $i=1,\ldots,n$.
They are put together as vectors $\bs q_i:=(q_{i0},q_{i1},q_{i2},\ldots),\bs p_{\bs \theta,i}:=(p_{\bs \theta,i0},p_{\bs \theta,i1},p_{\bs \theta,i2},\ldots)$, so that each of vectors $\bs q_i,\bs p_{\bs \theta,i}$ represents the distribution of $w_i \mid \bs x_i$.  
Then, we may estimate $\bs \theta$ by minimizing 
\begin{align}
\frac{1}{n}
    \sum_{i=1}^{n}
    D_{\varphi}(\bs q_i,\bs p_{\bs \theta,i}), 
    \label{eq:probability_matching}
\end{align}
where $\varphi$ is a user-specified generating function. 
However, the underlying conditional distributions $\bs q_1,\bs q_2,\ldots,\bs q_n$ used in (\ref{eq:probability_matching}) cannot be observed in practice; 
we instead consider the empirical conditional distribution $\hat{\bs q}_{i}=(\hat{q}_{i0},\hat{q}_{i1},\hat{q}_{i2},\ldots)$ whose $w_i$-th entry is $1$ and $0$ otherwise, for $i=1,2,\ldots,n$. Considering $\mathcal{I}=\mathbb{N}$, 
\begin{align*}
|\mathcal{I}|D_{\varphi}(\hat{\bs q}_i,\bs p_{\bs \theta,i}) 
&= 
\sum_{w \in \mathbb{N}} d_{\varphi}(\hat{q}_{iw},p_{\theta,iw}) 
=
\sum_{w \in \mathbb{N}} 
\{
\varphi(\hat{q}_{iw})
-
\varphi(p_{\theta,iw})
-
\varphi'(p_{\theta,iw})
(\hat{q}_{iw}-p_{\theta,iw})
\} \\
&=
\sum_{w \in \mathbb{N}} 
\{
\varphi'(p_{\theta,iw})p_{\theta,iw}
-
\varphi(p_{\theta,iw})
-
\varphi'(p_{\theta,iw})
\hat{q}_{iw}
\} + \text{Const.} \\
&=
\sum_{w \in \mathbb{N}} 
\{
\varphi'(p_{\bs \theta}(w \mid \bs x_i))p_{\bs \theta}(w \mid \bs x_i)
-
\varphi(p_{\bs \theta}(w \mid \bs x_i))
\}
-
\varphi'(p_{\bs \theta}(w_i \mid \bs x_i))
+
\text{Const.} \\
&\hspace{15em}
\left( \: \because 
p_{\bs \theta,iw}=p_{\bs \theta}(w \mid \bs x_i),
\hat{q}_{iw}=\begin{cases} 1 & (w=w_i) \\ 0 & (w \neq w_i) \\ \end{cases} \: \right)
\end{align*}
holds; minimizing (\ref{eq:probability_matching}) equipped with the empirical distributions $\{\hat{\bs q}_i\}_{i=1}^{n}$ is equivalent to minimizing
\begin{align}
    \frac{1}{n}\sum_{i=1}^{n}
    \bigg\{
    \underbrace{
    \sum_{w \in \mathbb{N}_0}
    \left(
        \varphi'(p_{\bs \theta}(w \mid \bs x_i))
        p_{\bs \theta}(w \mid \bs x_i)
        -
        \varphi(p_{\bs \theta}(w \mid \bs x_i))
    \right)
    }_{(\star)}
    -
    \varphi'(p_{\bs \theta}(w_i \mid \bs x_i))
    \bigg\}.
    \label{eq:probability_matching_empirical}
\end{align}
(\ref{eq:probability_matching_empirical}) appears in some existing studies, such as \citet{ghosh2013robust} for $\beta$-divergence in Table~\ref{table:bregman}. 
However, as \citet{okuno2019robust} Section 3.2 pointed out in a special case of HLR, 
the term ($\star$) in eq.~(\ref{eq:probability_matching_empirical}) is computationally intractable due to the infinite summation $\sum_{w \in \mathbb{N}_0}$; there remain a computational challenge in this approach. 
The fininite summation similarly appears in eq.~(4) of \citet{kawashima2019robust}, and they compute the term by the finite-sum approximation instead. 
Note that, the term ($\star$) reduces to $\sum_{w \in \mathbb{N}_0}p_{\bs \theta}(w \mid \bs x_i)=1$ if the generating function is specified as $\varphi(x)=x\log x-x$; the computational issue does not occur if KL-divergence is considered.

For solving the computational challenge, we also consider the second approach. 
This second approach simply matches the underlying expectation function $\mu_*(\bs x_i)=\mathbb{E}(w_i \mid \bs x_i)$ and the parametric model $\mu_{\bs \theta}(\bs x_i)$ without assuming any specific probability distribution for $w_i \mid \bs x_i$; 
we may obtain the estimator of $\bs \theta$ by minimizing
\begin{align}
    D_{\varphi}(\{\mu_*(\bs x_i)\}_{i=1}^{n},\{\mu_{\bs \theta}(\bs x_i)\}_{i=1}^{n}), 
    \label{eq:expectation_matching}
\end{align}
where $\varphi$ is a user-specified generating function whose domain $\text{dom}(\varphi)$ includes the set $\mathcal{S}$. 
However, the underlying expectation function $\mu_*$ cannot be observed in practice; we instead minimize
\begin{align}
    &D_{\varphi}(\{w_i\}_{i=1}^{n},\{\mu_{\bs \theta}(\bs x_i)\}_{i=1}^{n}) 
    =
    \frac{1}{n}\sum_{i=1}^{n}
    \left\{
        \varphi'(\mu_{\bs \theta}(\bs x_i))\mu_{\bs \theta}(\bs x_i)
        -
        \varphi(\mu_{\bs \theta}(\bs x_i))
        -
        w_i \varphi'(\mu_{\bs \theta}(\bs x_i))
    \right\}
    +
    C
    \label{eq:expectation_matching_empirical}
\end{align}
that approximates (\ref{eq:expectation_matching}) in the sense that the underlying true conditional expectation $\mu_*(\bs x_i)=\mathbb{E}(w_i \mid \bs x_i)$ is replaced with the observation $w_i$. 
$C:=\frac{1}{n}\sum_{i=1}^{n}\varphi(w_i)$ is a constant independent of the parameter $\bs \theta$. 
(\ref{eq:expectation_matching_empirical}) reduces to \citet{zhang2009new} eq.~(20), if the model is specified as $\mu_{\bs \theta}(\bs x)=g(\bs \theta^{\top}\bs x)$ for some non-linear function $g:\mathbb{R}\to\mathbb{R}$, whereas arbitrary similarity function $\mu_{\bs \theta}$ is considered in this study.

The second approach bypasses the computational challenge of the first approach, since (\ref{eq:expectation_matching_empirical}) does not include any infinite summatation; 
we consequently employ the second approach, and generalize it from $U=1$ to $U \in \mathbb{N}$ as shown in the next section.

\subsection{Proposed BHLR}
\label{subsec:HLR}

We here consider HLR with arbitrary $U \in \mathbb{N}$, for predicting the hyperlink weights $w_{\bs i}$ taking values in a set $\mathcal{S} \subset \mathbb{R}$ via a user-specified symmetric similarity function $\mu_{\bs \theta}:\mathcal{X}^U \to \mathcal{S}$. 
By generalizing the loss function (\ref{eq:expectation_matching_empirical}) from $U=1$ to $U \in \mathbb{N}$, we propose to minimize a simple loss function
\begin{align}
    L_{\varphi,n}(\bs \theta)
    &:=
    D_{\varphi}
    (
        \{w_{\bs i}\}_{\bs i \in \mathcal{I}_n^{(U)}}
        ,
        \{\mu_{\bs \theta}(\bs X_{\bs i})\}_{\bs i \in \mathcal{I}_n^{(U)}}
    ) \nonumber \\
    &=
    \frac{1}{|\mathcal{I}_n^{(U)}|}\sum_{\bs i \in \mathcal{I}_n^{(U)}}
    \left\{
        \varphi'(\mu_{\bs \theta}(\bs X_{\bs i}))\mu_{\bs \theta}(\bs X_{\bs i})
        -
        \varphi(\mu_{\bs \theta}(\bs X_{\bs i}))
        -
        w_{\bs i} \varphi'(\mu_{\bs \theta}(\bs X_{\bs i}))
    \right\} 
    +
    C,
    \label{eq:ell_phi}
\end{align}
where $\varphi$ is a user-specified generating function whose domain $\text{dom}(\varphi)$ includes the set $\mathcal{S}$, and $C:=\frac{1}{|\mathcal{I}_n^{(U)}|}\sum_{\bs i \in \mathcal{I}_n^{(U)}}\varphi(w_{\bs i})$ is a constant independent of the parameter $\bs \theta$. 
Subsequently, the estimator is defined as
\begin{align}
\hat{\bs \theta}_{\varphi,n}:=\argmin_{\bs \theta \in \bs \Theta}L_{\varphi,n}(\bs \theta).
\label{eq:estimator}
\end{align}
Once the estimator $\hat{\bs \theta}_{\varphi,n}$ is obtained, we may predict $w_{\bs i}$ by the estimated similarity function $\mu_{\hat{\bs \theta}_{\varphi,n}}(\bs X_{\bs i})$. 
We formally define predicting $w_{\bs i}$ by the function $\mu_{\hat{\bs \theta}_{\varphi,n}}(\bs X_{\bs i})$ as the BHLR. 

Since the hyperlink weights are symmetry, we assume that the function $\mu_{\bs \theta}$ also satisfies the symmetry
\begin{align}
    \mu_{\bs \theta}(\bs x_{i_1},\bs x_{i_2},\ldots,\bs x_{i_U})
    =
    \mu_{\bs \theta}(\bs x_{i_1'},\bs x_{i_2'},\ldots,\bs x_{i_U'})
    \label{eq:model_symmetry}
\end{align}
for any $\bs i'=(i_1',i_2',\ldots,i_U')$ obtained by permutating the elements of $\bs i=(i_1,i_2,\ldots,i_U) \in \mathcal{I}_n^{(U)}$. This symmetry should hold for all $\bs x_{i_1},\bs x_{i_2},\ldots,\bs x_{i_U} \in \mathcal{X}$ and $\bs \theta \in \bs \Theta$; 
the similarity function $\mu_{\bs \theta}$ in effect ignores the order of the vectors, 
as long as (9) is assumed. 
An example of such a symmetric similarity function is
\begin{align}
    \mu_{\bs \theta}(\bs x_{i_1},\bs x_{i_2},\ldots,\bs x_{i_U})
    =
    \eta\left(
        \langle 
            \bs f_{\bs \theta}(\bs x_{i_1}),
            \bs f_{\bs \theta}(\bs x_{i_2}),
            \cdots,
            \bs f_{\bs \theta}(\bs x_{i_U})
        \rangle
    \right),
    \label{eq:ex_similarity_model}
\end{align}
where $\bs f_{\bs \theta}:\mathcal{X} \to \mathbb{R}^K$ is a function parametrized by $\bs \theta$, e.g., vector-valued neural networks, $\eta:\mathbb{R} \to \mathcal{S}$ is a link function, e.g., exponential function for $\mathcal{S}=\mathbb{R}_{\ge 0}$ and sigmoid function for $\mathcal{S}=[0,1]$, 
and $\langle \bs y,\bs y',\bs y'',\ldots \rangle:=\sum_{k=1}^{K}y_k y_k'y_k''\cdots$. 
The above function (\ref{eq:ex_similarity_model}) is employed for our numerical experiments later in Section~\ref{sec:experiments}, and it reduces to tensor decomposition explained in Section~\ref{subsec:u_general} if $\bs f_{\bs \theta}(\bs x)=\bs \theta^{\top}\bs x$, $\eta(z)=z$, and $\bs x_i$ is $1$-hot vector.


The BHLR reduces to several existing methods, such as logistic regression~($U=1$), Poisson regression~($U=1$), and link prediction~($U=2$), by specifying $\mu_{\bs \theta}$ and $\varphi$. 
Furthermore, BHLR also reduces to several methods for representation learning, such as graph embedding~($U=2$), matrix factorization~($U=2$), tensor factorization~($U \geq 2$), and their variants equipped with arbitrary BD. 
We describe the relation between the BHLR and these existing methods in Section~\ref{sec:BHLR_family}.

In addition to the rich examples for the BHLR family, the BHLR possesses the following two favorable properties: 
(P-1) statistical consistency, and 
(P-2) computational tractability. 
We further explain these properties (P-1) and (P-2) in Section~\ref{subsec:expectation_consistency} and Section~\ref{subsec:propsoed_algorithm}, respectively, along with the proposal of a novel and generalized minibatch sampling procedure for hyper-relational data that can be used for efficient stochastic algorithms.

\subsection{BHLR is Equivalent to MLE through Corresponding Exponential Family Model}
\label{subsec:relation_bhlr_exponential}

In this section, we demonstrate that BHLR is interpreted as the maximum-likelihood estimation with a corresponding exponential family model.
In other words, specifying a generating function $\varphi$ for BD implicitly specifies a cpdf or cpmf for $w_{\bs i} \mid \bs X_{\bs i}$ of the form 
\begin{align}
    p_{\bs \zeta}(w \mid \mu)
    &:=
    \exp\left(
        w\zeta_1(\mu)
        +
        \zeta_2(\mu)
        +
        \zeta_3(w)
    \right)
    \label{eq:exponential_family_density_zeta}
\end{align}
with $\mu = \mu_{\bs \theta}(\bs X_{\bs i})$, 
where $\zeta_1(\mu):=\varphi'(\mu),\zeta_2(\mu):=\varphi(\mu)-\mu\varphi'(\mu)$, and
$\zeta_3(w)$ is specified such that $\int_\mathcal{S} p_{\bs \zeta}(w | \mu) \, \mathrm{d}w=1$ (cpdf) or 
$\sum_{w \in \mathcal{S}} p_{\bs \zeta}(w | \mu) =1$ (cpmf) holds.
This is easily understood as explained below.
Starting from (\ref{eq:ell_phi}), a simple calculation leads to
\begin{align*}
    \exp\left(
        -|\mathcal{I}_n^{(U)}|L_{\varphi,n}(\bs \theta)
    \right) 
&=
    \exp \left(
        -\sum_{\bs i \in \mathcal{I}_n^{(U)}}
        \left\{
        \varphi(w_{\bs i})
        -
        \varphi(\mu_{\bs \theta}(\bs X_{\bs i}))
        -
        \varphi'(\mu_{\bs \theta}(\bs X_{\bs i}))
        (w_{\bs i}-\mu_{\bs \theta}(\bs X_{\bs i}))
        \right\}
    \right) \\
    &=
    \prod_{\bs i \in \mathcal{I}_n^{(U)}}
        \exp \left(
        -
        \left\{
        \varphi(w_{\bs i})
        -
        \varphi(\mu_{\bs \theta}(\bs X_{\bs i}))
        -
        \varphi'(\mu_{\bs \theta}(\bs X_{\bs i}))
        (w_{\bs i}-\mu_{\bs \theta}(\bs X_{\bs i}))
        \right\}
    \right) \\
    &=
    D \cdot 
    \prod_{\bs i \in \mathcal{I}_n^{(U)}}
        \exp \left(
        w_{\bs i}\zeta_1(\mu_{\bs \theta}(\bs X_{\bs i}))
        +
        \zeta_2(\mu_{\bs \theta}(\bs X_{\bs i}))
        +
        \zeta_3(w_{\bs i})
    \right) \\
    &=:
    D \cdot \prod_{\bs i \in \mathcal{I}_n^{(U)}}
    p_{\bs \zeta}(w_{\bs i} \mid \mu_{\bs \theta}(\bs X_{\bs i})),
\end{align*}
where $D :=\prod_{\bs i \in \mathcal{I}_n^{(U)}}\exp(-\varphi(w_{\bs i})-\zeta_3(w_{\bs i}))$ is a constant independent of the parameter $\bs \theta$. 
The normalizing function $\zeta_3(w)$ is explicitly specified as
$\zeta_3(w) = -\log \int_\mathcal{S} \exp(w\zeta_1(\mu) + \zeta_2(\mu))\,\mathrm{d}w$ (cpdf) or
$\zeta_3(w) = -\log \sum_{w \in \mathcal{S}} \exp(w\zeta_1(\mu) + \zeta_2(\mu))$ (cpmf).
Therefore minimizing $L_{\varphi,n}(\bs \theta)$ in BHLR is formally equivalent to maximizing the likelihood function of the exponential family model $p_{\bs \zeta}(w_{\bs i} \mid \mu_{\bs \theta}(\bs X_{\bs i}))$.

When $U=1$, we associate the BHLR with the MLE of the generalized linear model~(GLM)~\citep{bishop2006pattern}. 
They are almost the same but do not exhibit inclusion in the following sense: 
(i) The GLM restricts $\zeta_1$ in (\ref{eq:exponential_family_density_zeta}) to be an identity function, and the function $\mu_{\bs \theta}(\bs X_{\bs i})$ is in the form of $g(\bs \theta^{\top}\bs x_{i_1})$ for some function $g$, whereas the BHLR is free from these constraints. 
(ii) Meanwhile, function $\zeta_2$ in (\ref{eq:exponential_family_density_zeta}) is constrained by the generating function $\varphi$, whereas this does not apply to GLM.

\section{BHLR Family Members and Related Works}
\label{sec:BHLR_family}

In this section, we describe the BHLR family members by specifying $U \in \mathbb{N}$ and the generating function $\varphi$ in Section~\ref{subsec:u1}--\ref{subsec:u_general} and Table~\ref{table:bhlr_family_members}. Other related works are explained in Section~\ref{subsec:other_related_works}.

Before explaining the BHLR family members, we first explicitly derive the corresponding loss functions $L_{\varphi,n}(\bs \theta)$ associated with some generating functions 
$\varphi_{\text{Logistic}}(x) := x \log x + (1-x) \log (1-x),
\varphi_{\text{KL}}(x) := x \log x - x, 
\varphi_{\text{Quad.}}(x) := x^2 - x$ and 
$\varphi_{\beta}(x):=\frac{x^{1+\beta}}{\beta(1+\beta)}-\frac{x}{\beta}$, that are listed in Table~\ref{table:bregman}. 
Subsequently, for an arbitrary $U \in \mathbb{N}$, we have
\begin{align}
    L_{\varphi_{\text{Logistic}},n}(\bs \theta)
    &=
\scalebox{0.9}{$\displaystyle 
    \frac{1}{|\mathcal{I}_n^{(U)}|}
    \sum_{\bs i \in \mathcal{I}_n^{(U)}}
    \left\{
        -w_{\bs i} \log \mu_{\bs \theta}(\bs X_{\bs i})
        -(1-w_{\bs i}) \log (1-\mu_{\bs \theta}(\bs X_{\bs i}))
    \right\}
    +
    C_{\text{Logistic}}^{(U)}
    $},
    \label{eq:l_logistic} \\
    L_{\varphi_{\text{KL}},n}(\bs \theta)
    &=
\scalebox{0.9}{$\displaystyle 
    \frac{1}{|\mathcal{I}_n^{(U)}|}
    \sum_{\bs i \in \mathcal{I}_n^{(U)}}
    \left\{
        -w_{\bs i} \log \mu_{\bs \theta}(\bs X_{\bs i})
        +
        \mu_{\bs \theta}(\bs X_{\bs i})
    \right\}
    +
    C_{\text{KL}}^{(U)}$}, 
    \label{eq:l_kl} \\
    L_{\varphi_{\text{Quad.}},n}(\bs \theta)
    &=
    \scalebox{0.9}{$\displaystyle 
    \frac{1}{|\mathcal{I}_n^{(U)}|}
    \sum_{\bs i \in \mathcal{I}_n^{(U)}}
    \left(
        w_{\bs i}-\mu_{\bs \theta}(\bs X_{\bs i})
    \right)^2$}, 
     \label{eq:l_quad} \\
    L_{\varphi_{\beta},n}(\bs \theta)
    &=
\scalebox{0.9}{$\displaystyle
    \frac{1}{|\mathcal{I}_n^{(U)}|}
    \sum_{\bs i \in \mathcal{I}_n^{(U)}}
    \left\{
        -\frac{1}{\beta}w_{\bs i}\mu_{\bs \theta}(\bs X_{\bs i})^{\beta}
        +
        \frac{1}{1+\beta}\mu_{\bs \theta}(\bs X_{\bs i})^{1+\beta}
    \right\}
    +C_{\beta}^{(U)}$},
    \label{eq:l_beta} 
\end{align}
respectively, where 
\begin{align*}
C_{\text{Logistic}}^{(U)}
&:=
\frac{1}{|\mathcal{I}_n^{(U)}|}\sum_{\bs i \in \mathcal{I}_n^{(U)}}\left\{
w_{\bs i} \log w_{\bs i} +
(1-w_{\bs i}) \log (1-w_{\bs i}) \right\}, \\
C_{\text{KL}}^{(U)}
&:=
\frac{1}{|\mathcal{I}_n^{(U)}|}\sum_{\bs i \in \mathcal{I}_n^{(U)}}\left\{
w_{\bs i} \log w_{\bs i} - w_{\bs i} \right\}, \quad
C_{\beta}^{(U)}
:=
\frac{1}{|\mathcal{I}_n^{(U)}|} \sum_{\bs i \in \mathcal{I}_n^{(U)}}\frac{w_{\bs i}^{1+\beta}}{\beta(1+\beta)}
\end{align*}
are constants independent of the parameter $\bs \theta$. 
By utilizing these loss functions (\ref{eq:l_logistic})--(\ref{eq:l_beta}), and sets 
\begin{align*}
    \mathcal{A}(p,K)
    &:=
    \{\bs \theta=(\theta_{ij}) \in \mathbb{R}^{p \times K} \mid \theta_{ij} \geq 0, \: \forall (i,j) \in [p] \times [K]\}, \\
    \mathcal{F}(p,K)
    &:=
    \{\bs \theta \mid \bs \theta \text{ is a parameter for the vector-valued neural network } \bs f_{\bs \theta}:\mathbb{R}^p \to \mathbb{R}^K \}, \\
    \mathcal{C}(n_1,n_2,\ldots,n_U)
    &:=
    \{\bs i=(i_1,i_2,\ldots,i_U) \mid i_1=1,2,\ldots,n_1; \\
    &\hspace{4em} i_2=n_1+1,n_1+2,\ldots,n_1+n_2; \cdots; i_U=\sum_{u=1}^{U-1}n_u+1,\ldots,\sum_{u=1}^{U}n_u\}, 
\end{align*}
various existing methods can be regarded as the BHLR family members, as shown in the following Table~\ref{table:bhlr_family_members}. 
A detailed explanation of the BHLR family members are provided in Section~\ref{subsec:u1} for $U=1$, Section~\ref{subsec:u2} for $U=2$, and Section~\ref{subsec:u_general} for $U \geq 2$. 
Other related works are explained in Section~\ref{subsec:other_related_works}.

\begin{table}[htbp]
\scalebox{0.75}{
\begin{tabular}{c|ccccccc}
\hlineB{2.5}
 & Method & $\mathcal{S}$ & $\varphi$ & $\mu_{\bs \theta}(\bs X_{\bs i})$ & $\bs \Theta$ & $\mathcal{I}_n^{(U)}$ & $\{\bs x_i\}_{i=1}^{n}$ \\
\hlineB{2.5}
\multirow{4}{*}{$U=1$}
 & Poisson reg.~\citep{cameron2007essentials} & $\mathbb{R}_{\geq 0}$ & $\varphi_{\text{KL}}$ & $\exp(\bs \theta^{\top}\bs x_{i_1})$ or $\exp(f_{\bs \theta}(\bs x_{i_1}))$ & $\mathbb{R}^p$ or $\mathcal{F}(p,1)$ & $[n]$ & observed \\
 & Logistic reg.~\citep{bishop2006pattern} & $[0,1]$ & $\varphi_{\text{Logistic}}$ & $\sigma(\bs \theta^{\top}\bs x_{i_1})$ or $\sigma(f_{\bs \theta}(\bs x_{i_1}))$ & $\mathbb{R}^p$ or $\mathcal{F}(p,1)$ & $[n]$ & observed \\
 & LS reg.~\citep{bishop2006pattern} & $\mathbb{R}$ & $\varphi_{\text{Quad.}}$ & $\bs \theta^{\top}\bs x_{i_1}$ or $f_{\bs \theta}(\bs x_{i_1})$ & $\mathbb{R}^p$ or $\mathcal{F}(p,1)$ & $[n]$ & observed \\
 & PBDR~\citep{zhang2009new} & any & any$^{\dagger}$ & $g(\bs \theta^{\top}\bs x_i)$ for some $g$ & $\mathbb{R}^p$ & $[n]$ & observed \\
\hline
\multirow{7}{*}{
$U=2$
}
 & Matrix Fact.~\citep{koren2009matrix} & any & any$^{\dagger}$ & $\langle \bs \theta^{\top}\bs x_{i_1},\bs \theta^{\top}\bs x_{i_2} \rangle$ & $\mathbb{R}^{(n_1+n_2) \times K}$ & $\mathcal{C}(n_1,n_2)$ & $1$-hot $\in \{0,1\}^{n_1+n_2}$ \\
 & NMF~\citep{cichocki2009nonnegative} & any & any$^{\dagger}$ & $\langle \bs \theta^{\top}\bs x_{i_1},\bs \theta^{\top}\bs x_{i_2} \rangle$ & $\mathcal{A}(n_1+n_2,K)$ & $\mathcal{C}(n_1,n_2)$ & $1$-hot $\in \{0,1\}^{n_1+n_2}$ \\
 & LINE~\citep{tang2015line} & $[0,1]$ & $\varphi_{\text{Logistic}}$ & $\sigma(\langle \bs f_{\bs \theta}(\bs x_{i_1}),\bs f_{\bs \theta}(\bs x_{i_2})\rangle)$ & $\mathcal{F}(p,K)$ & any & $1$-hot $\in \{0,1\}^{n}$ \\
 & KL-GE~\citep{okuno2018probabilistic} & $\mathbb{R}_{\geq 0}$ & $\varphi_{\text{KL}}$ & $\exp(\langle \bs f_{\bs \theta}(\bs x_{i_1}),\bs f_{\bs \theta}(\bs x_{i_2}) \rangle)$ & $\mathcal{F}(p,K)$ & any & observed \\
 & $\beta$-GE~\citep{okuno2019robust} & $\mathbb{R}_{\geq 0}$ & $\varphi_{\beta}$ & $\exp(\langle \bs f_{\bs \theta}(\bs x_{i_1}),\bs f_{\bs \theta}(\bs x_{i_2}) \rangle)$ & $\mathcal{F}(p,K)$ & any & observed \\
 & Poincar{\'e} Emb.~\citep{nickel2017poincare} & $[0,1]$ & $\varphi_{\text{Logistic}}$ & $\sigma(-d_{\text{Poincar{\'e}}}(\bs f_{\bs \theta}(\bs x_{i_1}),\bs f_{\bs \theta}(\bs x_{i_2})))$ & $\mathcal{F}(p,K)$ & any & 1-hot $ \in \{0,1\}^n$ \\
 & SBM~\citep{holland1983stochastic} & $[0,1]$ & $\varphi_{\text{Logistic}}$ & $\theta_1 \bs 1(x_{i_1}=x_{i_2})+\theta_2 \bs 1(x_{i_1} \neq x_{i_2})$ & $[0,1]^2$ & $[n]^2$ & cluster indicator $ \in [C]$ \\
\hline 
\multirow{2}{*}{$U \geq 2$}
 & PARAFAC~\citep{bro1997parafac} & any & any$^{\dagger}$ & $\langle \bs \theta^{\top}\bs x_{i_1},\bs \theta^{\top}\bs x_{i_2},\ldots,\bs \theta^{\top}\bs x_{i_U} \rangle$ & $\mathbb{R}^{(\sum_{u=1}^{U} n_u) \times K}$ & $\mathcal{C}(n_1,n_2,\cdots,n_U)$ & $1$-hot $\in \{0,1\}^{\sum_{u=1}^{U}n_u}$ \\
 & NTF~\citep{cichocki2009nonnegative} & any & any$^{\dagger}$ & $\langle \bs \theta^{\top}\bs x_{i_1},\bs \theta^{\top}\bs x_{i_2},\ldots,\bs \theta^{\top}\bs x_{i_U} \rangle$ & $\mathcal{A}(\sum_{u=1}^{U}n_u,K)$ & $\mathcal{C}(n_1,n_2,\cdots,n_U)$ & $1$-hot $\in \{0,1\}^{\sum_{u=1}^{U}n_u}$ \\
\hlineB{2.5}
\end{tabular}
} \\
{\small $^{\dagger}$domain of the generating function $\varphi$ should include the set $\mathcal{S}$.}
\caption{BHLR family members.}
\label{table:bhlr_family_members}
\end{table}

\subsection[U1]{$U=1$}
\label{subsec:u1}

\begin{itemize} 
\item \textbf{Least-squares~(LS) regression}~\citep{bishop2006pattern} minimizes $-\sum_{i_1 \in \mathcal{I}_n^{(1)}}
\log p_{\text{Norm}}(w_{i_1} \mid \mu_{\bs \theta}(\bs X_{i_1}))$ 
using the normal probability density function $p_{\text{Norm}}(w \mid \mu):=\frac{1}{\sqrt{2\pi}}\exp(-\frac{(w-\mu)^2}{2})$ for learning $\mu_{\bs \theta}(\bs X_{i_1})=f_{\bs \theta}(\bs x_{i_1})$. 
LS regression is equivalent to minimizing $L_{\varphi_{\text{Quad.}},n}(\bs \theta)$, and similarly, \textbf{logistic regression}~\citep{bishop2006pattern} and 
\textbf{Poisson regression}~\citep{cameron2007essentials} minimize $L_{\varphi_{\text{Logistic}}}(\bs \theta)$ and $L_{\varphi_{\text{KL}}}(\bs \theta)$, respectively. 
The regression function $f_{\bs \theta}:\mathbb{R}^p \to \mathbb{R}$ used in the regression methods above can be specified arbitrarily. 
Whereas linear transformation $\bs \theta^{\top}\bs x_i \in \mathbb{R}$ is typically used~\citep{zhang2010penalized}, NNs are incorporated currently for enhancing the expressive power of the regression function. 

\item 
\textbf{Parametric Bregman-divergence regression~(PBDR)}~\citep{zhang2009new} generalizes Poisson regression, logistic regression and least squares~(LS) regression; it is equivalent to the BHLR equipped with arbitrary generating functions $\varphi$ and functions $\mu_{\bs \theta}(\bs X_{\bs i})$ in the form of $g(\bs \theta^{\top}\bs x_{i_1})$ for some function $g$. 
The PBDR is a special case of the BHLR. 
However, PBDR considers only the limited form of functions $\mu_{\bs \theta}(\bs X_{\bs i})$, whereas BHLR can employ arbitrary function including neural networks. 
\end{itemize}

\subsection[U2]{$U=2$}
\label{subsec:u2}

\begin{itemize}

\item 
\textbf{Matrix factorization~(MF)}~\citep{koren2009matrix} decomposes a given matrix $\bs V=(v_{\bs j}) \in \mathbb{R}^{n_1 \times n_2}$ into matrices $\bs \xi^{(u)} \in \mathbb{R}^{n_u \times K} \: (u=1,2)$, 
by minimizing the BD between entries of $\bs V$ and those of $\bs \xi^{(1)} \bs \xi^{(2)\top}$. 
Subsequently, we can expect that $\bs V \approx \bs \xi^{(1)} \bs \xi^{(2)\top}$.

Here, we briefly explain that the BHLR includes MF as a special case, 
by considering link weights 
\begin{align}
    \bs W=(w_{\bs i})
    =
    \left(\begin{array}{cc}\bs O_{n_1 \times n_1} & \bs V \\
\bs V^{\top} & \bs O_{n_2 \times n_2} \end{array}\right),
\label{eq:def_V_to_W}
\end{align}
and ($n_1+n_2$)-dimensional $1$-hot data vectors $\{\bs x_i\}_{i=1}^{n_1+n_2}$.

Using the parameter $\bs \theta=(\bs \xi^{(1)\top},\bs \xi^{(2)\top})^{\top} \in \mathbb{R}^{(n_1+n_2) \times K}$ and an index set $\mathcal{C}(n_1,n_2):=\{(i_1,i_2) \mid i_1=1,2,\ldots,n_1;i_2=n_1+1,n_1+2,\ldots,n_1+n_2\}$, it holds that
\begin{align}
&D_{\varphi}(\{v_{\bs j}\}_{\bs j \in [n_1] \times [n_2]},\{(\bs \xi^{(1)} \bs \xi^{(2)\top})_{\bs j}\}_{\bs j \in [n_1] \times [n_2]}) \nonumber \\
&\hspace{10em}=
D_{\varphi}(\{w_{\bs i}\}_{\bs i \in \mathcal{C}(n_1,n_2)} , \{\langle \bs \theta^{\top} \bs x_{i_1},\bs \theta^{\top}\bs x_{i_2} \rangle \}_{\bs i \in \mathcal{C}(n_1,n_2)}),
 \label{eq:nmf_and_bhlr}
\end{align}
where $v_{\bs j}$ and $w_{\bs i}$ represent elements of the matrices $\bs V$ and $\bs W$ respectively.
Thus, MF minimizing the objective on the left-hand side is equivalent to the BHLR minimizing the objective on the right-hand side. 
Although MF employs the quadratic loss $L_{\varphi_{\text{Quad.}},n}(\bs \theta)$ in many cases, MF is in fact defined with an arbitrary BD~\citep{cichocki2009nonnegative}. 

MF~($U=2$) can be generalized to  $U \geq 2$, where the generalization is called tensor factorization~(TF). 
We describe TF in the following section, and its relation to the BHLR is described in detail in \ref{app:relation_to_ntf}.

Finally, MF is called a \textbf{non-negative MF~(NMF)}~\citep{cichocki2009nonnegative} if the entries of the decomposed matrices $\bs \xi^{(1)},\bs \xi^{(2)}$ are restricted to be non-negative.

\item \textbf{Graph embedding~(GE)}~\citep{tang2015line,okuno2018probabilistic,nickel2017poincare,okuno2019robust} is a method for representation learning, that trains the transformation $\bs f_{\bs \theta}:\mathcal{X}(\subset \mathbb{R}^p) \to \mathbb{R}^K$ with a user-specified dimension $K \in \mathbb{N}$, such that the link weight $w_{\bs i} \geq 0$ is predicted through $\mu_{\bs \theta}(\bs X_{\bs i})=g(\langle \bs f_{\bs \theta}(\bs x_{i_1}),\bs f_{\bs \theta}(\bs x_{i_2}) \rangle)$. $g:\mathbb{R}^K \times \mathbb{R}^K \to \mathbb{R}$ is a symmetric function, and $\bs \theta$ is a parameter vector to be estimated by minimizing $L_{\varphi_{\text{Logistic}},n}(\bs \theta)$ with sigmoid function $g(\cdot)=\sigma(\cdot)$ in \textbf{large-scale information network embedding~(LINE)}~\citep{tang2015line}, and 
$L_{\varphi_{\text{KL}},n}(\bs \theta)$ with $g(\cdot)=\exp(\cdot)$ in $1$-view version of probabilistic multi-view graph embedding~\citep{okuno2018probabilistic}, which we denote as KL-GE herein. 

While these GEs achieved outstanding success, the observed link weights may contain noise in practice that may degrade the GE's performance; \textbf{$\beta$-GE}~\citep{okuno2019robust} minimizes $L_{\varphi_{\beta},n}(\bs \theta)$ associated with $\beta$-divergence for learning the similarity function $\mu_{\bs \theta}(\bs X_{\bs i})$ robustly from  noisy link weights.

The GEs above are special cases of the BHLR. 
Once the estimator $\hat{\bs \theta}_{\varphi,n}$ for GE is obtained, we may compute \emph{feature vectors} $\bs y_i := \bs f_{\hat{\bs \theta}_{\varphi,n}}(\bs x_i)$, $(i=1,2,\ldots,n)$. 
Applying further statistical analysis methods such as visualization, clustering, and discriminant analysis to the obtained feature vectors $\{\bs y_i\}_{i=1}^{n}$ has demonstrated  empirically  better performance than using the original data vectors $\{\bs x_i\}_{i=1}^{n}$.

Many GEs employ the IPS model $\langle \bs f_{\bs \theta}(\bs x_{i_1}),\bs f_{\bs \theta}(\bs x_{i_2}) \rangle$ equipped with a vector-valued NN $\bs f_{\bs \theta}$ in their similarity function $\mu_{\bs \theta}$. 
In terms of its expressive power, \citet{okuno2018probabilistic} proved that the IPS approximates any PD similarity $g^{(\text{PD})}(\bs x_{i_1},\bs x_{i_2})$ arbitrarily well. 
However, non-PD similarities are not expressed by the IPS model, and thus some other similarity models are drawing attention. 
For instance, \citet{nickel2017poincare,nickel2018learning} employ negative Poincar{\'e} distance that can efficiently embed tree-structured graphs.
Furthermore, shifted IPS~(SIPS)~\citep{okuno2019graph} $\langle \bs f_{\bs \theta}(\bs x_{i_1}),\bs f_{\bs \theta}(\bs x_{i_2}) \rangle + u_{\bs \theta}(\bs x_{i_1})+u_{\bs \theta}(\bs x_{i_2})$ is proposed for GE by introducing the bias terms using a NN $u_{\bs \theta}:\mathcal{X} \to \mathbb{R}$, and it has been proven to approximate a wider class called conditionally PD similarities that include PD similarities and various non-PD similarities, such as negative Poincar{\'e} distance. 
Recently \citet{kim2019representation} proposed the weighted inner product similarity~(WIPS) for approximating general similarities including PD and conditionally PD similarities as special cases.

\item 
\textbf{Stochastic block model~(SBM)}~\citep{holland1983stochastic} considers a graph for which each node $i \in [n]$ is associated with the cluster index $x_i \in [C]$. The SBM learns $\theta_1, \theta_2 \in [0,1]$, representing probabilities that a link exists between two nodes belonging to the same cluster and different clusters, respectively. 
As the probability $\mathbb{P}(w_{\bs i}=1 \mid \bs X_{\bs i})$ is expressed as $\mu_{\bs \theta}(\bs X_{\bs i}):=\theta_1 \bs 1(x_{i_1}=x_{i_2})+\theta_2 \bs 1(x_{i_1} \neq x_{i_2})$ and the parameter $\bs \theta=(\theta_1,\theta_2)$ is learned by minimizing $L_{\varphi_{\text{Logistic}},n}(\bs \theta)$, the SBM is a special case of the BHLR.

\end{itemize}

\subsection[general]{$U \geq 2$} 
\label{subsec:u_general}

\begin{itemize}
\item 
\textbf{PARAFAC}~\citep{cichocki2009nonnegative,cong2015tensor}, that is also called TF, CP-decomposition, and CANDECOMP, decomposes a given tensor $\bs V := (v_{\bs j}) \in \mathbb{R}^{n_1 \times n_2 \times \cdots \times n_U}$ into matrices $\bs \xi^{(u)}:=(\xi^{(u)}_{jk}) \in \mathbb{R}^{n_u \times K} \: (u \in [U])$, by minimizing the BD between entries of $\bs V$ and $[\![\bs \xi^{(1)},\bs \xi^{(2)},\ldots,\bs \xi^{(U)}]\!]$ whose $\bs j=(j_1,j_2,\ldots,j_U)$-th entry is specified as  $\sum_{k=1}^{K}\xi^{(1)}_{j_1 k} \xi^{(2)}_{j_2 k} \cdots \xi^{(U)}_{j_U k}$. 
Subsequently, we can expect that $\bs V \approx [\![\bs \xi^{(1)},\bs \xi^{(2)},\ldots,\bs \xi^{(U)}]\!]$. 
TF $(U \geq 2)$ generalizes the MF~($U=2$) explained in Section~\ref{subsec:u2} because $[\![\bs \xi^{(1)},\bs \xi^{(2)}]\!]=\bs \xi^{(1)}\bs \xi^{(2)\top}$. Similar to MF, TF is a special case of the BHLR. See \ref{app:relation_to_ntf} for details.

\item PARAFAC is called a \textbf{non-negative tensor factorization~(NTF)}~\citep{cichocki2009nonnegative,kolda2009tensor} or non-negative PARAFAC, if the entries of the decomposed matrices $\bs \xi^{(u)} \: (u \in [U])$ are restricted to be non-negative. 
Although this PARAFAC-based NTF can be applied to general $U \in \mathbb{N}$~\citep{kolda2009tensor}, 
many different types of NTFs have been developed especially for $U=3$; 
by referring to \citet{cichocki2009nonnegative} p.54 Table 1.2, 
NTF1, NTF2~\citep{cichocki2006ntflab}, and shifted NTF~\citep{harshman2003shifted} decompose a given tensor into $2$ matrices and a tensor, and convolutive NTF~(CNTF) and C2NTF~\citep{morup2006sparse} decompose the tensor into a matrix and $2$ tensors. 
\end{itemize}

\subsection{Other Related Works}
\label{subsec:other_related_works}
In this section, some other related works are listed. Please also see \ref{app:remaining_related_works} for the remaining related works.

\begin{itemize}
\item[] For $U=1$, the MLE of a \textbf{generalized linear model}~\citep{bishop2006pattern} and the BHLR are almost the same; however, they do not exhibit inclusion, as explained in Section~\ref{subsec:relation_bhlr_exponential}.

\item[] For $U=2$, \textbf{Locality preserving projections~(LPP)}~\citep{he2004locality} computes a low-dimensional linearly transformed feature vectors $\bs y_{i}=\bs A^{\top}\bs x_i \: (i=1,2,\ldots,n)$ by considering link weights $w_{i_1 i_2} \geq 0$. 
\textbf{Cross-Domain Matching Correlation Analysis~(CDMCA)}~\citep{shimodaira2016cross} is a multiview extension of LPP.
Considering that 
(i) LPP can be regarded as $1$-view CDMCA and 
(ii) CDMCA is a quadratic approximation of multiview KL--GE equipped with linear transformations, as shown in \citet{okuno2018probabilistic} section 3.6, LPP is a quadratic approximations of KL--GE that is included in the BHLR. 
LPP reduces to \textbf{spectral graph embedding}~\citep{chung1997spectral} if the data vectors are $1$-hot.

\item[] For $U \ge 2$, 
\textbf{Hypergraph Incidence Matrix Factorization~(HIMFAC)}~\citep{nori2012multinomial} computes the linear transformation of given data vectors by considering the observed hyperlinks defined for $U$-tuples. 
HIMFAC consists of the following two steps: (i) for $i,i' \in [n]$, HIMFAC first counts the number $v_{ii'}$ of hyperlinks that both data vectors $\bs x_i,\bs x_{i'}$ belong; 
(ii) by regarding $\bs V=(v_{ii'})$ as a new adjacency matrix of data vectors, HIMFAC computes the LPP~\citep{he2004locality} if the link weight is defined among a single type of data, and CDMCA~\citep{shimodaira2016cross} for multiple types of data~(e.g., text, images, etc.). 
Similarly to LPP explained above~$(U=2)$, HIMFAC can be regarded as a quadratic approximation of BHLR~$(U=2)$, though the hyperlink weights $U \geq 2$ are converted into link weights $U=2$ through the preprocessing step (i). 

\end{itemize}

\section{BHLR Properties}
\label{sec:properties_of_bhlr}

In this section, we show two favorable properties of BHLR. 
The first property (P-1) statistical consistency: the BHLR asymptotically recovers the true conditional expectation of link weights, is explained in Section~\ref{subsec:expectation_consistency}. Additionally, we explain the second property (P-2) computational tractability: the BHLR can be efficiently computed by stochastic algorithms in Section~\ref{subsec:propsoed_algorithm}.

\subsection{BHLR Asymptotically Recovers True Conditional Expectations}
\label{subsec:expectation_consistency}

In this section, we demonstrate via Theorem~\ref{theo:consistency} that the similarity function $\mu_{\hat{\bs \theta}_{\varphi,n}}(\bs X_{\bs i})$ estimated by the BHLR asymptotically recovers the true conditional expectation $\mu_*(\bs X_{\bs i})=\mathbb{E}(w_{\bs i} \mid \bs X_{\bs i})$. 
For proving the asymptotic properties of BHLR in Proposition~\ref{prop:bregman_converge} and Theorem~\ref{theo:consistency}, only in this section, we specify the increasing order index set as
\begin{align}
\mathcal{J}_n^{(U)}=\{\bs i \in [n]^U \mid 1 \leq i_1 <i_2 <\cdots <i_U \leq n \},
\label{eq:inu}
\end{align}
such that it includes all the possible combinations of $U$ different entries $i_1,i_2,\ldots,i_U \in [n]$, 
whereas no two distinct indices $\bs i,\bs i' \in \mathcal{J}_n^{(U)}$ are obtained from each other by permutation.  Then, hyperlink weights $\{w_{\bs i}\}_{\bs i \in \mathcal{J}_n^{(U)}}$ are free from the symmetry constraints described in Section~\ref{subsec:problem_setting}; the underlying conditional distribution of $w_{\bs i} \mid \bs X_{\bs i}$ can be defined without the constraints, thus making the theoretical development easier.

In the following, we list conditions (C-1)--(C-5) needed for theoretical development. 
$w$ represents a random variable that follows a cpdf (or cpmf) $q$ of $w \mid \bs X$, for $\bs X=(\bs x,\bs x',\bs x'',\ldots) \in \mathcal{X}^U$. 

\begin{enumerate}
\item[(C-1)]
$\bs \Theta$ is compact.

\item[(C-2)]
Real-valued functions $\mu_{\bs \theta}(\bs X)$ and $\mu_{*}(\bs X):=\mathbb{E}(w \mid \bs X)$ are continuous on $\bs \Theta \times \mathcal{X}^U$ and $\mathcal{X}^U$, respectively. 
Especially, the function $\mu_{\bs \theta}(\bs X)$ is Lipschitz continuous on $\bs \Theta$ for each $\bs X \in \mathcal{X}^U$. 

\item[(C-3)]
Hyperlink weights $\{w_{\bs i}\}_{\bs i \in \mathcal{I}_n^{(U)}}$ follow a distribution whose cpdf (or cpmf) are specified as $\prod_{\bs i \in \mathcal{I}_n^{(U)}} q(w_{\bs i} \mid \bs X_{\bs i})$, and 
data vectors $\bs x_1,\bs x_2,\ldots,\bs x_n$ i.i.d. follow a pdf $q_X$, where the support of $q_X$ is compact.

\item[(C-4)]
$\mathbb{E}(w^2 \mid \bs X)<\infty$ and $\mathbb{E}(\varphi(w)^2 \mid \bs X)<\infty$ for all $\bs X \in \mathcal{X}^U$.

\item[(C-5)]
$\varphi$ is $C^2$ and strongly convex.
\end{enumerate}

It is noteworthy that all the functions listed in Table~\ref{table:bregman} satisfy the condition (C-5); all the conditions (C-1)--(C-5) are not difficult to satisfy in practice.  
Using these conditions, we demonstrate in the following Proposition~\ref{prop:bregman_converge} that $L_{\varphi,n}(\bs \theta)$ empirically approximates the expected value of $
    d_{\varphi}(\mu_*(\bs X),\mu_{\bs \theta}(\bs X))$ up to a constant.

\begin{prop}
\label{prop:bregman_converge} 
Let $U \in \mathbb{N}$, $\mathcal{I}_n^{(U)}=\mathcal{J}_n^{(U)}$ defined  in eq.~(\ref{eq:inu}) and suppose that (C-1)--(C-5) hold. Let $\mathbb{E}_{\mathcal{X}^U}$ represent the expectation with respect to the density of the $U$-tuple $\bs X=(\bs x,\bs x',\bs x'',\ldots) \in \mathcal{X}^U$; more specifically, $\bs x,\bs x',\bs x'',\ldots$ i.i.d. follow a pdf $q_X$.
Then, for $n \to \infty$, it holds that
\[L_{\varphi,n}(\bs \theta)
    =
\mathbb{E}_{\mathcal{X}^U}\left(
    d_{\varphi}(\mu_*(\bs X),\mu_{\bs \theta}(\bs X))
\right)
+
C_{\varphi}
+
O_p(1/\sqrt{n})
\]
for each $\bs \theta \in \bs \Theta$, where 
$C_{\varphi}:=\mathbb{E}_{\mathcal{X}^U}\left(
    \mathbb{E}(\varphi(w)\mid \bs X)
    -
    \varphi(\mu_*(\bs X))
\right)$ is a constant independent of the parameter $\bs \theta$.
\end{prop}
Proof is obtained by applying the law of large numbers for multiple indexed partially dependent random variables. 
See \ref{app:proof_of_prop:bregman_converge} for details.

As explained in Section~\ref{subsec:conditional_dist_hyperlink}, 
different tuples $\bs X_{\bs i},\bs X_{\bs i'}$ may be constrained as they may share some data vectors, 
even if data vectors $\bs x_i$ are i.i.d. generated; theories for HLR can be different from those of classical regression, that predicts response variables from i.i.d. explanatory variables. 
Due to the constraint, the convergence rate of the loss function for BHLR is $O(1/\sqrt{n})$ whereas the estimation leverages $|\mathcal{I}_n^{(U)}|=O(n^U)$ samples. The convergence rate is similar to $U$-statistic~\citep{lee1990u}, and is different from the rate $O(1/\sqrt{n^U})$ for classical regression using $O(n^U)$ i.i.d. data vectors. 
In addition, Proposition~\ref{prop:bregman_converge} with $\beta$-div. listed in Table~\ref{table:bregman} and $U=2$ corresponds to a special case $(\varepsilon=0)$ of Theorem~3.1 in \citet{okuno2019robust} that indicates the convergence of the GE's loss function using $\beta$-divergence.

Proposition~\ref{prop:bregman_converge} leads to the following Theorem~\ref{theo:consistency}, which claims that the estimated model $\mu_{\hat{\bs \theta}_{\varphi,n}}$ converges to $\mu_*$ in probability, 
by considering that $d_{\varphi}(\mu_*(\bs X),\mu_{\bs \theta}(\bs X))$ with fixed $\mu_*(\bs X)$ is minimized if $\mu_{\bs \theta}(\bs X)=\mu_*(\bs X)$.

\begin{theo}
\label{theo:consistency}
The symbols and conditions are the same as those of Proposition~\ref{prop:bregman_converge} except for the additional condition: there exists $\bs \theta_* \in \bs \Theta$ such that $\mu_{\bs \theta*}=\mu_*$. 
Using a norm $\|f\|:=\mathbb{E}_{\mathcal{X}^U}(f(\bs X)^2)^{1/2}$ defined for functions $f:\mathcal{X}^U \to \mathbb{R}$, it holds that
\begin{align}
    \|\mu_*-\mu_{\hat{\bs \theta}_{\varphi,n}}\|
    &\overset{p}{\to}
    0,
    \quad
    (n \to \infty),
    \label{eq:mu_star_approximates}
\end{align}
where $\hat{\bs \theta}_{\varphi,n}$ is the estimator~(\ref{eq:estimator}) computed with $n$ data vectors $\{\bs x_i\}_{i=1}^{n}$ and their hyperlink weights $\{w_{\bs i}\}_{\bs i \in \mathcal{I}_n^{(U)}}$. 
\end{theo}

Proof is provided in \ref{app:proof_of_theo:consistency}.
As indicated in Theorem~\ref{theo:consistency} above, the estimated similarity function $\mu_{\hat{\bs \theta}_{\varphi,n}}$ asymptotically recovers the underlying expectation function $\mu_*$ in probability, regardless of the choice of $\varphi$.
Thus, the BHLR is statistically consistent.

Interestingly, Theorem~\ref{theo:consistency} does not rely on the underlying conditional distribution of hyperlink weights; BHLR is also robust against the distributional misspecification for the weights, as long as the set of user-specified similarity functions $\{\mu_{\bs \theta}(\bs X)\}_{\bs \theta \in \bs \Theta}$ includes the conditional expectation $\mu_*(\bs X):=\mathbb{E}(w \mid \bs X)$ therein.

Note that a similar property is already known for exponential linear regression models~(e.g., Poisson regression model), that correspond to BHLR with $U=1$. 
See \citet{cameron2013regression} Section 2.4.2 and 3.2.3 for details.

\subsection{BHLR can be Efficiently Computed by Stochastic Algorithm}
\label{subsec:propsoed_algorithm}
In this section, we discuss the optimization for the BHLR. 
We first consider applying the classical fullbatch gradient descent (GD), i.e., GD using all data for computing gradients to obtain the estimator (\ref{eq:estimator}). Subsequently, we demonstrate that the fullbatch-based methods require considerable computational cost when considering $U \geq 2$. 
For reducing the computational complexity, we introduce an efficient algorithm based on minibatch stochastic GD~(SGD), i.e., GD using a sampled small dataset for computing gradients. 
Furthermore, we prove the asymptotics of the minibatch SGD, and demonstrate that it increases the ROC--AUC test score in our numerical experiments. 

For notational simplicity, $n,U \in \mathbb{N}$, generating function $\varphi$, index set $\mathcal{I}_n^{(U)}(\neq \emptyset) \subset [n]^U$, hyperlink weights $\{w_{\bs i}\}_{\bs i \in \mathcal{I}_n^{(U)}}$, and data vectors $\{\bs x_i\}_{i=1}^{n}$ are fixed in this section. 
It is noteworthy that the index set $\mathcal{I}_n^{(U)} \subset [n]^U$ can be arbitrary specified hereinafter, whereas the set  $\mathcal{I}_n^{(U)}$ was restricted to have a specific form $\mathcal{J}_n^{(U)}=(\ref{eq:inu})$ in the previous Section~\ref{subsec:expectation_consistency} for making the theory easier.
For example, both $(1,2)$ and $(2,1)$ can be included in $\mathcal{I}_n^{(2)}$ while only $(1,2)$ was included in $\mathcal{J}_n^{(2)}$.

We begin by obtaining the estimator (\ref{eq:estimator}) by applying the fullbatch GD with $T \in \mathbb{N}$ iterations started from a randomly initialized vector $\bs \theta^{(1)}$:
\begin{align}
    \bs \theta^{(t+1)}:=
    \mathcal{Q}_{\bs \Theta}
    \left(
    \bs \theta^{(t)} 
    -
    \gamma^{(t)} 
    g(\bs \theta^{(t)})
    \right)
    ,
    \quad
    t=1,2,\ldots,T,
    \label{eq:fullbatch_gd}
\end{align}
where $\{\gamma^{(t)}\}_{t=1,2,\ldots,T} \subset \mathbb{R}_{>0}$ are step sizes, 
$g(\bs \theta)$ is the gradient function, and
$\mathcal{Q}_{\bs \Theta}(\bs \theta):=\argmin_{\bs \theta' \in \bs \Theta}\|\bs \theta'-\bs \theta\|_2$
is the projection to the parameter space. 
This projection is required for ensuring that the estimator $\hat{\bs \theta}^{(t)}$ is included in the parameter space $\bs \Theta$; the projection can be ignored if $\bs \Theta=\mathbb{R}^p$. 
The gradient function is expressed as
\begin{align}
    g(\bs \theta)
    :=
    \frac{\partial L_{\varphi,n}(\bs \theta)}{\partial \bs \theta} 
    &=
    \frac{1}{|\mathcal{I}_n^{(U)}|}
    \bigg\{
    \sum_{\bs i \in \mathcal{I}_n^{(U)}}
    \mu_{\bs \theta}(\bs X_{\bs i})
    \varphi''(\mu_{\bs \theta}(\bs X_{\bs i}))
    \frac{\partial \mu_{\bs \theta}(\bs X_{\bs i})}{\partial \bs \theta} \nonumber \\
    &\hspace{10em}
    -
    \sum_{\bs i \in \mathcal{P}_n^{(U)}}
    w_{\bs i}
    \varphi''(\mu_{\bs \theta}(\bs X_{\bs i}))
    \frac{\partial \mu_{\bs \theta}(\bs X_{\bs i})}{\partial \bs \theta}
    \bigg\},
    \label{eq:derivative_of_ell}
\end{align}
where $\mathcal{P}_n^{(U)}:=\{\bs i \in \mathcal{I}_n^{(U)} \mid w_{\bs i} \neq 0\}$ is a set of indices whose corresponding weights are non-zero. 
After the $T$ iterations, $\bs \theta^{(T+1)}$ converges to the estimator (\ref{eq:estimator}) as $T \to \infty$ under some assumptions~\citep{dunn1981global}. 
However, computing the gradient (\ref{eq:derivative_of_ell}) requires considerable computational cost $O(|\mathcal{I}_n^{(U)}|)=O(n^U)$; the significant computational complexity is non-negligible especially for $U \geq 2$.

For efficiently computing the estimator (\ref{eq:estimator}), we alternatively employ minibatch SGD~\citep{ruder2016overview} that iteratively updates the parameter as
\begin{align}
    \tilde{\bs \theta}^{(t+1)} := 
    \mathcal{Q}_{\bs \Theta}
    \left(\tilde{\bs \theta}^{(t)} 
    -
    \gamma^{(t)} 
    \tilde{g}_{\eta}^{(t)}(\tilde{\bs \theta}^{(t)})\right),
    \quad
    t=1,2,\ldots,T,
    \label{eq:minibatch_sgd}
\end{align}
where $\tilde{g}_{\eta}^{(t)}(\bs \theta)$ is a stochastic gradient as will be defined in (\ref{eq:minibatch_gradient}) using the sampled small dataset called \emph{minibatch}.

\begin{figure}[!ht]
\centering
\includegraphics[width=0.7\textwidth]{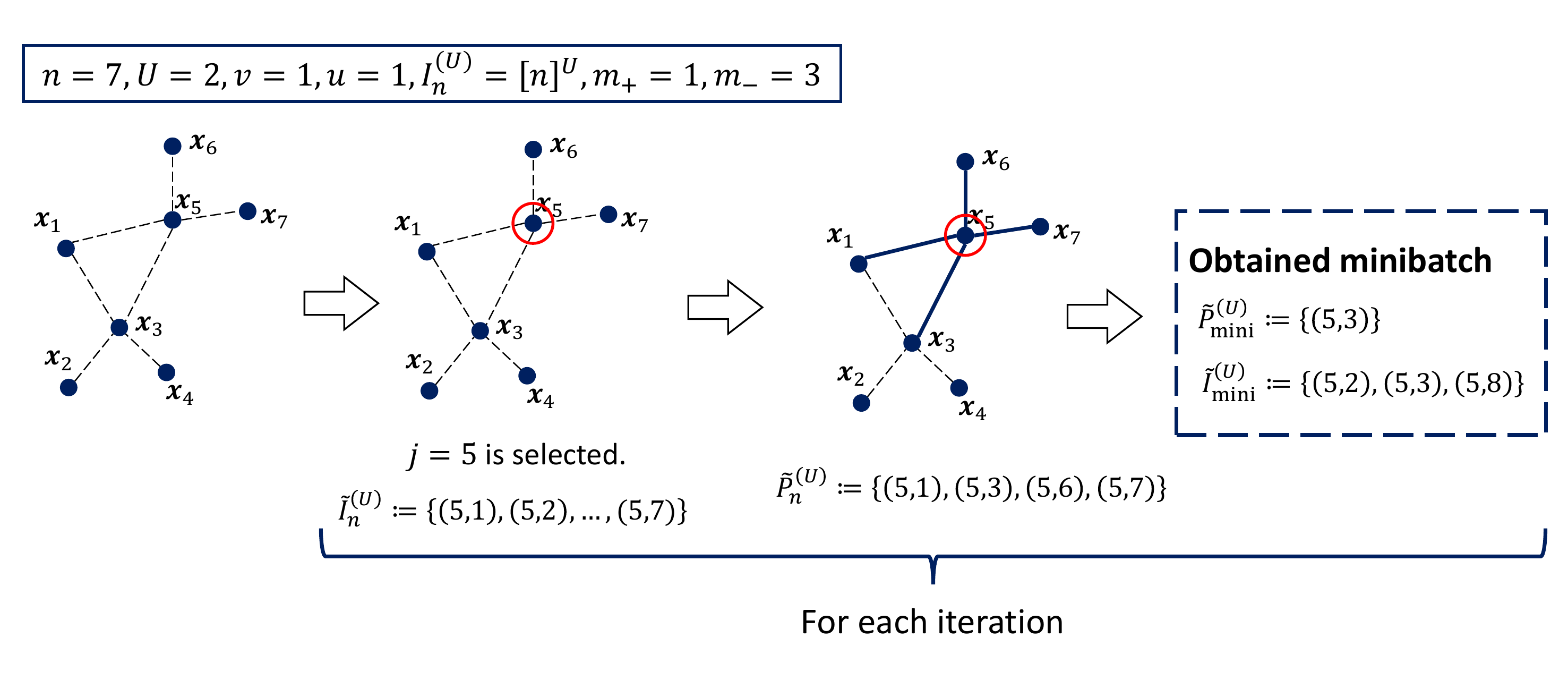}
\caption{
Negative-sampling used in skip-gram~\citep{mikolov2013distributed}, that is  illustrated with $n=7,u=1,\mathcal{I}_n^{(U)}=[n]^U$ in this figure, is a special case ($U=2,v=1$) of the proposed procedure. 
For each iteration, 
(i) $j \in [n]^v(=\{1,2,\ldots,7\})$ is randomly selected; $j=5$ is selected herein. 
(ii) $\tilde{\mathcal{I}}_{n}^{(U)}$ is a set of all the possible indices whose $u(=1)$-th entry is fixed as $j(=5)$, i.e., $\tilde{\mathcal{I}}_{n}^{(U)}=\{(5,1),(5,2),(5,3),...,(5,7)\}$, and $\tilde{\mathcal{P}}_{n}^{(U)}$ represents a set of indices whose corresponding weights are non-zero (shown as dot lines in this figure), i.e., $\tilde{\mathcal{P}}_{n}^{(U)}=\{(5,1),(5,3),(5,6),(5,7)\}$. 
(iii) $m_+,m_-$ entries are randomly selected from sets $\tilde{\mathcal{P}}_n^{(U)},\tilde{\mathcal{I}}_n^{(U)}$, and denote the sets as 
$\tilde{\mathcal{P}}_{\text{mini}}^{(U)},\tilde{\mathcal{I}}_{\text{mini}}^{(U)}$; they are called ``minibatch". 
We further generalize the negative sampling from $U=2$ to arbitrary $U \in \mathbb{N}$.
}
\label{fig:negative_sampling}
\end{figure}

Although minibatch sampling can be easily formulated in the case of $U=1$, several different sampling patterns may occur when $U \geq 2$. 
For instance, when $U=2$, the negative-sampling used in skip-gram~\citep{mikolov2013distributed} first randomly fixes the first entry $i_1$ in the index $\bs i=(i_1,i_2)$ and subsequently samples a minibatch as shown in Figure~\ref{fig:negative_sampling}, whereas the minibatch SGD used in \citet{okuno2018probabilistic} and \citet{okuno2019robust} samples a minibatch without fixing any entries in the index. 
Thus, we unify both of these existing methods in this study, and propose a general procedure for sampling a minibatch that can be used for both $U=1,2$ and $U \geq 3$. 
The proposed general procedure is explained in the following and Algorithm~\ref{alg:minibatch}.

\bigskip
In the proposed procedure, that generalizes negative sampling~($U=2,v=1$) used in skip-gram~\citep{mikolov2013distributed}, we first specify $v \in \{0,1,2,\ldots,U-1\}$, that represents the number of entries in the index $\bs i$ to be fixed. 
$v=0$ indicates that no entry is fixed; we herein consider $v \geq 1$. 
For fixing the entries, we specify $\bs u$ in a set
\begin{align}
\{\bs u=(u_1,u_2,\ldots,u_v) \in [U]^v \mid u_1<u_2<\cdots<u_v\}.
\label{eq:specifying_u}
\end{align}
Then, the proposed procedure is summarized in Algorithm~\ref{alg:minibatch} using a set of $\bs i \in \mathcal{I}_n^{(U)}$ whose $\bs u=(u_1,u_2,\ldots,u_v)$-th entry is fixed as $\bs j=(j_1,j_2,\ldots,j_v) \in [n]^v $, that is
\begin{align}
    \mathcal{I}_{n,\bs u}^{(U)}(\bs j) 
    &:=
    \{\bs i:=(i_1,i_2,\ldots,i_U) \mid \bs i \in \mathcal{I}_n^{(U)},i_{u_1}=j_1,\ldots,i_{u_v}=j_v\}, \quad (\bs j \in [n]^v),
    \label{eq:def_Inuv} 
\end{align}
and a set
\begin{align}
    \mathcal{K}_{\bs u}&:=\{\bs j \in [n]^v \mid \mathcal{I}^{(U)}_{n,\bs u}(\bs j) \neq \emptyset\},
    \label{eq:def_K}
\end{align}
that decomposes the index set as $\mathcal{I}_n^{(U)}=\bigcup_{\bs j \in \mathcal{K}_{\bs u}}\mathcal{I}_{n,\bs u}^{(U)}(\bs j)$ without any overlap. $p_{\bs j}$ represents the probability to choose $\bs j$ from the set $\mathcal{K}_{\bs u}$; 
we employ $p_{\bs j}=1/|\mathcal{K}_{\bs u}|$ later in Theorem~\ref{theo:sgd_convergence}, whereas it can be arbitrarily specified by users in practice. 
The proposed minibatch sampling for hyper-relations is also illustrated in Example~\ref{ex:minibatch}.

\begin{algorithm}[!ht]
\caption{Proposed minibatch sampling procedure $\mathcal{M}_v(\mathcal{I}_n^{(U)},\bs u,\{p_{\bs j}\}_{\bs j \in \mathcal{K}_{\bs u}},m_+,m_-)$.} 
\label{alg:minibatch}
\begin{algorithmic}   
\REQUIRE An index set $\mathcal{I}_n^{(U)} \subset [n]^U$, numbers of minibatch samples $m_+,m_- \in \mathbb{N}$, 
a vector $\bs u=(u_1,u_2,\ldots,u_v)$ in the set (\ref{eq:specifying_u}), 
and probability $p_{\bs j}$ that samples $\bs j$ from the set $\mathcal{K}_{\bs u}$. 
Note that $\bs u,\{p_{\bs j}\}_{\bs j \in \mathcal{K}_{\bs u}}$ are not required for $v=0$.
\IF{$v \geq 1$}
    \STATE Randomly choose a $\bs j$ from the set $\mathcal{K}_{\bs u}$ defined in (\ref{eq:def_K}), with the probability $p_{\bs j}$. 
    \STATE $\tilde{\mathcal{I}}_n^{(U)}:=\mathcal{I}_{n,\bs u}^{(U)}(\bs j)$ defined in (\ref{eq:def_Inuv}).
\ELSIF{$v = 0$}
    \STATE $\tilde{\mathcal{I}}_n^{(U)}:=\mathcal{I}_n^{(U)}$, as $v=0$ indicates no fixed entry in the index $\bs i$. 
\ENDIF
\STATE $\tilde{\mathcal{P}}_n^{(U)}:=\{\bs i \mid \bs i \in \tilde{\mathcal{I}}_n^{(U)},w_{\bs i} \neq 0\}$.
\STATE Choose $m_+,m_-$ entries uniformly and randomly from $\tilde{\mathcal{P}}_n^{(U)},\tilde{\mathcal{I}}_n^{(U)}$, and denote the sets as $\tilde{\mathcal{P}}_{\text{mini}}^{(U)},\tilde{\mathcal{I}}_{\text{mini}}^{(U)}$.
\STATE $s_+:=|\tilde{\mathcal{P}}_{n}^{(U)}|/m_+,s_-:=|\tilde{\mathcal{I}}_{n}^{(U)}|/m_-$.
\ENSURE $(\tilde{\mathcal{P}}_{\text{mini}}^{(U)},\tilde{\mathcal{I}}_{\text{mini}}^{(U)},s_+,s_-)$.
\end{algorithmic}
\end{algorithm}

\begin{ex}[Minibatch sampling for hyper-relations]
\label{ex:minibatch}
    We consider $n=7,U=4,v=2,\mathcal{I}_n^{(U)}=[n]^U,\bs u=(1,3)$, and $\bs j=(2,5)$ is herein randomly selected. 
    We define a set of indices whose $\bs u=(1,3)$-th entry is fixed as $\bs j=(2,5)$, i.e.,  
    \[ 
    \tilde{\mathcal{I}}_n^{(U)}
    =
    \mathcal{I}_{n,\bs u}^{(U)}(\bs j)
    :=
    \{(2,a,5,b) \in \mathcal{I}_n^{(U)} \mid a,b \in \{1,2,\ldots,7\}\},
    \]
    and a set of indices whose corresponding hyperlink weights are non-zero, i.e., $\tilde{\mathcal{P}}_n^{(U)}=\{\bs i \in \tilde{\mathcal{I}}_n^{(U)} \mid w_{\bs i} \neq 0\}$; they are sets of candidate indices to be resampled. 
    We uniformly and randomly choose $m_+,m_-$ indices from sets $\tilde{\mathcal{P}}_n^{(U)},\tilde{\mathcal{I}}_n^{(U)}$, and denote the sets as $\mathcal{P}_{\text{mini}}^{(U)},\mathcal{I}_{\text{mini}}^{(U)}$; they are used for computing the gradient~(\ref{eq:minibatch_gradient}) and update the parameter by (\ref{eq:minibatch_sgd}). 
\end{ex}

It is noteworthy that the sampling procedure in Algorithm~\ref{alg:minibatch} can efficiently pick up non-zero weights even if most of the weights $\{w_{\bs i}\}_{\bs i \in \mathcal{I}_n^{(U)}}$ are zero. 
Similarly to \citet{mikolov2013distributed} and \citet{okuno2019robust}, 
the gradient $g(\bs \theta)$ at the iteration $t$ can be stochastically approximated by
\begin{align}
\tilde{g}_{\eta}^{(t)}(\bs \theta)
&:=
    s^{(t)}_-
    \sum_{\bs i \in \tilde{\mathcal{I}}_{\text{mini}}^{(t)}}
    \mu_{\bs \theta}(\bs X_{\bs i})
    \varphi''(\mu_{\bs \theta}(\bs X_{\bs i}))
    \frac{\partial \mu_{\bs \theta}(\bs X_{\bs i})}{\partial \bs \theta} 
    -
    \eta
    \cdot 
    s^{(t)}_+
    \sum_{\bs i \in \tilde{\mathcal{P}}_{\text{mini}}^{(t)}}
    w_{\bs i}
    \varphi''(\mu_{\bs \theta}(\bs X_{\bs i}))
    \frac{\partial \mu_{\bs \theta}(\bs X_{\bs i})}{\partial \bs \theta},
    \label{eq:minibatch_gradient}
\end{align}
where the minibatch $\mathcal{M}^{(t)}:=(\tilde{\mathcal{P}}_{\text{mini}}^{(t)},\tilde{\mathcal{I}}_{\text{mini}}^{(t)},s_+^{(t)},s_-^{(t)})$ is obtained via Algorithm~\ref{alg:minibatch} and $\eta>0$ is a user-specified parameter. 
The coefficient $s^{(t)}_-=|\tilde{\mathcal{I}}_n^{(U)}|/|\tilde{\mathcal{I}}_{\text{mini}}^{(t)}|$ is needed for adjusting the first term in the stochastic gradient (\ref{eq:minibatch_gradient}), since only the fixed size of minibatch $\tilde{\mathcal{I}}_{\text{mini}}^{(t)}$ is sampled from the set $\tilde{\mathcal{I}}_n^{(U)}$ whose size may depend on the selected $\bs j \in \mathcal{K}_{\bs u}$. 
Similarly, $s^{(t)}_+=|\tilde{\mathcal{P}}_n^{(U)}|/|\tilde{\mathcal{P}}_{\text{mini}}^{(t)}|$ is needed for adjusting the second term. 
Although these coefficients $s^{(t)}_+,s^{(t)}_-$ are required for theoretical development, they may be ignored in practice as explained later.

\bigskip
The computational complexity for the stochastic gradient (\ref{eq:minibatch_gradient}) is $O(m_+ + m_-)$, and it can be significantly less than the complexity $O(n^{U})$ of the fullbatch gradient (\ref{eq:derivative_of_ell}), at least for each iteration. 
Moreover, the minibatch SGD~(\ref{eq:minibatch_sgd}) using (\ref{eq:minibatch_gradient}) reaches approximately the optimal value within a reasonable number of iterations, as will be empirically demonstrated at the last of this section; 
BHLR can be efficiently computed by the minibatch SGD.

The minibatch SGD equipped with Algorithm~\ref{alg:minibatch} and (\ref{eq:minibatch_gradient}), can be applied to general $U \geq 2$ and $v \geq 0$ whereas it encompasses several existing methods; in our context, it reduces to the minibatch SGD using the negative sampling for skip-gram~\citep{mikolov2013distributed} if $(U,v,\varphi,m_+)=(2,1,\varphi_{\text{Logistic}},1)$, 
and it also reduces to \citet{okuno2018probabilistic} and \citet{okuno2019robust} if $(U,v,\varphi)=(2,0,\varphi_{\text{KL}}),(2,0,\varphi_{\beta})$, respectively, where their sampling procedures are called ``negative sampling: unigram''~$(v=1)$ and ``uniform link sampling''~$(v=0)$ in \citet{victor2019empirical}. 
Other major stochastic algorithms such as AdaGrad~\citep{duchi2011adaptive} and Adam~\citep{kingma2014adam} can be employed as well, once the minibatch-based stochastic gradient~(\ref{eq:minibatch_gradient}) is formally defined with Algorithm~\ref{alg:minibatch}.

\bigskip
Hereinafter, we discuss the asymptotics of the minibatch SGD when the number of iterations is sufficiently large, by employing \citet{ghadimi2013stochastic} Theorem 2.1~(a).

Whereas the standard stochastic optimization algorithms preliminary determine the number of iterations $T$, 
for theoretical purposes, 
\citet{ghadimi2013stochastic} randomly choose the number of iterations $\tau$ from the set $[T]=\{1,2,\ldots,T\}$ with the probability $\mathbb{P}(\tau)$, and update the parameter $\bs \theta$ within $\tau$ iterations. 
In this setting, the expectation of the stochastic gradient $\tilde{g}_{\eta}^{(\tau)}(\tilde{\bs \theta}^{(\tau)})$ is proved to approach $\bs 0$ as $T \to \infty$; 
considering the Bregman divergence between the 
hyperlink weights multiplied by a user-specified constant $\eta>0$ and the similarities $\{\mu_{\bs \theta}(\bs X_{\bs i})\}_{\bs i \in \mathcal{I}_n^{(U)}}$, i.e., 
\begin{align}
    Q_{\eta}(\bs \theta)
    &:=
    D_{\varphi}(
\{\eta w_{\bs i}\}_{\bs i \in \mathcal{I}_n^{(U)}},
\{\mu_{\bs \theta}(\bs X_{\bs i})\}_{\bs i \in \mathcal{I}_n^{(U)}}),
\label{eq:q_eta}
\end{align}
we apply \citet{ghadimi2013stochastic} to our setting, and show in the following Theorem~\ref{theo:sgd_convergence} that the gradient of $Q_{\eta}(\bs \theta)$ approaches to $\bs 0$ as $T$ increases.

For applying \citet{ghadimi2013stochastic}, we further assume following conditions (D-1)--(D-3):

\begin{enumerate}[{(D-1)}]
    \item \textbf{Differentiability of $Q_{\eta}(\bs \theta)$:} the loss function $Q_{\eta}(\bs \theta)$ defined in eq.~(\ref{eq:q_eta}) is differentiable with respect to $\bs \theta$.
    \item \textbf{Lipschitz continuity for the gradient of $Q_{\eta}(\bs \theta)$:} using the coefficient $\alpha:=
\begin{cases}
|\mathcal{I}_n^{(U)}|/|\mathcal{K}_{\bs u}| & (v=1) \\
|\mathcal{I}_n^{(U)}| & (v=0) \\
\end{cases}$, the gradient $\alpha \frac{\partial}{\partial \bs \theta}Q_{\eta}(\bs \theta)$ is $H$-Lipschitz continuous for some $H>0$, i.e., $\|\alpha \frac{\partial}{\partial \bs \theta}Q_{\eta}(\bs \theta)-\alpha \frac{\partial}{\partial \bs \theta}Q_{\eta}(\bs \theta')\|_2 \le H\|\bs \theta-\bs \theta'\|_2, \: (\forall \bs \theta,\bs \theta' \in \bs \Theta)$.
\item \textbf{Bounded variance for the stochastic gradient:} variance of the minibatch-based stochastic gradient $\tilde{g}^{(1)}_{\eta}(\bs \theta)$ is uniformly bounded with respect to resampling the minibatch, i.e., 
$\sup_{\bs \theta \in \bs \Theta}\text{tr} \mathbb{V}_{\mathcal{M}^{(1)}}(\tilde{g}_{\eta}^{(1)}(\bs \theta))<\infty$.
\end{enumerate}
Symbols $\mathbb{E}_{\mathcal{M}^{(t)}}(\cdot),\mathbb{V}_{\mathcal{M}^{(t)}}(\cdot)$ represent the expectation and the variance-covariance matrix with respect to resampling the minibatch $\mathcal{M}^{(t)}=(\tilde{\mathcal{P}}_{\text{mini}}^{(t)},\tilde{\mathcal{I}}_{\text{mini}}^{(t)},s_+^{(t)},s_-^{(t)})$, 
and $E_{\tau}(\cdot)$ takes expectation with respect to selecting $\tau \in [T]$. 
$\text{tr}\bs Z$ represents the trace of the matrix $\bs Z=(z_{ij}) \in \mathbb{R}^{p \times p}$, i.e., $\text{tr}\bs Z=\sum_{i=1}^{p}z_{ii}$.

(D-1)--(D-3) are assumed in \citet{ghadimi2013stochastic}, and they are not unusually strong assumptions in our setting; 
when assuming (C-1) compactness of the parameter set $\bs \Theta$, 
$Q_{\eta}(\bs \theta)$ using any generating function listed in Table~\ref{table:bregman} and the similarity function~(\ref{eq:ex_similarity_model}) equipped with vector-valued neural networks $\bs f_{\bs \theta}:\mathcal{X}^U \to \mathbb{R}^K$ activated by sigmoid function, satisfies the assumptions (D-1)--(D-2).
Then, (D-3) also holds since the stochastic gradient $\tilde{g}_{\eta}^{(1)}(\bs \theta)$ is $C^1$ on the compact set $\bs \Theta$ and the minibatch $\mathcal{M}^{(t)}$ is a realization of random variable taking value in a finite set.

\begin{theo}
\label{theo:sgd_convergence}
Let $m_+,m_-,q,T,U \in \mathbb{N},v \in \{0,1,\ldots,U-1\},\eta>0,\bs \Theta:=\mathbb{R}^q$, and $\{\tilde{\bs \theta}^{(t)}\}_{t=1}^{T}$ is a sequence of the minibatch SGD~(\ref{eq:minibatch_sgd}), and the conditions (D-1)--(D-3) are assumed. 
If $v \geq 1$, let $\bs u$ be a vector in the set $(\ref{eq:specifying_u})$, and $p_{\bs j}:=1/|\mathcal{K}_{\bs u}|$ for all $\bs j \in \mathcal{K}_{\bs u}$. 
By specifying $\gamma^{(t)}=\gamma t^{-1}$ with $\gamma \in (0,2/H)$, 
and choosing the number of iterations $\tau \in [T]$ with the probability $\mathbb{P}(\tau=t)=\frac{2\gamma/t-H\gamma^2/t^2}{\sum_{t=1}^{T}(2\gamma/t-H\gamma^2/t^2)}$, 
it holds that
\begin{align*}
\mathbb{E}_{\tau}\left(
\mathbb{E}_{\{\mathcal{M}^{(t)}\}_{t \in [\tau]}}\left(
\:
\bigg\|
\frac{\partial }{\partial \bs \theta}
Q_{\eta}(\tilde{\bs \theta}^{(\tau)})
\bigg\|_2^2
\:
\right)
\right)
=
O(1/\log T)
\to 
0, \quad (T \to \infty).
\end{align*} 
\end{theo}
See \ref{app:proof_of_theo:sgd_convergence} for the proof. 


Theorem~\ref{theo:sgd_convergence} indicates that the gradient $\frac{\partial }{\partial \bs \theta}
Q_{\eta}(\tilde{\bs \theta}^{(\tau)})
=
\frac{\partial}{\partial \bs \theta}
D_{\varphi}(
\{\eta w_{\bs i}\}_{\bs i \in \mathcal{I}_n^{(U)}},
\{\mu_{\bs \theta}(\bs X_{\bs i})\}_{\bs i \in \mathcal{I}_n^{(U)}})\bigg|_{\bs \theta=\tilde{\bs \theta}^{(\tau)}}$ approaches $\bs 0$ as $T \to \infty$. 
Considering $\lim_{T \to \infty}\mathbb{P}(\tau \le T')=0$ for any fixed constant $T' \in \mathbb{N}$, indicating that large $\tau$ tends to be selected when $T$ is sufficiently large, the estimator $\tilde{\bs \theta}^{(t)}$ computed through the iterative update (\ref{eq:minibatch_sgd}) approaches a set of stationary points of the function $D_{\varphi}(\{\eta w_{\bs i}\}_{\bs i \in \mathcal{I}_n^{(U)}},\{\mu_{\bs \theta}(\bs X_{\bs i})\}_{\bs i \in \mathcal{I}_n^{(U)}})$ as $t$ increases. 
Although the estimator can be trapped in local minimizers or saddle points during the iterative update, gradient descent using randomly perturbed gradients is proved to escape saddle points efficiently~\citep{jin2017escape}. 
The similar is expected for minibatch SGD; the estimator may approach a good minimizer efficiently, depending on the situation. 
When the estimator approaches a global minimizer, 
under some assumptions, we can expect that 
\begin{align} \label{eq:mu-eta}
    \eta \mu_*(\bs X)
    \approx 
    \mu_{\tilde{\bs \theta}^{(t)}}(\bs X), \quad 
    (\forall \bs X \in \mathcal{X}^{U})
\end{align}
for some sufficiently large $n,t \in \mathbb{N}$, by considering Theorem~\ref{theo:consistency} with $\mathbb{E}(\eta w_{\bs i} \mid \bs X_{\bs i})=\eta \mu_*(\bs X_{\bs i})$. 
Although specifying $\eta=1$ appears better in terms of exactly recovering the underlying true similarity function $\mu_*$, it is not necessarily so in practice; 
only the ratio $\mu_{\bs \theta}(\bs X_{\bs i})/\mu_{\bs \theta}(\bs X_{\bs i'})$ is required to infer which of the tuples $\bs X_{\bs i}, \bs X_{\bs i'}$ exhibits a stronger relation. 
Thus $\eta$ can be arbitrarily specified by users. 
In practice, we may set $s_+^{(t)}=s_-^{(t)}=1,\eta=1$ in (\ref{eq:minibatch_gradient}), which is justified if the ratio $|\tilde{\mathcal{I}}_n^{(U)}|/|\tilde{\mathcal{P}}_n^{(U)}|$ is constant;
this in effect specifies 
$\eta=(|\tilde{\mathcal{I}}_n^{(U)}| m_+)/(|\tilde{\mathcal{P}}_n^{(U)}| m_-)$ in (\ref{eq:mu-eta})
and $\gamma^{(t)}$ being multiplied by 
$|\tilde{\mathcal{I}}_n^{(U)}|/m_-$
in (\ref{eq:minibatch_sgd}).

It is noteworthy that \citet{okuno2019robust} Theorem 3.2 already shows the convergence of the estimator $\tilde{\bs \theta}^{(t)}$ when $(U,v,\varphi)=(2,0,\varphi_{\beta})$, by assuming that the loss function is locally strongly convex. 
However, Theorem~\ref{theo:sgd_convergence} admits non-convex loss functions by considering not the convergence of the estimator $\tilde{\bs \theta}^{(t)}$ but that of the gradient $\frac{\partial}{\partial \bs \theta} Q(\tilde{\bs \theta}^{(t)})$. 
As the objective function $Q(\bs \theta)$ is typically unidentifiable when NNs therein, implying that the strong convexity is rarely satisfied, Theorem~\ref{theo:sgd_convergence} satisfies the practical situations more than \citet{okuno2019robust} Theorem 3.2. 
Furthermore, Theorem~\ref{theo:sgd_convergence} can be applied to general $U \in \mathbb{N}$, whereas only a few theoretical aspects of  stochastic algorithms have been investigated even for $U=2$~\citep{victor2019empirical}.

\begin{figure}[!ht]
\centering 
\begin{tabular}{c|cccc}
 & Kullback Leibler & $\beta$-div. ($\beta=1$) & Dual Logistic & Logistic \\
\hline 
$U=2$ &  
\begin{minipage}{0.2\hsize}
    \includegraphics[scale=0.2]{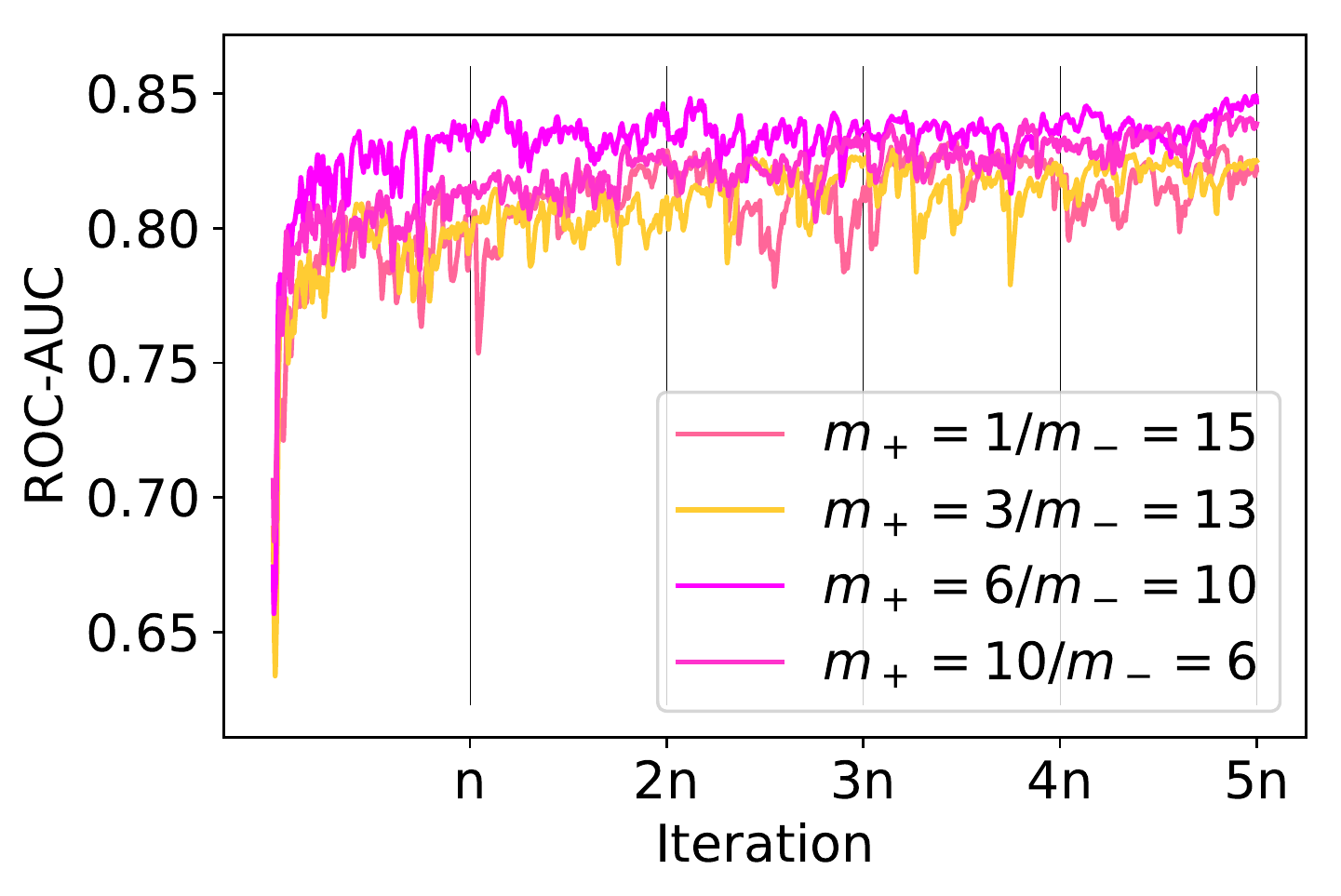}
    \label{fig:T2_KL}
  \end{minipage} & 
  \begin{minipage}{0.2\hsize}
    \includegraphics[scale=0.2]{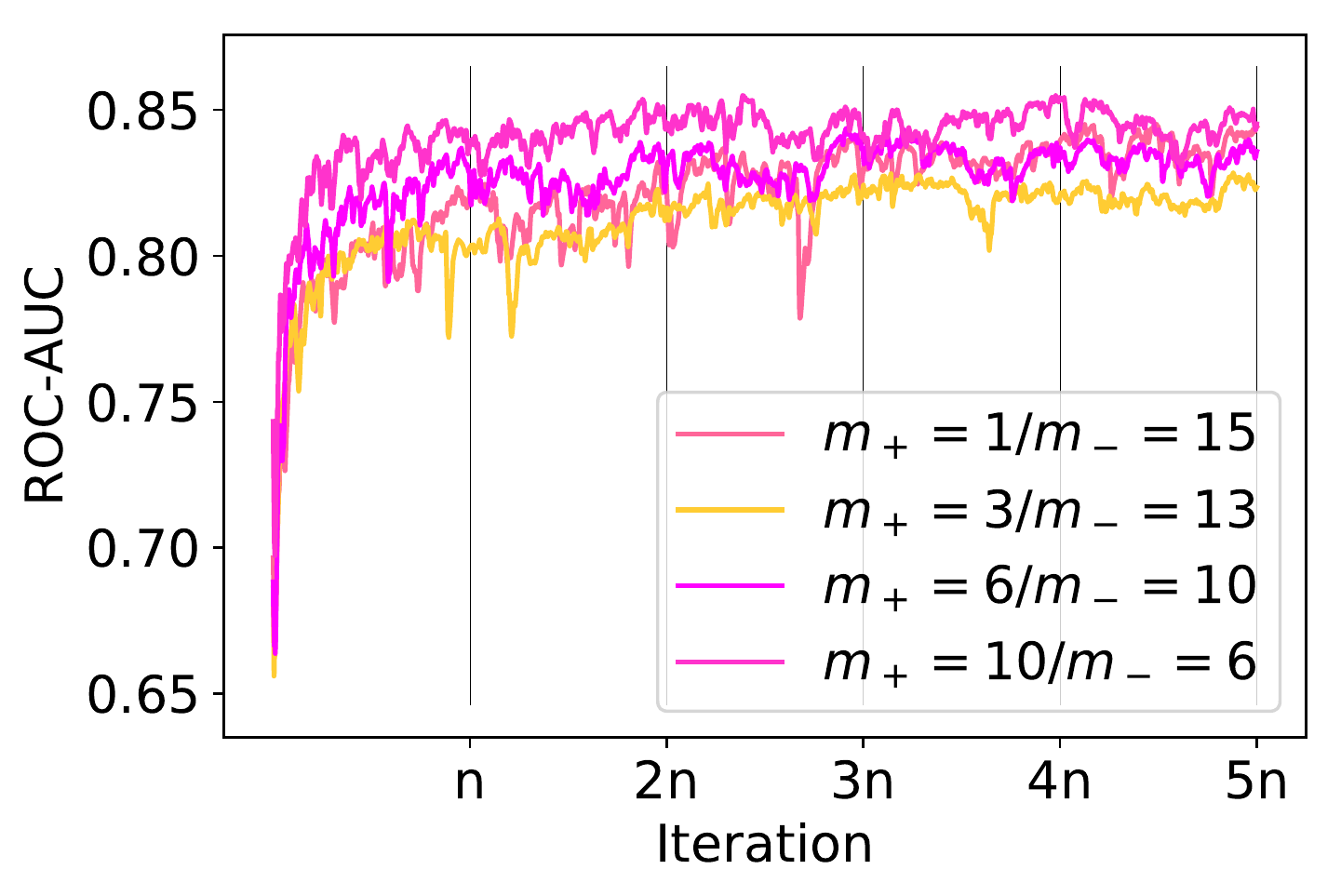}
    \label{fig:T2_beta1}
  \end{minipage}&
  \begin{minipage}{0.2\hsize}
    \includegraphics[scale=0.2]{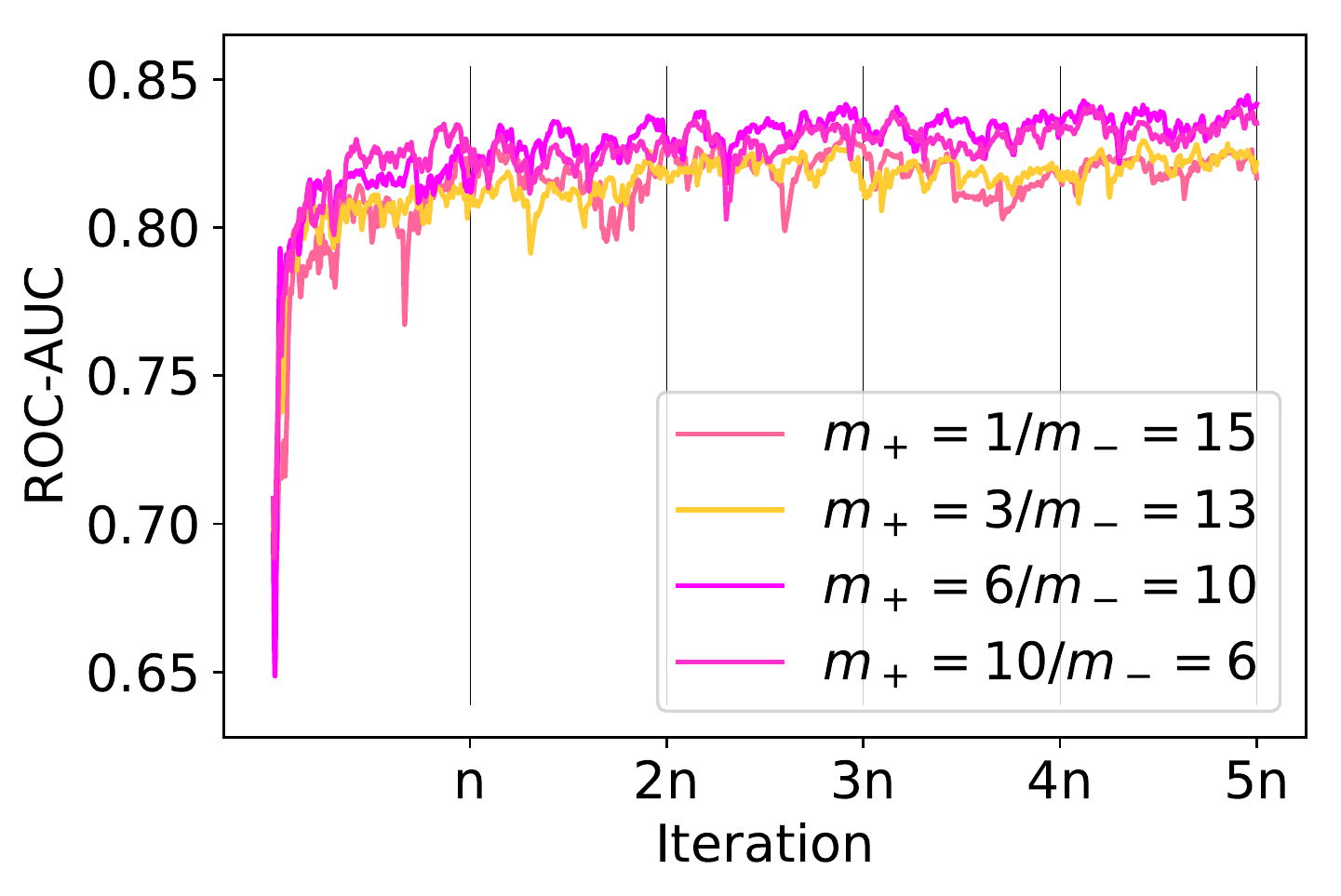}
    \label{fig:T2_DualLogistic}
  \end{minipage}&
  \begin{minipage}{0.2\hsize}
    \includegraphics[scale=0.2]{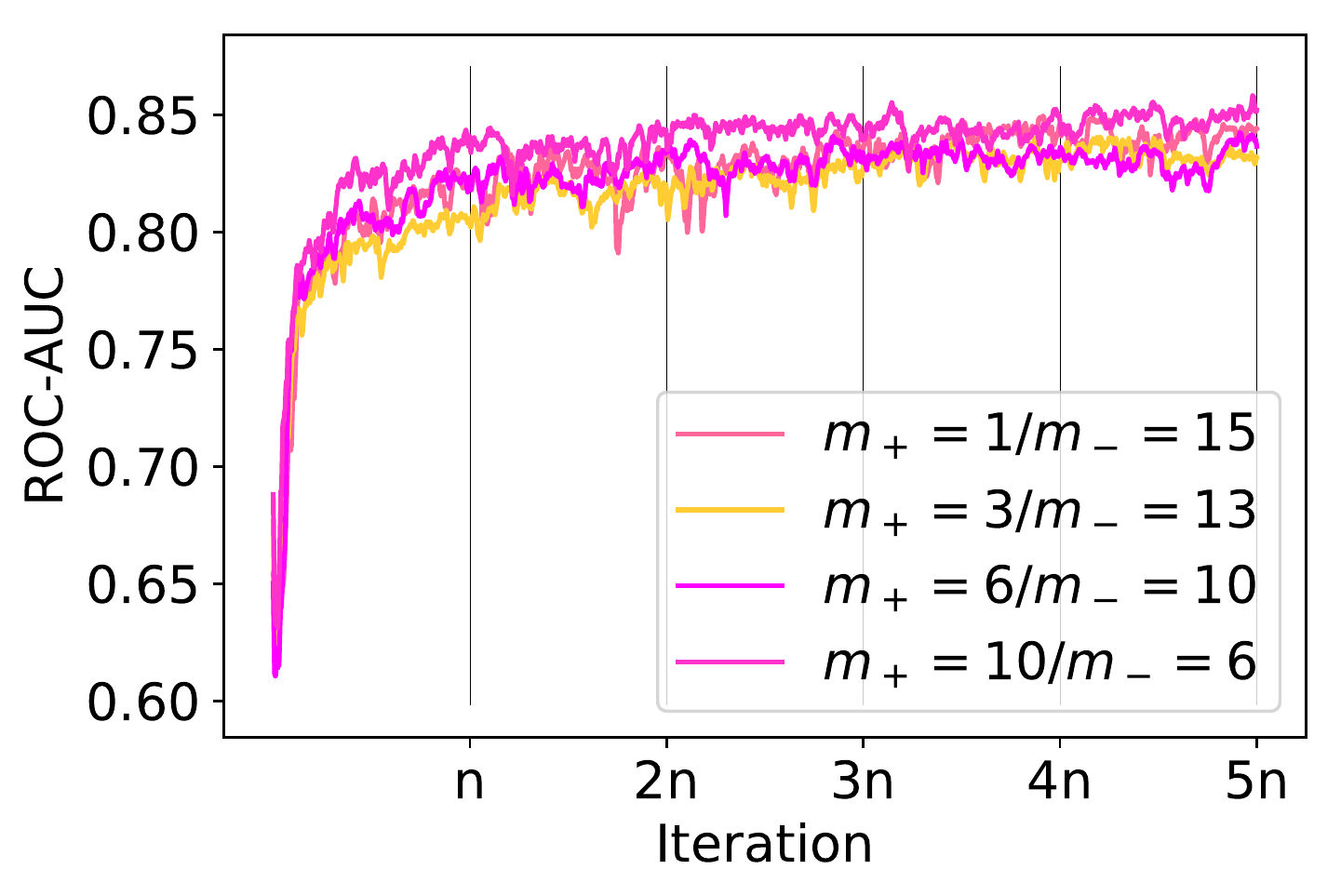}
    \label{fig:T2_Logistic}
  \end{minipage} \\
  $U=3$ &  \begin{minipage}{0.2\hsize}
    \includegraphics[scale=0.2]{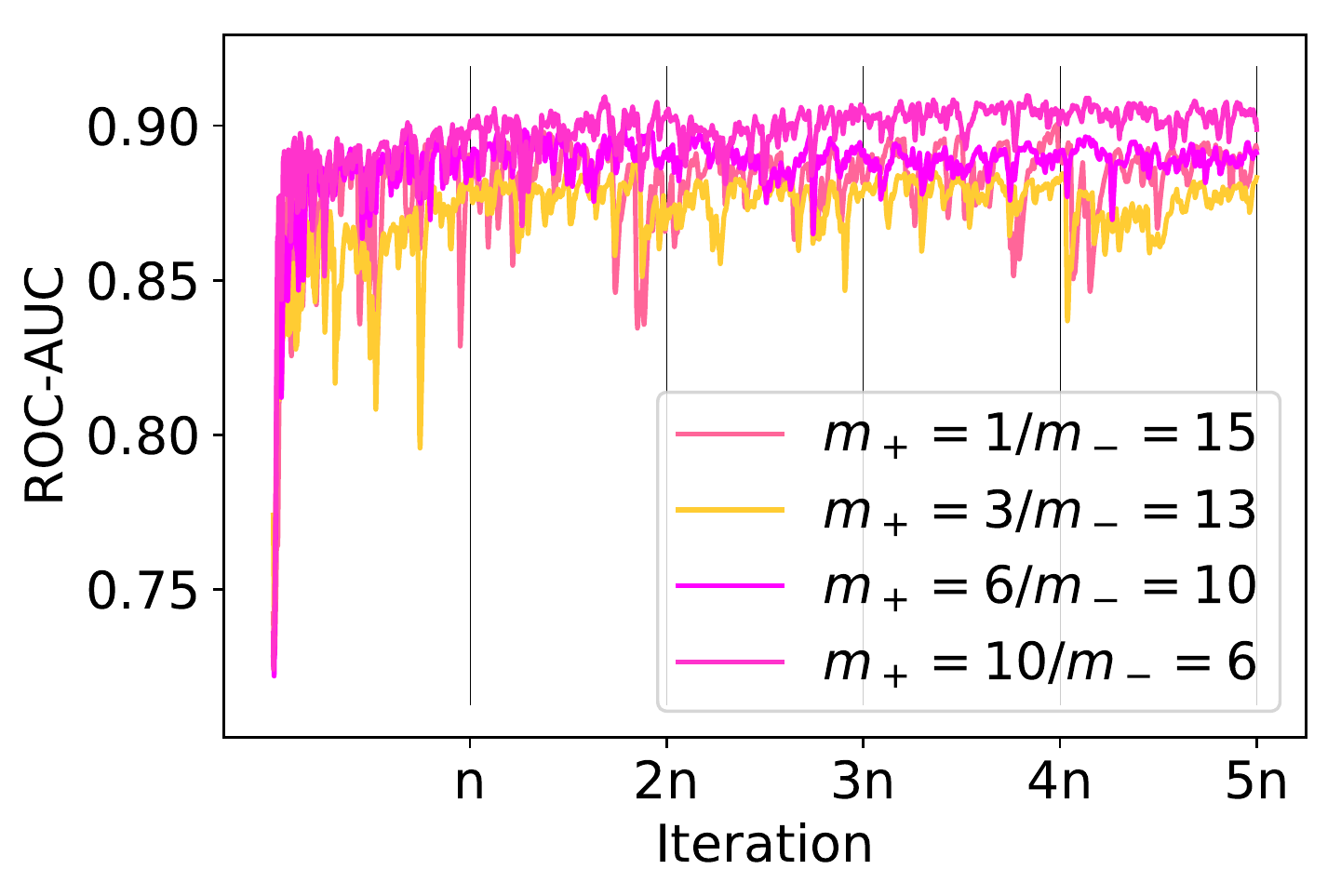}
    \label{fig:T3_KL}
  \end{minipage} & 
  \begin{minipage}{0.2\hsize}
    \includegraphics[scale=0.2]{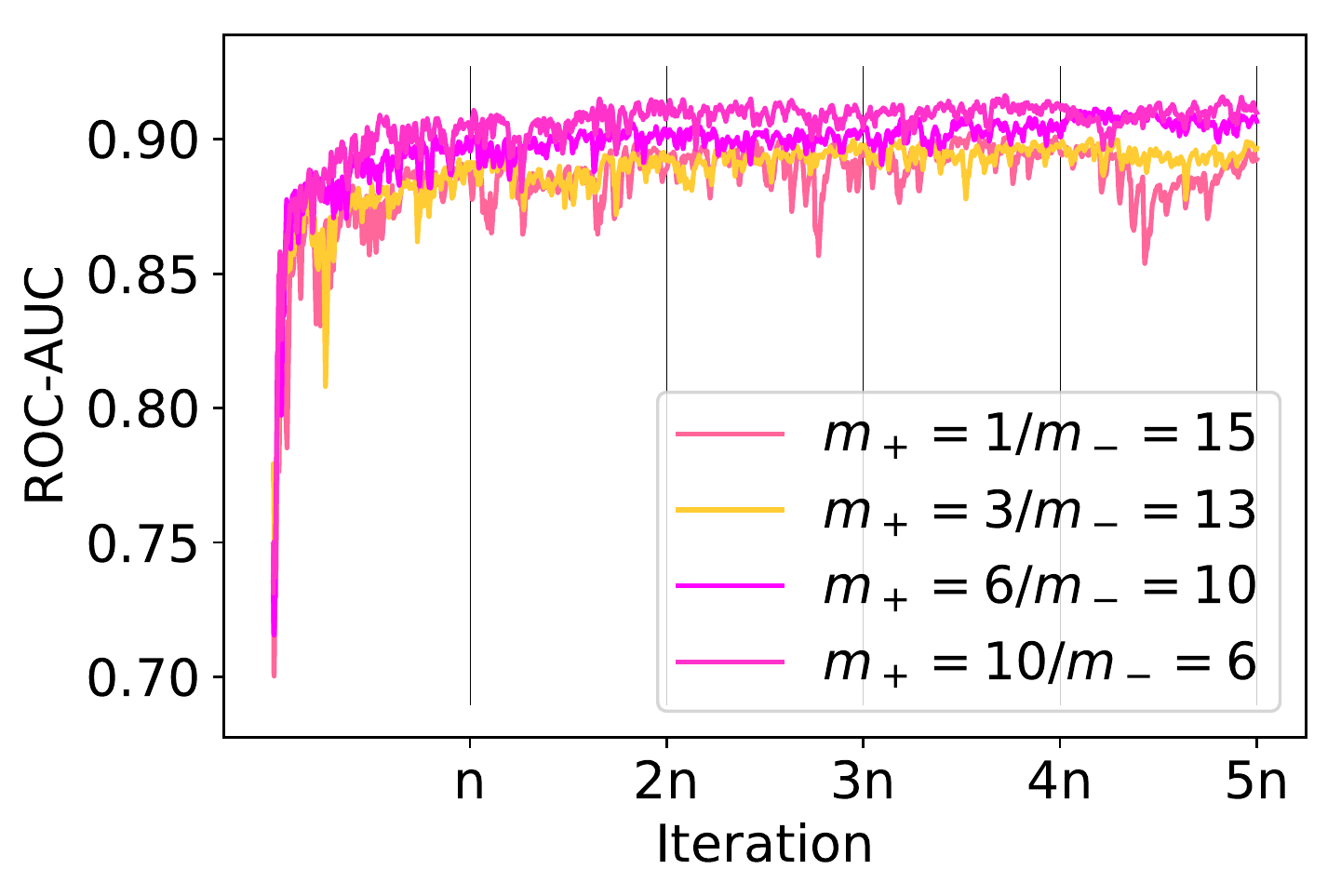}
    \label{fig:T3_beta1}
  \end{minipage}&
  \begin{minipage}{0.2\hsize}
    \includegraphics[scale=0.2]{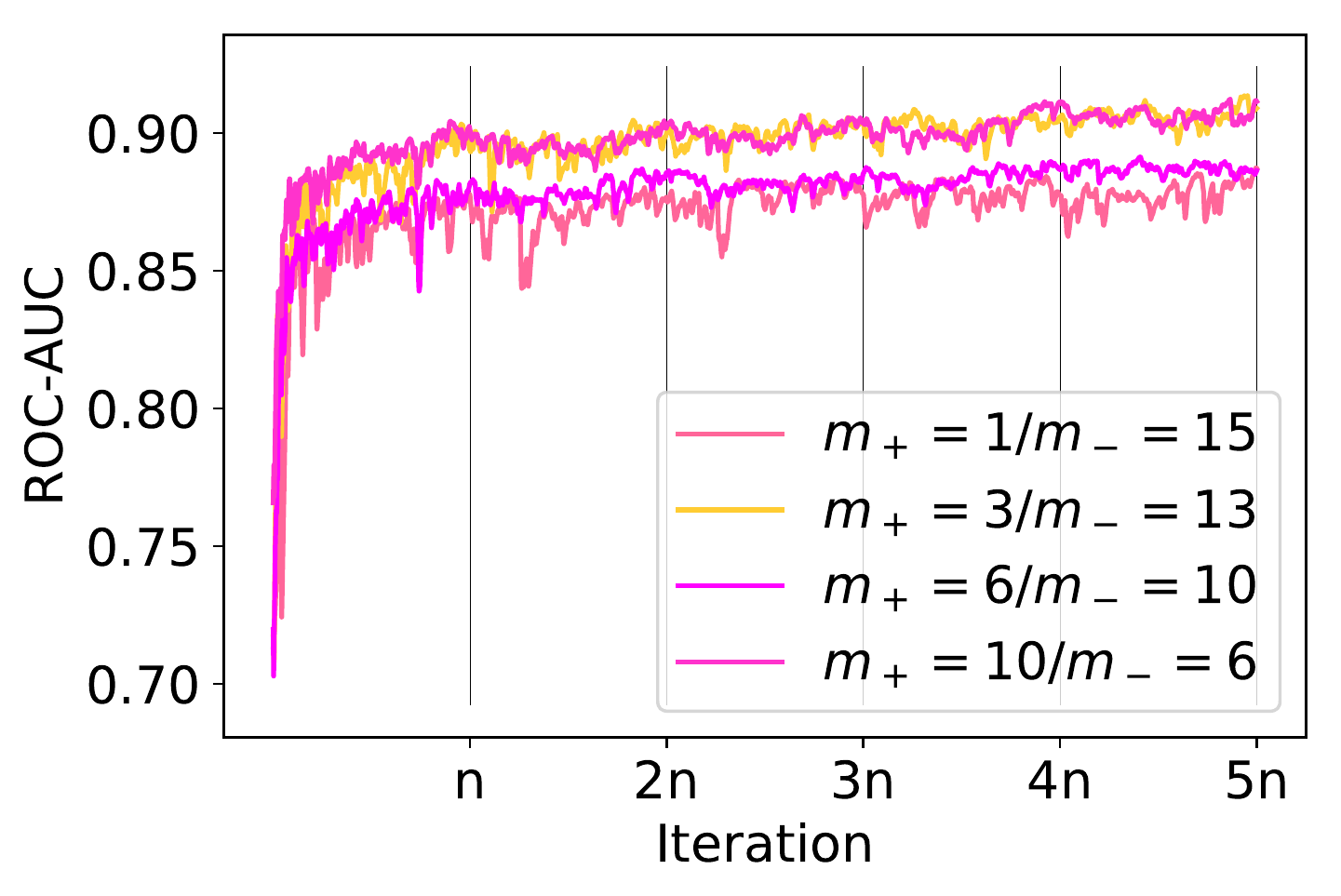}
    \label{fig:T3_DualLogistic}
  \end{minipage}&
  \begin{minipage}{0.2\hsize}
    \includegraphics[scale=0.2]{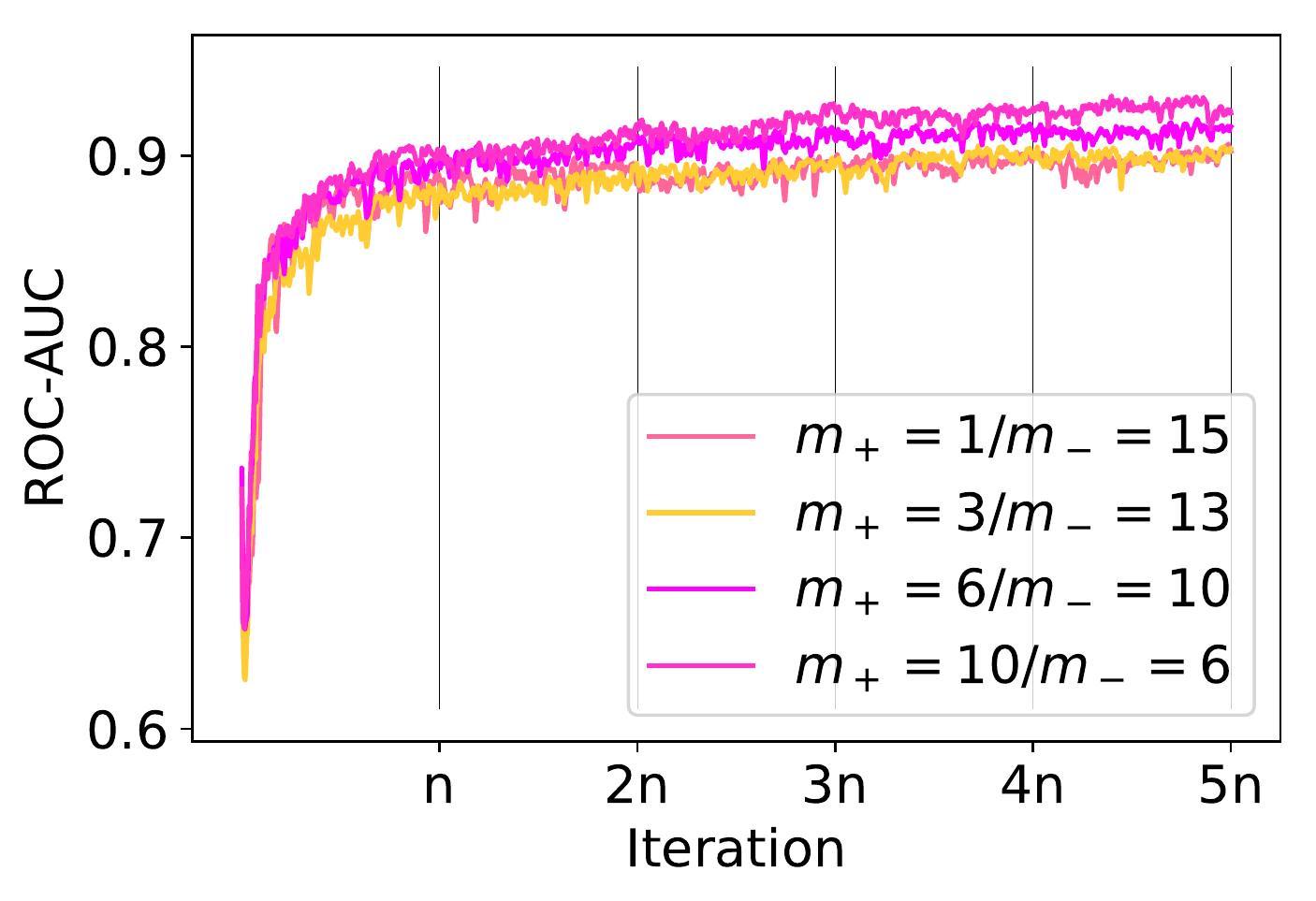}
    \label{fig:T3_Logistic}
  \end{minipage}
\end{tabular}
  \caption{
  For $U=2,3$, we plot the changes in the ROC--AUC test score over the Adam iterations~\citep{kingma2014adam} using Algorithm~\ref{alg:minibatch} with $v=1$, initial step size $10^{-3}$, and weight decay $10^{-2}$. 
  The $x$-axis represents the iteration number, where $n$ is the number of data vectors in the training dataset, and 
  the $y$-axis represents the ROC--AUC test score. 
  The results indicate that the ROC--AUC test score reaches approximately the maximum value within approximately $2n$ iterations. 
  The experimental details are same as those in Section~\ref{subsec:experiments_link_regression} and \ref{subsec:experiments_hyperlink_regression}~(a) with $K=40$. 
  }
  \label{fig:adam_process}
\end{figure}

Here, we empirically demonstrate that a stochastic optimization algorithm called Adam~\citep{kingma2014adam} equipped with the proposed minibatch sampling procedure shown in Algorithm~\ref{alg:minibatch} appropriately optimizes the similarity function within the reasonable number of iterations, in Figure~\ref{fig:adam_process}.


\section{Experiments}
\label{sec:experiments}

In this section, we describe the numerical experiments that we conducted on real-world datasets. 
In Section~\ref{subsec:experiments_poisson_regression}, we utilized the Boston housing dataset to perform the BHLR with $U=1$, that corresponds to the Poisson regression. 
In Section~\ref{subsec:experiments_link_regression} and \ref{subsec:experiments_hyperlink_regression}, we employed the attributed DBLP co-authorship network dataset~\citep{desmier2012cohesive} for performing the BHLR with $U=2$ and $U=3$, corresponding to link regression and hyperlink regression, respectively.

Hereinafter, we incorporate a regularization $\varphi_{\text{KL}}(z)=z \log (z+\varepsilon)$ with a small constant $\varepsilon:=10^{-4}$ into the KL divergence, for numerically stabilizing the experimental results.

\subsection[Poisson regression]{Poisson regression ($U=1$)}
\label{subsec:experiments_poisson_regression}

\begin{itemize}
\item \textbf{Dataset:} We employ the Boston housing dataset\footnote{\href{http://lib.stat.cmu.edu/datasets/boston}{http://lib.stat.cmu.edu/datasets/boston} (visited on June 13th, 2019)} that contains $n=506$ samples, comprising $p=13$ dimensional standardized explanatory variables $\{\bs x_i\}_{i=1}^{506} \subset \mathbb{R}^{13}$ and non-negative-valued target variables $\{y_i\}_{i=1}^{506} \subset \mathbb{R}_{\geq 0}$.

\item \textbf{Architecture of $\mu_{\bs \theta}$:} 1-hidden-layer multilayer perceptron~(see, e.g., \citet{bishop2006pattern} Chapter~5) with $1{,}000$ hidden units activated by Rectified Linear Unit~(ReLU), i.e., $\text{ReLU}(z):=\max\{0,z\}$, and unactivated $1$-dimensional output unit, are used for $f_{\bs \theta}:\mathbb{R}^{13} \to \mathbb{R}$. 
Using the NN $f_{\bs \theta}$, we define two different functions 
$\mu_{\bs \theta}(\bs x_i):=\exp(f_{\bs \theta}(\bs x_i))$ and $\mu_{\bs \theta}(\bs x_i):=f_{\bs \theta}(\bs x_i)$, where the former is restricted to positive values whereas the latter is not.

\item \textbf{Learning $\mu_{\bs \theta}$:} 
The NN in the function $\mu_{\bs \theta}$ is trained through the BHLR with $U=1$ using fullbatch gradient descent with the training dataset.

\item \textbf{Evaluation:} 
The dataset is randomly duivided into $3$ non-overlapping sets for training, validation, and test, whose numbers are $304~(60\%)$, $101~(20\%)$, and $101~(20\%)$, respectively. 
We first predict the target variables for validation and test datasets, and the mean squared error between the predicted values $\{\mu_{\hat{\bs \theta}_{\varphi,n}}(\bs X_{\bs i})\}$ and the observed values $w_{\bs i}$ are recorded at each iteration of GD. 
At the end of the iteration, the test score whose validation score is the best, is recorded as ``optimal" test score. 
We repeat the experiment 100 times, and compute the sample average and the standard error of the optimal test scores, for each setting.

\item \textbf{Baselines:} 
We perform Poisson regression using a linear model and a simple linear regression that are already implemented in a Python \verb|statsmodels| module~\citep{seabold2010statsmodels}. 
We also perform Poisson regression using a neural network~\citep{fallah2009nonlinear}. 
(\textbf{Random}) We first compute the sample average $\hat{\mu}$ and the sample standard deviation $\hat{\sigma}$ for the target variables in each of the $100$ test datasets. For each, we generate random numbers from a normal distribution whose mean and standard deviation are $\hat{\mu},\hat{\sigma}$, respectively, and evaluate the mean-squared error between the target variables in the test dataset and the generated random numbers. We repeat this evaluation 100 times for each of the $100$ test datasets, and compute the sample average and standard error. 

\end{itemize}

\textbf{Results:} 
The experimental results are shown in Table~\ref{table:experiment_u1}. 
Although the linear methods are much better than the baseline~(Random), 
NN-based methods outperformed the linear methods. 
Among the NN-based methods, using $\varphi_{\beta}$ with $\beta \geq 1$, which corresponds to using $\beta$-divergence, 
demonstrated better performance than $\varphi_{\text{KL}}$. 
This result indicates that, the classical loss function for the Poisson regression $L_{\varphi_{\text{KL}},n}(\bs \theta)$ is not always the best choice for learning the function $\mu_{\bs \theta}$.

\begin{table}[htbp]
\centering
\begin{tabular}{llcc}
\hlineB{2.5}
 & Generating function & $\mu_{\bs \theta}(\bs x):=\exp(f_{\bs \theta}(\bs x))$ & $\mu_{\bs \theta}(\bs x):=f_{\bs \theta}(\bs x)$ \\
\hlineB{2.5}
\multirow{6}{*}{Neural Network}
& \textbf{BHLR + $\beta$-div.} ($\beta=2.0$) & $\second{14.57} \pm 0.65$ & $\best{14.03} \pm 0.62$ \\
& \textbf{BHLR + $\beta$-div.} ($\beta=1.5$) & $\best{14.12} \pm 0.60$ & $\second{14.20} \pm 0.70$ \\
& \textbf{BHLR + $\beta$-div.} ($\beta=1.0$) & $14.32 \pm 0.70$ & $15.30 \pm 0.50$\\
& \textbf{BHLR + $\beta$-div.} ($\beta=0.5$) & $14.90 \pm 0.64$ & $15.31 \pm 0.64$\\
& \textbf{BHLR + $\beta$-div.} ($\beta=0.1$) & $16.12 \pm 0.70$ & $16.07 \pm 0.62$ \\
& Poisson regression$^{\dagger}$~\citep{fallah2009nonlinear} & $16.08 \pm 0.58$  & $16.86 \pm 0.73$ \\
\hline
\multirow{2}{*}{Linear} & Poisson regression$^{\dagger}$~\citep{cameron2013regression} & \multicolumn{2}{c}{$18.86 \pm 0.56$} \\
& LS regression$^{\dagger}$~\citep{bishop2006pattern} & \multicolumn{2}{c}{$24.58 \pm 0.64$} \\
\hline
Random$^{\dagger}$ & &  \multicolumn{2}{c}{$170.01 \pm 3.51$} \\
\hlineB{2.5}
\end{tabular} \\
\hspace*{-30em}{\small $^{\dagger}$Baselines}

\caption{
Poisson regression~($U=1$) is conducted on a randomly sampled Boston housing dataset, and the sample average and standard error of the mean squared error for $100$ experiments are listed. 
\textbf{A smaller score is better}. 
The best score is \best{bolded}, and the second best score is \second{underlined}. 
}
\label{table:experiment_u1}
\end{table}

\subsection[Link regression]{Link regression ($U=2$)}
\label{subsec:experiments_link_regression}

\begin{itemize}
    \item \textbf{Dataset:} We utilize a network comprising $n=2{,}723$ attributed nodes and $37{,}322$ positive binary link weights, that aggregates $9$ snapshots of the DBLP dynamic co-authorship network dataset~\citep{desmier2012cohesive}. 
In the aggregated network, 
each binary link weight represents whether the corresponding authors have at least one co-authorship relation in the $9$ snapshots; $w_{i_1i_2}=1$ if the authors $i_1$ and $i_2$ have the relation, and $0$ otherwise. 
Each node has $p=43$ dimensional data vectors, representing the number of publications, summed up over the $9$ snapshots, in each of the selected 43 journals/conferences.

\item \textbf{Similarity function architecture:} 
Vector-valued NN $\bs f_{\bs \theta}:\mathbb{R}^{43} \to \mathbb{R}^{K}$ is a $1$-hidden-layer multilayer perceptron with $1{,}000$ hidden units activated by the ReLU and $K$ unactivated output units. 
Using $\bs f_{\bs \theta}$, we exploit a similarity function $\mu_{\bs \theta}(\bs X_{\bs i}):=\sigma(\langle \bs f_{\bs \theta}(\bs x_{i_1}),\bs f_{\bs \theta}(\bs x_{i_2}) \rangle)$, where $\sigma(z):=(1+\exp(-z))^{-1}$ is a sigmoid function.

\item \textbf{Learning similarity functions:} 
NN $\bs f_{\bs \theta}$ in the similarity function is trained by Adam optimizer~\citep{kingma2014adam} using Algorithm~\ref{alg:minibatch} for minibatch sampling. 
For computing the stochastic gradient~(\ref{eq:minibatch_gradient}), we utilize $s_+^{(t)}=s_-^{(t)}=1,\eta=1$, and batch sizes $(m_+,m_-)$ are selected the set $\{(1,15),(3,13),(6,10),(10,6)\}$. 
For each of batch sizes $(m_+,m_-)$, the weight decay is grid searched over $\{10^{-2},10^{-3}\}$.

\item \textbf{Evaluation:}
The set of data vectors is randomly divided into $3$ non-overlapping sets for training, validation, and test, whose numbers are $n_{\text{train}}=1,907~(70\%),\, n_{\text{valid}}=408~(15\%),\, n_{\text{test}}=408~(15\%)$. 
In the test dataset, $10$ pairs are sampled from the set $\{\bs i=(i_1,i_2) \mid w_{\bs i}^{(\text{test})}=0\}$ for each $i_1=1,2,\ldots,n_{\text{test}}$, and combined with positive pairs $\{\bs i=(i_1,i_2) \mid w_{\bs i}^{(\text{test})}>0\}$; 
we compute the ROC-AUC score~\citep{bradley1997use} using these link weights, and record the scores for each of the $50$ iterations. 
Similarly, we compute the ROC--AUC score for the validation dataset. 
At the end of the iteration~($T=3n_{\text{train}}$), we record the test score whose validation score is the best. 
We repeat this experiment $40$ times, and compute the sample average and the standard error for each $(m_+,m_-)$; the best validated score amongst all $(m_+,m_-)$ is also computed.

\item \textbf{Baselines:} We employ LINE~\citep{tang2015line}, KL-GE~\citep{okuno2018probabilistic}, and $\beta$-GE~\citep{okuno2019robust} that correspond to the BHLR equipped with $L_{\varphi_{\text{Logistic}},n}(\bs \theta)$, $L_{\varphi_{\text{KL}},n}(\bs \theta)$, and $L_{\varphi_{\beta},n}(\bs \theta)$, respectively. 
LPPs~\citep{he2004locality} are also conducted for obtaining the linearly transformed feature vectors $\tilde{\bs y}_i:=\hat{\bs A}^{\top}\bs x_i \: (i \in [n])$. Subsequently, similarities for the feature vectors are computed by $\mu_{\bs \theta}(\bs X_{\bs i})=\sigma(\langle \tilde{\bs y}_{i_1},\tilde{\bs y}_{i_2} \rangle)$.

\end{itemize}

\textbf{Results:} 
The experimental results are shown in Table~\ref{table:experiment_u2}. 
Overall, the NN-based methods outperformed the LPPs as the NN is highly expressive whereas the LPP is linear. 
In addition, NN-based methods demonstrated better performance by increasing the dimension $K$ of the feature vectors, unlike the LPPs that imposes a quadratic constraint on the feature vectors $\{\bs y_i\}_{i=1}^{n}$. 
Overall, the exponential divergence and logistic loss demonstrated good performances; particularly, the exponential divergence demonstrated the best performance among the KL divergence, $\beta$-divergence, logistic loss, dual logistic loss, and exponential divergence employed in this experiment. 
In terms of selecting $m_+$ and $m_-$, in this case, using more than one positive minibatch sample $(m_+>1)$ is better.

\begin{table}[htbp]
\centering
\scalebox{0.85}{
\begin{tabular}{llcccc|c}
\hlineB{2.5}
\multirow{2}{*}{\scalebox{1.2}{$K=10$}}
& \multirow{2}{*}{Method} & \multicolumn{4}{c|}{$m_+$/$m_-$} & \multirow{2}{*}{Best~(validated)}\\
& & 1/15 & 3/13 & 6/10 & 10/6 & \\
\hlineB{2.5}
\multirow{7}{*}{Neural network} & \textbf{BHLR + exponential div.} & $\best{81.5} \pm 0.4$  & $\second{82.5} \pm 0.2$  & $\best{82.7} \pm 0.4$  & $\second{82.7} \pm 0.3$  & $\best{83.0} \pm 0.4$  \\
& \textbf{BHLR + dual logistic loss} & $80.0 \pm 0.1$  & $81.4 \pm 0.2$  & $81.7 \pm 0.2$  & $81.5 \pm 0.1$  & $81.7 \pm 0.2$  \\
& KL-GE$^{\dagger,1}$~\citep{okuno2018probabilistic} & $80.1 \pm 0.2$  & $81.5 \pm 0.3$  & $82.1 \pm 0.2$  & $82.1 \pm 0.2$  & $82.2 \pm 0.3$  \\
& $\beta$-GE$^{\dagger,2}$~\citep{okuno2019robust} ($\beta=0.1$) & $\second{81.4} \pm 0.1$  & $82.3 \pm 0.2$  & $82.3 \pm 0.2$  & $\second{82.7} \pm 0.2$  & $82.3 \pm 0.3$  \\
& $\beta$-GE$^{\dagger,2}$~\citep{okuno2019robust} ($\beta=0.5$)  & $80.6 \pm 0.3$  & $82.2 \pm 0.2$  & $\second{82.5} \pm 0.2$  & $\best{82.9} \pm 0.2$  & $82.2 \pm 0.3$  \\
& $\beta$-GE$^{\dagger,2}$~\citep{okuno2019robust} ($\beta=1$)  & $81.2 \pm 0.3$  & $82.2 \pm 0.2$  & $82.4 \pm 0.3$  & $82.4 \pm 0.2$  & $82.2 \pm 0.3$  \\
& LINE$^{\dagger,3}$~\citep{tang2015line} & $\second{81.4} \pm 0.2$  & $\best{82.6} \pm 0.1$  & $82.0 \pm 0.2$  & $82.3 \pm 0.3$  & $\second{82.8} \pm 0.2$  \\
\hline
Linear & LPP$^{\dagger}$~\citep{he2004locality} & \multicolumn{5}{c}{$78.9 \pm 0.3$} \\
\hlineB{2.5}
\end{tabular}}
\\[1mm]
\scalebox{0.85}{
\begin{tabular}{llcccc|c}
\hlineB{2.5}
\multirow{2}{*}{\scalebox{1.2}{$K=40$}}
& \multirow{2}{*}{Method}
& \multicolumn{4}{c|}{$m_+$/$m_-$} & \multirow{2}{*}{Best~(validated)}\\
& & 1/15 & 3/13 & 6/10 & 10/6 & \\
\hline 
\multirow{7}{*}{Neural network} & \textbf{BHLR + exponential div.} & $\second{82.7} \pm 0.2$  & $\second{83.4} \pm 0.3$  & $\best{83.8} \pm 0.2$  & $\second{83.3} \pm 0.2$ & $\best{83.6} \pm 0.3$ \\
&\textbf{BHLR + dual logistic loss} & $82.2 \pm 0.2$  & $81.8 \pm 0.2$  & $82.4 \pm 0.3$  & $82.1 \pm 0.2$  & $82.2 \pm 0.3$  \\
&KL-GE$^{\dagger,1}$~\citep{okuno2018probabilistic} & $82.0 \pm 0.2$  & $82.4 \pm 0.2$  & $83.1 \pm 0.2$  & $82.7 \pm 0.2$  & $82.9 \pm 0.3$  \\
&$\beta$-GE$^{\dagger,2}$~\citep{okuno2019robust} ($\beta=0.1$) & $81.9 \pm 0.2$  & $82.6 \pm 0.2$  & $82.7 \pm 0.2$  & $\best{83.5} \pm 0.1$  & $82.8 \pm 0.3$  \\
&$\beta$-GE$^{\dagger,2}$~\citep{okuno2019robust} ($\beta=0.5$)  & $81.5 \pm 0.2$  & $82.5 \pm 0.2$  & $82.8 \pm 0.2$  & $83.1 \pm 0.2$  & $82.7 \pm 0.2$  \\
&$\beta$-GE$^{\dagger,2}$~\citep{okuno2019robust} ($\beta=1$)  & $82.5 \pm 0.3$  & $\second{83.3} \pm 0.2$  & $\second{83.3} \pm 0.2$  & $83.2 \pm 0.3$  & $83.3 \pm 0.2$  \\
&LINE$^{\dagger,3}$~\citep{tang2015line} & $\best{83.0} \pm 0.2$  & $\best{83.5} \pm 0.2$  & $83.1 \pm 0.2$  & $83.0 \pm 0.2$  & $\second{83.4} \pm 0.2$  \\
\hline
Linear &LPP$^{\dagger}$~\citep{he2004locality} & \multicolumn{5}{c}{$73.8 \pm 0.4$} \\
\hlineB{2.5}
\end{tabular}
} \\
{\small $^{\dagger}$Baselines, $^{1}$BHLR + KL-div., $^{2}$BHLR + $\beta$-div., $^{3}$BHLR + logistic loss.}

\caption{Link prediction~($U=2$) is conducted on the attributed DBLP co-authorship network dataset~\citep{desmier2012cohesive}, and the sample average and standard error of the ROC--AUC test scores for 40 experiments are listed. 
\textbf{A higher score is better}. The best score is \best{bolded}, and the second best score is \second{underlined}.}
\label{table:experiment_u2}
\end{table}

\subsection[Hyperlink regression]{Hyperlink regression ($U=3$)}
\label{subsec:experiments_hyperlink_regression}

Experimental settings are almost similar to those of $U=2$. 
We employ the same dataset used in Section~\ref{subsec:experiments_link_regression}, and compute synthetic hyperlink weights from their link weights.

\begin{itemize} 
\item \textbf{Similarity function architecture:} using $\bs f_{\bs \theta}$ defined in Section~\ref{subsec:experiments_link_regression}, we exploit a similarity function: 
$
\mu_{\bs \theta}(\bs X_{\bs i})
    :=
    \sigma
    \left(
        \langle \bs f_{\bs \theta}(\bs x_{i_1}),\bs f_{\bs \theta}(\bs x_{i_2}),\bs f_{\bs \theta}(\bs x_{i_3}) \rangle
    \right)
$, where $\langle \bs y,\bs y',\bs y'' \rangle=\sum_{k=1}^{K}y_k y'_k y''_k$. Similarity functions are trained and evaluated similarly to those of $U=2$.

\item \textbf{Evaluation:} 
We first divide the set of data vectors into training, validation, and test sets, similarly to $U=2$. 
However, these datasets contain only the link weights~($U=2$) but not hyperlink weights~($U=3$); 
in each of the datasets, 
we compute synthetic hyperlink weights $\bs W:=(w_{\bs i})$ in two different ways: 
\begin{enumerate}[{(a)}]
\item $w_{\bs i}=w_{i_1i_2i_3}=1$ if $\bs x_{i_1},\bs x_{i_2},\bs x_{i_3}$ are connected, i.e., a path exists between any of the two in $\bs i=(i_1,i_2,i_3)$, and $w_{\bs i}=0$ otherwise. 
\item $w_{\bs i}=w_{i_1 i_2 i_3}=1$ if $\bs x_{i_1},\bs x_{i_2},\bs x_{i_3}$ are fully connected, i.e., all of two in $\bs i=(i_1,i_2,i_3)$ are connected, and $w_{\bs i}=0$ otherwise. 
\end{enumerate}
In the test dataset, $15$ tuples are sampled from the set $\{\bs i=(i_1,i_2,i_3) \mid w^{\text{(test)}}_{\bs i}=0\}$ for each $i_1=1,2,\ldots,n_{\text{test}}$, and combine them with positive tuples $\{\bs i=(i_1,i_2,i_3) \mid w^{(\text{test})}_{\bs i}>0\}$. 
Using these tuples, we evaluated the experimental results by ROC-AUC score, similarly to $U=2$.

\item \textbf{Baseline:} We employ HIMFAC~\citep{nori2012multinomial} for obtaining the linearly transformed feature vectors $\tilde{\bs y}_i:=\hat{\bs A}^{\top}\bs x_i \: (i \in [n])$. Subsequently, similarities for the feature vectors are computed by 
(i) $\mu_{\bs \theta}(\bs X_{\bs i}):=\sigma(\langle \tilde{\bs y}_{i_1},\tilde{\bs y}_{i_2},\tilde{\bs y}_{i_3} \rangle)$ and 
(ii) $\mu_{\bs \theta}(\bs X_{\bs i}):=\sigma(\sum_{1 \leq k<l \leq 3}\langle \tilde{\bs y}_{i_k},\tilde{\bs y}_{i_l}\rangle)$.

\end{itemize}

\textbf{Results:} The experimental results are shown in Table~\ref{table:experiment_u3_a} for the setting (a) and Table~\ref{table:experiment_u3_b} for (b). 
Overall, the NN-based methods outperformed HIMFAC, since the NN is highly expressive whereas HIMFAC is linear. 
NN-based methods demonstrated a slight improvement by increasing the dimension $K$ of the feature vectors. 
There is significant difference between the settings (a) and (b) for HIMFAC, unlike NN-based methods. 
Regarding the setting (a), the logistic loss, exponential divergence and $\beta$-divergence with $\beta=1$ demonstrated good performances for $K=10$. 
On the other hand, the $\beta$-divergence with $\beta=0.5$ and KL-divergence, whose scores for $K=10$ were not that high, demonstrated good performance for $K=40$; experimental results depend on the choice of $K$. 
HIMFAC with (i) demonstrates a low performance, since their feature vectors are consequently obtained via LPP, that is based on the simple inner product $\langle \bs y,\bs y' \rangle$ whereas (i) is based on the similarity for triplets $\langle \bs y,\bs y',\bs y''\rangle$.
On the other hand, HIMFAC with (ii) demonstrates much higher performance than (i), since HIMFAC is compatible with the simple inner product. 
In terms of selecting $m_+$ and $m_-$, in this case, using more than one positive minibatch sample $(m_+>1)$ is better. 
Regarding the setting (b), tendency of the results are almost similar to the setting (a).

\begin{table}[htbp]
\centering
\scalebox{0.85}{
\begin{tabular}{llcccc|c}
\hlineB{2.5}
\multirow{2}{*}{\scalebox{1.2}{$K=10$}}
& \multirow{2}{*}{Method} & \multicolumn{4}{c|}{$m_+$/$m_-$} & \multirow{2}{*}{Best~(validated)}\\
& & 1/15 & 3/13 & 6/10 & 10/6 & \\
\hlineB{2.5}
\multirow{7}{*}{Neural network} & \textbf{BHLR + exponential div.} & $\best{86.1} \pm 0.3$  & $\best{87.3} \pm 0.3$  & $\second{87.5} \pm 0.3$  & $\best{87.2} \pm 0.3$  & $\second{87.3} \pm 0.3$  \\
& \textbf{BHLR + dual logistic loss} & $85.4 \pm 0.3$  & $86.1 \pm 0.2$  & $86.0 \pm 0.3$  & $86.7 \pm 0.3$  & $86.2 \pm 0.3$   \\
& \textbf{BHLR + KL-div.} & $85.2 \pm 0.3$  & $85.1 \pm 0.2$  & $85.3 \pm 0.3$  & $85.6 \pm 0.2$  & $85.4 \pm 0.3$  \\
& \textbf{BHLR + $\beta$-div.}~($\beta=0.1$) & $85.3 \pm 0.3$  & $85.5 \pm 0.3$  & $85.8 \pm 0.3$  & $85.8 \pm 0.2$  & $85.8 \pm 0.3$  \\
& \textbf{BHLR + $\beta$-div.}~($\beta=0.5$)  & $85.3 \pm 0.3$  & $86.1 \pm 0.2$  & $86.9 \pm 0.3$  & $86.3 \pm 0.3$  & $86.6 \pm 0.3$  \\
& \textbf{BHLR + $\beta$-div.} ~($\beta=1$)  & $85.7 \pm 0.3$  & $\second{86.5} \pm 0.2$  & $86.8 \pm 0.3$  & $\second{87.0} \pm 0.3$  & $\second{87.3} \pm 0.2$  \\
& \textbf{BHLR + logistic loss} & $\second{86.0} \pm 0.3$  & $\best{87.3} \pm 0.3$  & $\best{87.9} \pm 0.2$  & $\best{87.2} \pm 0.2$  & $\best{87.4} \pm 0.3$  \\
\hline 
\multirow{2}{*}{Linear} & HIMFAC$^{\dagger}$~\citep{nori2012multinomial} + (i) & \multicolumn{5}{c}{$48.4 \pm 0.5$} \\
& HIMFAC$^{\dagger}$~\citep{nori2012multinomial} + (ii) & \multicolumn{5}{c}{$76.9 \pm 0.3$}  \\
\hlineB{2.5}
\end{tabular}}\\[1mm]
\scalebox{0.85}{
\begin{tabular}{llcccc|c}
\hlineB{2.5}
\multirow{2}{*}{\scalebox{1.2}{$K=40$}}
& \multirow{2}{*}{Method} & \multicolumn{4}{c|}{$m_+$/$m_-$} & \multirow{2}{*}{Best~(validated)}\\
& & 1/15 & 3/13 & 6/10 & 10/6 & \\
\hlineB{2.5}
\multirow{7}{*}{Neural network} & \textbf{BHLR + exponential div.} & $88.7 \pm 0.3$  & $\second{89.4} \pm 0.3$  & $88.7 \pm 0.3$  & $89.3 \pm 0.3$  & $89.8 \pm 0.2$  \\
& \textbf{BHLR + dual logistic loss} & $87.3 \pm 0.3$  & $87.8 \pm 0.3$  & $\second{89.1} \pm 0.2$  & $88.0 \pm 0.2$  & $89.0 \pm 0.3$  \\ 
& \textbf{BHLR + KL-div.} & $87.9 \pm 0.3$  & $88.4 \pm 0.2$  & $\best{89.2} \pm 0.3$  & $89.6 \pm 0.3$  & $\second{90.6} \pm 0.2$ \\
& \textbf{BHLR + $\beta$-div.}~($\beta=0.1$) & $\best{89.3} \pm 0.2$  & $89.0 \pm 0.2$  & $89.0 \pm 0.3$  & $89.3 \pm 0.3$  & $90.4 \pm 0.2$ \\
& \textbf{BHLR + $\beta$-div.}~($\beta=0.5$) & $\second{88.9} \pm 0.2$  & $\second{89.4} \pm 0.2$  & $89.6 \pm 0.3$  & $\second{90.0} \pm 0.3$  & $\best{90.8} \pm 0.2$  \\
& \textbf{BHLR + $\beta$-div.} ~($\beta=1$)  & $88.4 \pm 0.3$  & $\best{89.7} \pm 0.2$  & $89.0 \pm 0.2$  & $89.4 \pm 0.2$  & $90.5 \pm 0.2$  \\
& \textbf{BHLR + logistic loss} & $88.3 \pm 0.2$  & $88.9 \pm 0.3$  & $89.3 \pm 0.2$  & $\best{90.2} \pm 0.2$  & $89.9 \pm 0.2$ \\
\hline
\multirow{2}{*}{Linear} & HIMFAC$^{\dagger}$~\citep{nori2012multinomial} + (i) & \multicolumn{5}{c}{$49.6 \pm 0.5$} \\
& HIMFAC$^{\dagger}$~\citep{nori2012multinomial} + (ii) & \multicolumn{5}{c}{$75.4 \pm 0.4$}  \\
\hlineB{2.5}
\end{tabular}} \\
{\small $^{\dagger}$Baselines}
\caption{Hyperlink prediction~($U=3$) with the setting (a) is conducted on the attributed DBLP co-authorship network dataset~\citep{desmier2012cohesive}, and the sample average and  standard error of the ROC-AUC test scores for 40 experiments are listed. 
\textbf{A higher score is better}. The best score is \best{bolded}, and the second best score is \second{underlined}.}
\label{table:experiment_u3_a}
\end{table}

\begin{table}[htbp]
\centering
\scalebox{0.85}{
\begin{tabular}{llcccc|c}
\hlineB{2.5}
\multirow{2}{*}{\scalebox{1.2}{$K=10$}}
& \multirow{2}{*}{Method} & \multicolumn{4}{c|}{$m_+$/$m_-$} & \multirow{2}{*}{Best~(validated)}\\
& & 1/15 & 3/13 & 6/10 & 10/6 & \\
\hlineB{2.5}
\multirow{7}{*}{Neural network} & \textbf{BHLR + exponential div.} & $\second{85.7} \pm 0.3$  & $\second{86.6}\pm 0.3$  & $\second{86.7} \pm 0.3$  & $\second{86.9} \pm 0.3$  & $\second{86.7} \pm 0.3$ \\
& \textbf{BHLR + dual logistic loss} & $\second{85.7} \pm 0.4$  & $85.9 \pm 0.4$  & $86.0 \pm 0.3$  & $86.2 \pm 0.3$  & $86.4 \pm 0.3$  \\
& \textbf{BHLR + KL-div.} & $84.5 \pm 0.4$  & $85.0 \pm 0.4$  & $85.6 \pm 0.4$  & $85.3 \pm 0.5$  & $86.1 \pm 0.5$  \\
& \textbf{BHLR + $\beta$-div.}~($\beta=0.1$) & $84.9 \pm 0.4$  & $85.7 \pm 0.3$  & $85.5 \pm 0.3$  & $85.8 \pm 0.3$  & $85.9 \pm 0.4$  \\
& \textbf{BHLR + $\beta$-div.}~($\beta=0.5$)  & $85.0 \pm 0.4$  & $85.7 \pm 0.3$  & $85.9 \pm 0.3$  & $86.3 \pm 0.4$  & $86.5 \pm 0.4$ \\
& \textbf{BHLR + $\beta$-div.} ~($\beta=1$)  & $85.4 \pm 0.4$  & $86.0 \pm 0.4$  & $\second{86.7} \pm 0.3$  & $86.4 \pm 0.3$  & $86.6 \pm 0.3$ \\
& \textbf{BHLR + logistic loss} & $\best{85.9} \pm 0.3$  & $\best{86.8} \pm 0.3$  & $\best{87.2} \pm 0.3$  & $\best{87.3} \pm 0.3$  & $\best{86.8} \pm 0.3$ \\
\hline 
\multirow{2}{*}{Linear} & HIMFAC$^{\dagger}$~\citep{nori2012multinomial} + (i) & \multicolumn{5}{c}{$49.1 \pm 1.3$} \\
& HIMFAC$^{\dagger}$~\citep{nori2012multinomial} + (ii) & \multicolumn{5}{c}{$82.6 \pm 0.4$}  \\
\hlineB{2.5}
\end{tabular}}\\[1mm]
\scalebox{0.85}{
\begin{tabular}{llcccc|c}
\hlineB{2.5}
\multirow{2}{*}{\scalebox{1.2}{$K=40$}}
& \multirow{2}{*}{Method} & \multicolumn{4}{c|}{$m_+$/$m_-$} & \multirow{2}{*}{Best~(validated)}\\
& & 1/15 & 3/13 & 6/10 & 10/6 & \\
\hlineB{2.5}
\multirow{7}{*}{Neural network} & \textbf{BHLR + exponential div.} & $\second{88 .1} \pm 0.3$  & $\second{89.2} \pm 0.2$  & $88.6 \pm 0.3$  & $\best{89.4} \pm 0.2$  & $89.7 \pm 0.2$   \\
& \textbf{BHLR + dual logistic loss} & $87.0 \pm 0.3$  & $88.1 \pm 0.3$  & $88.0 \pm 0.2$  & $87.9 \pm 0.2$  & $89.0 \pm 0.2$  \\ 
& \textbf{BHLR + KL-div.} & $87.7 \pm 0.2$  & $88.8 \pm 0.3$  & $\second{89.3} \pm 0.3$  & $\second{89.1} \pm 0.3$  & $90.0 \pm 0.2$ \\
& \textbf{BHLR + $\beta$-div.}~($\beta=0.1$) & $\best{88.2} \pm 0.2$  & $\second{89.2} \pm 0.2$  & $\best{89.9} \pm 0.2$  & $88.8 \pm 0.3$  & $\best{90.3} \pm 0.1$ \\
& \textbf{BHLR + $\beta$-div.}~($\beta=0.5$) & $87.8 \pm 0.2$  & $\best{89.6} \pm 0.2$  & $88.9 \pm 0.2$  & $88.5 \pm 0.3$  & $\second{90.1} \pm 0.2$  \\
& \textbf{BHLR + $\beta$-div.} ~($\beta=1$)  & $\best{88.2} \pm 0.3$  & $88.5 \pm 0.2$  & $89.0 \pm 0.2$  & $88.4 \pm 0.3$  & $89.5 \pm 0.2$ \\
& \textbf{BHLR + logistic loss} & $\second{88.1} \pm 0.2$  & $88.6 \pm 0.2$  & $88.9 \pm 0.1$  & $88.7 \pm 0.2$  & $89.3 \pm 0.1$ \\
\hline
\multirow{2}{*}{Linear} & HIMFAC$^{\dagger}$~\citep{nori2012multinomial} + (i) & \multicolumn{5}{c}{$48.6 \pm 0.8$} \\
& HIMFAC$^{\dagger}$~\citep{nori2012multinomial} + (ii) & \multicolumn{5}{c}{$82.9 \pm 0.6$}  \\
\hlineB{2.5}
\end{tabular}} \\
{\small $^{\dagger}$Baselines}
\caption{
Hyperlink prediction~($U=3$) with the setting (b) is conducted on the attributed DBLP co-authorship network dataset~\citep{desmier2012cohesive}, and the sample average and  standard error of the ROC-AUC test scores for 40 experiments are listed. 
\textbf{A higher score is better}. The best score is \best{bolded}, and the second best score is \second{underlined}.
}
\label{table:experiment_u3_b}
\end{table}

\section{Conclusion and future works}
\label{sec:conclusion}

In this study, we considered hyperlink weight $w_{\bs i}$ defined for $U$-tuple $\bs X_{\bs i}$ that is a collection of $U$ data vectors $(\bs x_{i_1},\bs x_{i_2},\ldots,\bs x_{i_U})$. 
The hyperlink weights are assumed to be symmetric with respect to permutation of the entries $i_1,i_2,\ldots,i_U$ in the index. 
We proposed the BHLR that learns a user-specified symmetric similarity function $\mu_{\bs \theta}(\bs X_{\bs i})$ such that it predicts a tuple's hyperlink weight $w_{\bs i}$ through data vectors $(\bs x_{i_1},\bs x_{i_2},\ldots,\bs x_{i_U})$ stored in the corresponding $U$-tuple $\bs X_{\bs i}$. 
The BHLR encompassed various existing methods such as logistic regression~($U=1$), Poisson regression~($U=1$), graph embedding~($U=2$), matrix factorization~($U=2$), stochastic block model~($U=2$), tensor factorization~($U \geq 2$), and their variants equipped with arbitrary BD. 
We provided theoretical guarantees for BHLR including several existing methods, in the sense that general BHLR possessed the following two favorable properties: (P-1) statistical consistency and (P-2) computational tractability. 
Novel minibatch-sampling procedure for hyper-relations and theoretical guarantee for the entire stochastic optimization was also provided.

For future work, it would be worthwhile to simultaneously learn several BLHRs with different sizes of tuples; it is straightforward to modify our method to incorporate several $U$ values.
Because a single BHLR first fixes the tuple size $U \in \mathbb{N}$, the association strengths for the different sizes of tuples cannot be measured by the similarity function. 
Although we empirically demonstrated the BHLR only for $U=1,2,3$ in this study, a BHLR with a larger $U$ can be conducted, and it would be natural to learn tuples with several sizes at the same time.


Another interesting direction is designing a better similarity function for $U$-tuples. 
Although we employed limited forms of similarity functions in our numerical experiments in the current study, arbitrary similarity functions can be employed for the BHLR. 
We are especially interested in identifying highly expressive similarity functions for capturing the underlying complicated data structure. 
Some recent studies~\citep{okuno2018probabilistic,okuno2019graph,kim2019representation} demonstrated that the inner product similarity used in graph embedding~($U=2$) exhibited a limited representation capability, and more expressive similarities have been proposed; their results may be simply generalized to the setting of the BHLR with general $U \in \mathbb{N}$.

The last direction is to apply the proposed BHLR to larger-scale hypernetworks. 
Although the BHLR is already demonstrated on several thousands of nodes in our numerical experiments, a more efficient implementation is required for conducting the BHLR on much larger hypernetworks.

\section*{Acknowledgement}
This work was partially supported by JSPS KAKENHI grant 16H02789 to HS, and 17J03623 to AO.

\appendix

\section{Remaining related works}
\label{app:remaining_related_works}

In this section, we describe the remaining related works, that are not listed in Section~\ref{subsec:other_related_works}.

\bigskip
For $U=2$, 
\begin{itemize}
\item 
    \textbf{Metric learning}~\citep{bellet2013survey} is a type of similarity learning that captures the discrepancy between two data vectors $\bs x_{i_1},\bs x_{i_2}$ by some metric function. Many existing methods consider the Mahalanobis distance and Mahalanobis inner product $\bs x_{i_1}^{\top}\bs M\bs x_{i_2}$ where $\bs M \in \mathbb{R}^{p \times p}$ is a non-negative definite matrix to be estimated. Owing to the decomposition $\bs M=\bs \theta\bs \theta^{\top}$ with $\bs \theta \in \mathbb{R}^{p\times K}$, the Mahalanobis inner product measures the inner product similarity between $\bs \theta^{\top}\bs x_{i_1}$ and $\bs \theta^{\top}\bs x_{i_2}$; obtaining such a linear transformation $\bs x \mapsto \bs \theta^{\top}\bs x$ is also known as graph embedding. 
    Although the Mahalanobis metric/similarity learning above is an HLR similarly to graph embedding, it is not exactly a BHLR as most of the existing studies employ loss functions that are not exactly consistent with the BD, such as triplet loss and margin-based loss functions. 
    However, some margin-based loss functions can be written in the form of BD by removing the strict convexity assumption of $\varphi$, as explained in Section~\ref{sec:bregman_divergence}. 
\end{itemize}

For $U \ge 2$, 
\begin{itemize}
        \item \textbf{Hyperlink prediction using latent social features~(HPLSF)}~\citep{xu2013hyperlink} first computes entropy of data vectors.
Let $\bs z_{\bs i}=(z_{\bs i1},z_{\bs i2},\ldots,z_{\bs ip}) \in \mathbb{R}^p$ be a vector of entropy for each tuple $\bs X_{\bs i}$ such that the $j$-th entry $z_{\bs i j}$ ($j=1,\ldots, p$) is defined as the entropy of $\{x_{i_1 j},x_{i_2 j},\ldots,x_{i_U j}\} \subset \mathbb{R}$, where $\bs x_{i}:=(x_{i1},x_{i2},\ldots,x_{ip}) \in \mathbb{R}^p, \: i \in [n]$. 
Subsequently, hyperlink weight $w_{\bs i}$ can be predicted through the single vector $\bs z_{\bs i}$; applying a structural SVM results in a hyperlink prediction. 
As the SVM finally predicts the target label $w_{\bs i}$ through the similarity function $\mu_{\bs \theta}(\bs X_{\bs i}):=\langle \bs \theta,\bs \Psi(\bs z_i) \rangle$ with a high-dimensional feature map $\bs \Psi:\mathbb{R}^p \to \mathbb{R}^{p'}$, the HPLSF is an HLR. However, the similarity function is typically trained with some loss functions that are not consistent with the BD; the HPLSF is not exactly included in the BHLR. 

\item \textbf{Coordinated matrix minimization~(CMM)}~\citep{zhang2018beyond} efficiently infers a subset of user-specified candidate hyperlinks that are the most suitable to fill the training hypernetworks using a low-rank approximation. 
However, CMM can find hyperlinks only among the training nodes, implying that it cannot be used for obtaining hyperlinks among test nodes outside the training dataset. CMM is neither an HLR or a BHLR.

\item \textbf{Deep Sets}~\citep{zaheer2017deep} provides a permutation invariant expressive similarity function $\tilde{\mu}_{\bs \theta}:2^{\mathcal{X}} \to \mathbb{R}$ defined for sets of data vectors. 
The function $\tilde{\mu}_{\bs \theta}$ is trained by leveraging KL-divergence and logistic loss, whereas BHLR is equipped with arbitrary Bregman divergence. 
Although the similarity function of Deep Sets can be used for BHLR, the functional form is more restrictive than those considered in our setting.
For paying the price of arbitrary size of vector sets, 
their Theorem~2 proves that a function $\tilde{\mu}_{\bs \theta}:2^{\mathcal{X}} \to \mathbb{R}$ is permutation invariant if and only if $\tilde{\mu}_{\bs \theta}$ is in the form of $\tilde{\mu}_{\bs \theta}(\bs x_{i_1},\bs x_{i_2},\ldots)=\rho_{\bs \theta}(\sum_{u}\phi_{\bs \theta}(\bs x_{i_u}))$ for some functions $\rho_{\bs \theta}$ and $\phi_{\bs \theta}$, by assuming that the set $\mathcal{X}$ is countable, or the dimension of $\mathcal{X}$ is $1$. 
\end{itemize}

\section{Tensor factorization~(TF) is a special case of BHLR}
\label{app:relation_to_ntf}
As explained in Section~\ref{subsec:u_general}, tensor factorization~(TF)~\citep{cichocki2009nonnegative} decomposes a given tensor $\bs V=(v_{\bs j}) \in \mathbb{R}^{n_1 \times n_2 \times \cdots \times n_U}$ into matrices $\bs \xi^{(u)}=(\xi^{(u)}_{ik}) \in \mathbb{R}^{n_u \times K}$, by minimizing the BD between the entries of 
$\bs V$ and $[\![\bs \xi^{(1)},\bs \xi^{(2)},\ldots,\bs \xi^{(U)}]\!]$ whose $\bs j=(j_1,j_2,\ldots,j_U)$-th entry is specified as  $\sum_{k=1}^{K}\xi^{(1)}_{j_1 k} \xi^{(2)}_{j_2 k} \cdots \xi^{(U)}_{j_U k}$. 
Namely, TF minimizes the BD 
\begin{align}
D_{\varphi}(
\{v_{\bs j}\}_{\bs j \in [n_1] \times [n_2] \times \cdots \times [n_U]}
,
\{\langle \bs \xi_{j_1}^{(1)},
        \bs \xi_{j_2}^{(2)}
        ,
        \ldots
        ,
        \bs \xi_{j_U}^{(U)} \rangle\}_{\bs j \in [n_1] \times [n_2] \times \cdots \times [n_U]})
    \label{eq:original_ntf_objective}
\end{align}
where $\langle \bs y,\bs y',\bs y''\ldots \rangle:=\sum_{k=1}^{K} y_k y_k' y_k'' \cdots$, and $\bs \xi_l^{(u)}=(\xi_{l1}^{(u)},\xi_{l2}^{(u)},\ldots,\xi_{lK}^{(u)}) \: (l \in [n_u])$ are column vectors of the matrix $\bs \xi^{(u)}$. 
Subsequently, we can expect that 
$v_{\bs j} \approx 
    \langle 
        \bs \xi_{j_1}^{(1)},
        \bs \xi_{j_2}^{(2)}
        ,
        \ldots
        ,
        \bs \xi_{j_U}^{(U)}
    \rangle$ for all $\bs j \in [n_1] \times [n_2] \times \cdots \times [n_U]$.

For showing that BHLR includes TF~($U \geq 2$), 
we first briefly review the relation between BHLR and MF~($U=2$), that is explained in Section~\ref{subsec:u2}. 
In the case of $U=2$, factorizing the matrix $\bs V$ corresponds to BHLR using
\begin{align}
    \bs W=\left(
        \begin{array}{cc}
            \bs O_{n_1 \times n_1} & \bs V \\
            \bs V^{\top} & \bs O_{n_2 \times n_2} \\
        \end{array}
    \right),
    \label{eq:V_to_W}
\end{align}
that is defined in eq.~(\ref{eq:def_V_to_W}). 
The link weights~(\ref{eq:V_to_W}) indicate $v_{\bs j}=v_{j_1,j_2}=w_{j_1,n_1+j_2}=w_{\bs i}$; indices of the matrix $\bs V=(v_{\bs j})$ are formally transformed into those of the matrix $\bs W=(w_{\bs i})$, by utilizing the conversion $\mathcal{F}:(j_1,j_2) \mapsto (j_1,n_1+j_2)=:(i_1,i_2)$. 
Although this conversion only considers the correspondence between $\bs V$ and the upper-right part of the matrix $\bs W$, the lower-left part is specified by the symmetry of $\bs W$. 
In the case of $U \geq 2$, we generalize the conversion as
\begin{align*}
    \mathcal{F}:
    (j_1,j_2,\ldots,j_U)
    &\mapsto 
    \left(
    j_1, \: n_1+j_2, \: (n_1+n_2)+j_3, \: \ldots, \: \big(\sum_{u=1}^{U-1}n_u\big)+j_U \right) 
    =:
    (i_1,i_2,\ldots,i_U),
\end{align*}
whose inverse $\mathcal{F}^{-1}$ can be defined over a set
\begin{align*}
    \mathcal{C}(n_1,n_2,\ldots,n_U)
&:=
\{
    \bs i \mid 
    \bs i=\mathcal{F}(\bs j),
    \bs j \in [n_1] \times [n_2] \times \cdots \times [n_U]
\} \nonumber \\
&=
\{\bs i=(i_1,i_2,\ldots,i_U) \mid i_1=1,2,\ldots,n_1; \nonumber \\
&\hspace{3em} i_2=n_1+1,n_1+2,\ldots,n_1+n_2; \cdots; i_U=\sum_{u=1}^{U-1}n_u+1,\ldots,\sum_{u=1}^{U}n_u\}, 
\end{align*}
such that $\mathcal{F}^{-1}:\mathcal{C}(n_1,n_2,\ldots,n_U) \ni \bs i \mapsto \bs j \in [n_1] \times [n_2] \times \cdots \times [n_U]$. 
Since $\mathcal{F}^{-1}$ converts the indices of $\bs W=(w_{\bs i})$ to those of $\bs V=(v_{\bs j})$, we may specify the hyperlink weights as $w_{\bs i}:=v_{\mathcal{F}^{-1}(\bs i)}$ for all $\bs i \in \mathcal{C}(n_1,n_2,\ldots,n_U)$, similarly to $U=2$.

Although the above specification is essentially sufficient for describing the relation between BHLR and TF, the hyperlink weights $\bs W=(w_{\bs i})$ are assumed to be symmetric as explained in Section~\ref{subsec:problem_setting}. 
The symmetry can be realized by considering the non-decreasing order permutation $r(\bs i)$ defined for any $\bs i$; 
a tensor $\bs W=(w_{\bs i}) \in \mathbb{R}^{N^U}$ ($N:=\sum_{u=1}^{U}n_u$), whose entries are specified as  
\begin{align}
    w_{\bs i}
    &:=
    \begin{cases}
        v_{\mathcal{F}^{-1}(r(\bs i))} & (r(\bs i) \in \mathcal{C}(n_1,n_2,\ldots,n_U)) \\
        0 & (\text{otherwise})
    \end{cases}
    \quad 
    (
        \forall \bs i \in [N]^U 
    ),
    \label{eq:V_to_W_tensor}
\end{align}
simultaneously satisfies the symmetry $w_{\bs i}=w_{\bs i'}$ for any $\bs i' \in [N]^U$ obtained by permutating the entries of $\bs i \in [N]^U$, and 
the above specification $w_{\bs i}=v_{\mathcal{F}^{-1}(\bs i)}$ for any $\bs i \in \mathcal{C}(n_1,n_2,\ldots,n_U)$. 
Therefore, (\ref{eq:V_to_W_tensor}) generalizes (\ref{eq:V_to_W}) from the case of $U=2$ to $U \geq 2$.

Using the hyperlink weights~(\ref{eq:V_to_W_tensor}), 
the parameter $\bs \theta=(\bs \xi^{(1)\top},\bs \xi^{(2)\top},\ldots,\bs \xi^{(U)\top})^{\top} \in \mathbb{R}^{N \times K}$, and one-hot vector $\bs x_i \in \{0,1\}^{N}$ whose $i$-th entry is $1$ and $0$ otherwise ($i \in [N]$), 
we have
\begin{align}
(\ref{eq:original_ntf_objective})
=
D_{\varphi}(\{w_{\bs i}\}_{\bs i \in \mathcal{C}(n_1,n_2,\ldots,n_U)} 
    , 
    \{
    \underbrace{
    \langle 
    \bs \theta^{\top}\bs x_{i_1},
    \bs \theta^{\top}\bs x_{i_2},
    \ldots,
    \bs \theta^{\top}\bs x_{i_U}
\rangle}_{=:\mu_{\bs \theta}(\bs X_{\bs i})}\}_{\bs i \in \mathcal{C}(n_1,n_2,\ldots,n_U)}),
\label{eq:final_ntf_objective}
\end{align}
generalizing eq.~(\ref{eq:nmf_and_bhlr}) from $U=2$ to $U \geq 2$. 
Therefore, TF that minimizes $(\ref{eq:original_ntf_objective})$ is equivalent to BHLR minimizing $(\ref{eq:final_ntf_objective})$; TF is a special case of BHLR.

\section{Proofs}

In \ref{subsec:prelimiary_for_proofs}, we first show and prove Theorem~\ref{theo:mlln}, that is the law of large numbers for multiply-indexed partially-dependent random variables. 
In \ref{app:proof_of_prop:bregman_converge}, 
we prove Proposition~\ref{prop:bregman_converge} by applying Theorem~\ref{theo:mlln}. 
In \ref{app:proof_of_theo:consistency}, we prove Theorem~\ref{theo:consistency}, indicating that BHLR asymptotically recovers the underlying conditional expectation of link weights as $n \to \infty$. 
In \ref{app:proof_of_theo:sgd_convergence}, we last prove Theorem~\ref{theo:sgd_convergence}, showing the asymptotics of the minibatch SGD using the proposed Algorithm~\ref{alg:minibatch}, as $T \to \infty$. 

\subsection{Preliminary for proofs}
\label{subsec:prelimiary_for_proofs}

\begin{theo}
\normalfont
\label{theo:mlln}
Let $\bs Z:=(Z_{\bs i})$ be an array of random variables $Z_{\bs i} \in \mathcal{Z}$,  $\bs i \in \mathcal{I}_n^{(U)}
=
\mathcal{J}_n^{(U)}
\overset{(\ref{eq:inu})}{:=}
\{(i_1,i_2,\ldots,i_U) \mid 1 \leq i_1 < i_2 < \cdots < i_U \leq n\}$, and $h:\mathcal{Z} \to \mathbb{R}$ be a continuous function. 
We assume that $Z_{\bs i}$ is independent of $Z_{\bs j}$ if $\bs j \in \mathcal{R}_n^{(U)}(\bs i):=\{(j_1,j_2,\ldots,j_U) \in \mathcal{I}_n^{(U)} \mid j_1,j_2,\ldots,j_U \in \{1,\ldots, n\} \setminus \{i_1,i_2,\ldots,i_U\} \}$, and 
$\mathbb{E}_{\bs Z}(h(Z_{\bs i})^2)<\infty$, for all $\bs i \in \mathcal{I}_n^{(U)}$. 
Then the average of $h(Z_{\bs i})$ over $\mathcal{I}_n^{(U)}$ converges to the expectation in probability as $n\to\infty$; that is
\begin{align*}
	\frac{1}{|\mathcal{I}_n^{(U)}|}\sum_{\bs i \in \mathcal{I}_n^{(U)}} h(Z_{\bs i})
	=
	\frac{1}{|\mathcal{I}_n^{(U)}|}\sum_{\bs i \in \mathcal{I}_n^{(U)}} \mathbb{E}_{\bs Z}(h(Z_{\bs i}))
	+
	O_p(1/\sqrt{n}).	
\end{align*}
\end{theo}

\textbf{Proof of Theorem~\ref{theo:mlln}.} 
Proof is almost the same as that of \citet{okuno2019robust} Theorem~A.1, that indicates the same assertion for $U=2$. 
Regarding the variance of the average, we have
\begin{align}
&\mathbb{V}_{\bs Z}\left( \frac{1}{|\mathcal{I}_n^{(U)}|}\sum_{\bs i \in \mathcal{I}_n^{(U)}}h(Z_{\bs i}) \right) \nonumber \\
&=
\mathbb{E}_{\bs Z}\left(
	\left(\frac{1}{|\mathcal{I}_n^{(U)}|}\sum_{\bs i \in \mathcal{I}_n^{(U)}}h(Z_{\bs i})\right)^2
\right)
-
\mathbb{E}_{\bs Z}\left(
	\frac{1}{|\mathcal{I}_n^{(U)}|}\sum_{\bs i \in \mathcal{I}_n^{(U)}}h(Z_{\bs i})
\right)^2 \nonumber \\
&=
\frac{1}{|\mathcal{I}_n^{(U)}|^2}
\left(
\sum_{\bs i \in \mathcal{I}_n^{(U)}}
\sum_{\bs j \in \mathcal{I}_n^{(U)}}
\mathbb{E}_{\bs Z}\left(
	h(Z_{\bs i})h(Z_{\bs j})
\right)
-
\left(
\sum_{\bs i \in \mathcal{I}_n^{(U)}}
\mathbb{E}_{\bs Z}\left(h(Z_{\bs i})\right)
\right)
^2
\right) \nonumber \\
&=
\frac{1}{|\mathcal{I}_n^{(U)}|^2}
\sum_{\bs i \in \mathcal{I}_n^{(U)}}
\sum_{\bs j \in \mathcal{I}_n^{(U)} \setminus \mathcal{R}_n^{(U)}(\bs i)}
\left(
	\mathbb{E}_{\bs Z}\left(h(Z_{\bs i})h(Z_{\bs j})\right)
	-
	\mathbb{E}_{\bs Z} (h(Z_{\bs i})) 
	\mathbb{E}_{\bs Z} (h(Z_{\bs j}))
\right), \label{eq:vh_last}
\end{align}
where $\mathbb{E}_{\bs Z},\mathbb{V}_{\bs Z}$ represent expectation and variance with respect to $\bs Z$. 
Considering $\mathbb{E}_{\bs Z}(|h(Z_{\bs i})|)\le \mathbb{E}_{\bs Z}(h(Z_{\bs i})^2)^{1/2}<\infty,
\mathbb{E}_{\bs Z}(|h(Z_{\bs i})h(Z_{\bs j})|)\le \sqrt{\mathbb{E}_{\bs Z}(h(Z_{\bs i})^2)\mathbb{E}_{\bs Z}(h(Z_{\bs j})^2)}<\infty$, $|\mathcal{I}_n^{(U)}|=O(n^{U})$, and  
\begin{align*}
    |\mathcal{I}_n^{(U)} \setminus \mathcal{R}_n^{(U)}(\bs i)|
    &=
    \bigg|
        \left\{ 
        (j_1,j_2,\ldots,j_U) \in \mathcal{I}_n^{(U)} \big| 
        \exists u \in \{1,2,\ldots,U\} \text{ s.t. }
        j_u \in \{i_1,i_2,\ldots,i_U\}
        \right\}
    \bigg| \\
    &\leq 
    \sum_{u=1}^{U}
    \bigg|
        \left\{ 
        (j_1,j_2,\ldots,j_U) \in \mathcal{I}_n^{(U)} \big| 
        j_u \in \{i_1,i_2,\ldots,i_U\}
        \right\}
    \bigg| \\
    &=
    \sum_{u=1}^{U}
    \sum_{l=1}^{U}
    \bigg|
        \left\{ 
        (j_1,\ldots,j_{u-1},i_l,j_{u+1},\ldots,j_U) \in \mathcal{I}_n^{(U)} 
        \right\}
    \bigg| \\
    &=
    O(U^2 n^{U-1})=O(n^{U-1})
\end{align*} 
for any fixed $\bs i=(i_1,i_2,\ldots,i_U) \in \mathcal{I}_n^{(U)}$, the formula (\ref{eq:vh_last}) is of order $O(n^{-2U} \cdot n^{U} \cdot n^{U-1})=O(n^{-1})$. Therefore, 
\begin{align}
\mathbb{V}_{\bs Z}\left( \frac{1}{|\mathcal{I}_n^{(U)}|}\sum_{\bs i \in \mathcal{I}_n^{(U)}}h(Z_{\bs i}) \right)
=
O(n^{-1}).
\label{eq:vz}
\end{align}
(\ref{eq:vz}) and Chebyshev's inequality indicate the assertion. 
\qed

This theorem generalizes \citet{okuno2019robust} Theorem~A.1, that proves the same assertion for $U=2$. 
We note that the convergence rate is $O_p(n^{-1/2})$ but not $O_p(1/|\mathcal{I}_n^{(U)}|^{1/2})=O_p(n^{-U/2})$, even though we leverage $|\mathcal{I}_n^{(U)}|=O(n^{U})$ observations $\{Z_{\bs i}\}_{\bs i \in \mathcal{I}_n^{(U)}}$. 

\subsection{Proof of Proposition~\ref{prop:bregman_converge}}
\label{app:proof_of_prop:bregman_converge}
By a simple calculation, we have
    \begin{align}
        L_{\varphi,n}(\bs \theta) 
        &=
        \frac{1}{|\mathcal{I}_n^{(U)}|}
        \sum_{\bs i \in \mathcal{I}_n^{(U)}}
        \left\{
            \varphi(w_{\bs i})
            -
            \varphi(\mu_{\bs \theta}(\bs X_{\bs i}))
            -
            \varphi'(\mu_{\bs \theta}(\bs X_{\bs i}))
            (w_{\bs i}-\mu_{\bs \theta}(\bs X_{\bs i}))
        \right\} \nonumber \\
        &=
        \frac{1}{|\mathcal{I}_n^{(U)}|}
        \sum_{\bs i \in \mathcal{I}_n^{(U)}}
        \underbrace{
        \{ \varphi(\mu_*(\bs X_{\bs i}))-\varphi(\mu_{\bs \theta}(\bs X_{\bs i}))
        -
        \varphi'(\mu_{\bs \theta}(\bs X_{\bs i}))
        (\mu_*(\bs X_{\bs i})-\mu_{\bs \theta}(\bs X_{\bs i})) \}
        }_{=d_{\varphi}(\mu_*(\bs X_{\bs i}),\mu_{\bs \theta}(\bs X_{\bs i}))}
        \label{eq:ell_1} \\
        &\hspace{2em}+
        \frac{1}{|\mathcal{I}_n^{(U)}|}
        \sum_{\bs i \in \mathcal{I}_n^{(U)}}
        \left\{
            \varphi(w_{\bs i})
            -
            \varphi(\mu_{*}(\bs X_{\bs i}))
        \right\} \label{eq:ell_2}\\
        &\hspace{4em}
        +
        \frac{1}{|\mathcal{I}_n^{(U)}|}
        \sum_{\bs i \in \mathcal{I}_n^{(U)}}
        \left\{
            \varphi'(\mu_{\bs \theta}(\bs X_{\bs i}))
            (\mu_{*}(\bs X_{\bs i})-w_{\bs i})
        \right\}. \label{eq:ell_3}
    \end{align}
    Under the conditions (C-1)--(C-5), Theorem~\ref{theo:mlln} can be applied to the terms~(\ref{eq:ell_1})--(\ref{eq:ell_3}) as shown in the following:

    specifying $Z_{\bs i}:=\bs X_{\bs i},h(Z_{\bs i}):=d_{\varphi}(\mu_*(\bs X_{\bs i}),\mu_{\bs \theta}(\bs X_{\bs i}))$ leads to 
    \begin{align*}
    (\ref{eq:ell_1})
    &\overset{\text{Theorem~\ref{theo:mlln}}}{=}
    \frac{1}{|\mathcal{I}_n^{(U)}|}\sum_{\bs i \in \mathcal{I}_n^{(U)}}
    \mathbb{E}_{\mathcal{X}^U}(d_{\varphi}(\mu_*(\bs X_{\bs i}),\mu_{\bs \theta}(\bs X_{\bs i})))+O_p(1/\sqrt{n}) \\
    &=
    \mathbb{E}_{\mathcal{X}^U}(d_{\varphi}(\mu_*(\bs X),\mu_{\bs \theta}(\bs X)))+O_p(1/\sqrt{n}), 
    \end{align*}

    specifying $Z_{\bs i}:=(w_{\bs i},\bs X_{\bs i}),h(Z_{\bs i}):=\varphi(w_{\bs i})-\varphi(\mu_*(\bs X_{\bs i}))$ leads to 
    \begin{align*}
    (\ref{eq:ell_2})
    &\overset{\text{Theorem~\ref{theo:mlln}}}{=}
    \frac{1}{|\mathcal{I}_n^{(U)}|}\sum_{\bs i \in \mathcal{I}_n^{(U)}}
    \mathbb{E}_{Z_{\bs i}}\left(
        \varphi(w_{\bs i})-\varphi(\mu_*(\bs X_{\bs i}))
    \right) + O_p(1/\sqrt{n}) \\
    &=
    \frac{1}{|\mathcal{I}_n^{(U)}|}\sum_{\bs i \in \mathcal{I}_n^{(U)}}
    \mathbb{E}_{\mathcal{X}^U}\left(
        \mathbb{E}(\varphi(w_{\bs i}) \mid \bs X_{\bs i})-\varphi(\mu_*(\bs X_{\bs i}))
    \right) + O_p(1/\sqrt{n}) \\
    &=
    \mathbb{E}_{\mathcal{X}^U}\left(
        \mathbb{E}(\varphi(w) \mid \bs X)-\varphi(\mu_*(\bs X))
    \right) + O_p(1/\sqrt{n}) \\
    &=
    C_{\varphi}+O_p(1/\sqrt{n}), 
    \end{align*}
    
    and specifying $Z_{\bs i}:=(w_{\bs i},\bs X_{\bs i}),h(Z_{\bs i}):=\varphi'(\mu_{\bs \theta}(\bs X_{\bs i}))(\mu_{*}(\bs X_{\bs i})-w_{\bs i})$ leads to
    \begin{align*}
    (\ref{eq:ell_3})
    &\overset{\text{Theorem~\ref{theo:mlln}}}{=}
    \mathbb{E}_{Z_{\bs i}}\left(
        \varphi'(\mu_{\bs \theta}(\bs X_{\bs i}))(\mu_{*}(\bs X_{\bs i})-w_{\bs i})
    \right) + O_p(1/\sqrt{n}) \\
    &=
    \mathbb{E}_{\mathcal{X}^U}(
        \varphi'(\mu_{\bs \theta}(\bs X_{\bs i}))(
            \underbrace{\mu_{*}(\bs X_{\bs i})-\underbrace{\mathbb{E}(w_{\bs i} \mid \bs X_{\bs i})}_{=\mu_*(\bs X_{\bs i})}
            }_{=0}
            )) + O_p(1/\sqrt{n}) \\ 
    &=
    O_p(1/\sqrt{n}).
    \end{align*}
    
    Thus proving the assertion 
    \begin{align*}
        L_{\varphi,n}(\bs \theta) 
        &=
        (\ref{eq:ell_1})
        +
        (\ref{eq:ell_2})
        +
        (\ref{eq:ell_3}) \\
        &=
        \mathbb{E}_{\mathcal{X}^U}(d_{\varphi}(\mu_*(\bs X),\mu_{\bs \theta}(\bs X)))
        +
        C_{\varphi}
        +
        O_p(1/\sqrt{n}).
    \end{align*}
    \qed

\subsection{Proof of Theorem~\ref{theo:consistency}}
\label{app:proof_of_theo:consistency}

Definition of the estimator (\ref{eq:estimator}) leads to
\begin{align}
    L_{\varphi,n}(\bs \theta_*)-C_{\varphi}
    \geq 
    \min_{\bs \theta \in \bs \Theta}L_{\varphi,n}(\bs \theta)-C_{\varphi}
    =
    L_{\varphi,n}(\hat{\bs \theta}_{\varphi,n})-C_{\varphi}.
    \label{eq:inequality_L}
\end{align}
We evaluate both sides of the inequality (\ref{eq:inequality_L}), for proving the assertion.  

\begin{itemize}
    \item Regarding the left-hand side of the inequality (\ref{eq:inequality_L}), Proposition~\ref{prop:bregman_converge} indicates that
    \begin{align}
        L_{\varphi,n}(\bs \theta_*)-C_{\varphi}
        &=
        L_{\varphi,n}(\bs \theta)\bigg|_{\bs \theta=\bs \theta_*}-C_{\varphi} \nonumber \\
        &\overset{\text{Proposition}~\ref{prop:bregman_converge}}{=} 
        \left( 
        \mathbb{E}_{\mathcal{X}^U}(
            d_{\varphi}( \mu_*(\bs X),\mu_{\bs \theta}(\bs X)))
        \bigg|_{\bs \theta=\bs \theta_*}
        +C_{\varphi} + \varepsilon^{(1)}_n\right)
        -
        C_{\varphi} \nonumber \\
        &=
        \underbrace{
        \mathbb{E}_{\mathcal{X}^U}(
            d_{\varphi}( \mu_*(\bs X),\mu_{\bs \theta_*}(\bs X)))
        }_{=0}+\varepsilon^{(1)}_n \\
        &=
        \varepsilon_n^{(1)}
        \label{eq:lhs}
    \end{align}
    where $\varepsilon^{(1)}_n:=
    L_{\varphi,n}(\bs \theta_*)
    -
    \left(
        \mathbb{E}_{\mathcal{X}^U}(\mu_*(\bs X),\mu_{\bs \theta_*}(\bs X))
        +
        C_{\varphi}
    \right)
    =
    O_p(1/\sqrt{n})$.
    
    \item We here consider the right-hand side of the inequality (\ref{eq:inequality_L}). 
    Since the function $\varphi$ is strongly convex, 
    the definition indicates the existence of $M_{\varphi}>0$ such that 
    \begin{align*}
        d_{\varphi}(a,b)
        =
        \varphi(a)-(\varphi(b)+\varphi'(b)(a-b))
        \overset{(\because \: \text{srtongly convex})}{\geq} 
        M_{\varphi} \cdot (a-b)^2,
    \end{align*}
    for all $a,b \in \text{dom}(\varphi)$. 
    This inequality indicates that the squared difference is bounded by the function $d_{\varphi}$. 
    By substituting $\mu_*(\bs X),\mu_{\bs \theta}(\bs X)$ into $a,b$, respectively, we have an inequality
    \begin{align}
        d_{\varphi}( \mu_*(\bs X),\mu_{\bs \theta}(\bs X) )
        \geq 
        M_{\varphi} \cdot ( \mu_*(\bs X)-\mu_{\bs \theta}(\bs X) )^2,
        \quad 
        (\forall \bs \theta \in \bs \Theta).
        \label{eq:bregman_inequality}
    \end{align}
    Using the above inequality~(\ref{eq:bregman_inequality}), the right-hand side of the inequality~(\ref{eq:inequality_L}) is evaluated as
     \begin{align}
        L_{\varphi,n}(\hat{\bs \theta}_{\varphi,n})-C_{\varphi}
        &=
        L_{\varphi,n}(\bs \theta) \bigg|_{\bs \theta=\hat{\bs \theta}_{\varphi,n}}-C_{\varphi} \nonumber \\
        &\overset{\text{Proposition}~\ref{prop:bregman_converge}}{=}
        \left(\mathbb{E}_{\mathcal{X}^U}(d_{\varphi}( \mu_*(\bs X),\mu_{\bs \theta}(\bs X) )) +
        C_{\varphi}+\varepsilon^{(2)}_n(\bs \theta)
        \right)
        \bigg|_{\bs \theta=\hat{\bs \theta}_{\varphi,n}}
        -
        C_{\varphi} \nonumber \\
        &=
        \mathbb{E}_{\mathcal{X}^U}(d_{\varphi}( \mu_*(\bs X),\mu_{\bs \theta}(\bs X) )) \bigg|_{\bs \theta=\hat{\bs \theta}_{\varphi,n}} + \varepsilon^{(2)}_n (\hat{\bs \theta}_{\varphi,n}) \nonumber \\
        &\overset{\text{Ineq.}~(\ref{eq:bregman_inequality})}{\geq} 
        M_{\varphi} \cdot \mathbb{E}_{\mathcal{X}^U}(( \mu_*(\bs X)-\mu_{\bs \theta}(\bs X) )^2)\bigg|_{\bs \theta=\hat{\bs \theta}_{\varphi,n}} + 
        \varepsilon^{(2)}_n (\hat{\bs \theta}_{\varphi,n}), \nonumber \\
        &=
        M_{\varphi} \cdot \|\mu_*-\mu_{\hat{\bs \theta}_{\varphi,n}}\|^2
        +
        \varepsilon^{(2)}_n (\hat{\bs \theta}_{\varphi,n})
        \label{eq:rhs}
    \end{align}
    where 
    $\|f\|:=\mathbb{E}_{\mathcal{X}^U}(f(\bs X)^2)^{1/2}$ for functions $f:\mathcal{X}^U \to \mathbb{R}$ and $\varepsilon_n^{(2)}(\bs \theta):=L_{\varphi,n}(\bs \theta)-\left\{\mathbb{E}_{\mathcal{X}^U}\left(
        d_{\varphi}(\mu_*(\bs X),\mu_{\bs \theta}(\bs X)
    \right)+C_{\varphi}\right\}$ represents the residual in Proposition~\ref{prop:bregman_converge} using the parameter $\bs \theta$, 
    that satisfies $\varepsilon_n^{(2)}(\bs \theta)=O_p(1/\sqrt{n})$ for each $\bs \theta \in \bs \Theta$.
\end{itemize}

By substituting (\ref{eq:lhs}) and (\ref{eq:rhs}) into (\ref{eq:inequality_L}), we have 
\begin{align*}
    \varepsilon_n^{(1)}
    \geq 
    M_{\varphi} \cdot \|\mu_*-\mu_{\hat{\bs \theta}_{\varphi,n}}\|^2
    +
    \varepsilon^{(2)}_n (\hat{\bs \theta}_{\varphi,n}),
\end{align*}
indicating that
\begin{align}
    \varepsilon^{(1)}_n-\varepsilon^{(2)}_n (\hat{\bs \theta}_{\varphi,n})
    \geq 
    M_{\varphi} \cdot \|\mu_*-\mu_{\hat{\bs \theta}_{\varphi,n}}\|^2
    \geq 0,
    \label{eq:upper_bound_mse}
\end{align}
where $\varepsilon_n^{(1)}=O_p(1/\sqrt{n})=o_p(1)$. 
The term $\varepsilon^{(2)}_n (\hat{\bs \theta}_{\varphi,n})$ is proved to be $o_p(1)$, as shown in the remaining of this proof; then, (\ref{eq:upper_bound_mse}) immediately proves Theorem~\ref{theo:consistency}.

Hereinafter, we last prove $\varepsilon^{(2)}_n (\hat{\bs \theta}_{\varphi,n})=o_p(1)$, 
by employing \citet{newey1991uniform} Corollary 2.2, indicating that $\sup_{\bs \theta \in \bs \Theta} |\varepsilon_n^{(2)}(\bs \theta)|=o_p(1)$ under the following assumptions: (i) $\bs \Theta$ is compact, 
(ii) $\varepsilon_n^{(2)}(\bs \theta) =o_p(1)$ for each $\bs \theta \in \bs \Theta$, and 
(iii) $\exists B_n=O_p(1)$ such that $|\varepsilon_n^{(2)}(\bs \theta)-\varepsilon_n^{(2)}(\bs \theta')| \leq B_n\|\bs \theta-\bs \theta'\|_2$ for all $\bs \theta,\bs \theta' \in \bs \Theta$. 
Above assumptions (i), (ii) and (iii) correspond to assumptions 1, 2 and 3A, in \citet{newey1991uniform}. 
In our setting, the assumption (i) is assumed, (ii) is proved by Proposition~\ref{prop:bregman_converge}. 
(iii) is obtained similarly to Proof B.1 in Supplement of \citet{okuno2018probabilistic}; 
since the product of two bounded Lipschitz continuous~(LC) functions is LC, 
$C^1$-function applied to LC function is LC, and the expectation of LC function is also LC, 
there exist $M_1,M_2>0$ such that
\begin{align*}
    |\varepsilon_n^{(2)}(\bs \theta)-\varepsilon_n^{(2)}(\bs \theta')|
    &\leq
    \bigg|
    L_{\varphi,n}(\bs \theta)
    -
    L_{\varphi,n}(\bs \theta')
    \bigg|
    +
    \bigg|
        \mathbb{E}_{\mathcal{X}^U}(d_{\varphi}(\mu_*(\bs X),\mu_{\bs \theta}(\bs X)))
        -
        \mathbb{E}_{\mathcal{X}^U}(d_{\varphi}(\mu_*(\bs X),\mu_{\bs \theta'}(\bs X)))
    \bigg|\\
    &\leq
    \frac{1}{|\mathcal{I}_n^{(U)}|}
    \sum_{\bs i \in \mathcal{I}_n^{(U)}}
    \bigg|
            \varphi'(\mu_{\bs \theta}(\bs X_{\bs i}))w_{\bs i}
            -
            \varphi'(\mu_{\bs \theta'}(\bs X_{\bs i}))w_{\bs i}
    \bigg| \\
    &\hspace{2em}+
    \frac{1}{|\mathcal{I}_n^{(U)}|}
    \sum_{\bs i \in \mathcal{I}_n^{(U)}}
    \bigg|
    \underbrace{
        \varphi'(\mu_{\bs \theta}(\bs X_{\bs i}))\mu_{\bs \theta}(\bs X_{\bs i})
    }_{(\text{Lipschitz})}
        -
        \varphi'(\mu_{\bs \theta'}(\bs X_{\bs i}))\mu_{\bs \theta'}(\bs X_{\bs i})
    \bigg|\\  
    &\hspace{4em}+
    \frac{1}{|\mathcal{I}_n^{(U)}|}
    \sum_{\bs i \in \mathcal{I}_n^{(U)}}
    \bigg|
    \underbrace{
        \varphi(\mu_{\bs \theta}(\bs X_{\bs i}))
    }_{(\text{Lipschitz})}
        -
        \varphi(\mu_{\bs \theta'}(\bs X_{\bs i}))
    \bigg| \\
    &\hspace{6em}+
    \bigg|
    \underbrace{
    \mathbb{E}_{\mathcal{X}^U}\left(
        \varphi'(\mu_{\bs \theta}(\bs X))
            \mu_{*}(\bs X)
    \right)
    }_{(\text{Lipschitz})}
        -
    \mathbb{E}_{\mathcal{X}^U}\left(
            \varphi'(\mu_{\bs \theta'}(\bs X))
            \mu_{*}(\bs X)
    \right)
    \bigg| \\
    &\hspace{8em}+
    \bigg|
    \underbrace{
    \mathbb{E}_{\mathcal{X}^U}\left(
            \varphi'(\mu_{\bs \theta}(\bs X))
            \mu_{\bs \theta}(\bs X)
    \right)
    }_{(\text{Lipschitz})}
        -
    \mathbb{E}_{\mathcal{X}^U}\left(
            \varphi'(\mu_{\bs \theta'}(\bs X))
            \mu_{\bs \theta'}(\bs X)
    \right) 
    \bigg| \\
    &\hspace{10em}+
    \bigg|
    \underbrace{
    \mathbb{E}_{\mathcal{X}^U}\left(
            \varphi(\mu_{\bs \theta}(\bs X))
    \right)
    }_{(\text{Lipschitz})}
        -
    \mathbb{E}_{\mathcal{X}^U}\left(
            \varphi(\mu_{\bs \theta'}(\bs X))
    \right)
    \bigg| \\
    &\leq 
    \frac{1}{|\mathcal{I}_n^{(U)}|}
    \sum_{\bs i \in \mathcal{I}_n^{(U)}}
    |w_{\bs i}| 
    \bigg|
        \underbrace{
            \varphi'(\mu_{\bs \theta}(\bs X_{\bs i}))
        }_{(\text{Lipschitz})}
            -
            \varphi'(\mu_{\bs \theta'}(\bs X_{\bs i}))
    \bigg|
    +
    M_2\|\bs \theta-\bs \theta'\|_2\\
    &\leq 
    M_1
    \left( 
    \frac{1}{|\mathcal{I}_n^{(U)}|}
    \sum_{\bs i \in \mathcal{I}_n^{(U)}}
    |w_{\bs i}|
    \right)
    \cdot 
    \|\bs \theta-\bs \theta'\|_2
    +
    M_2\|\bs \theta-\bs \theta'\|_2.
\end{align*}
Denoting by $B_n:=M_1\left( 
    \frac{1}{|\mathcal{I}_n^{(U)}|}
    \sum_{\bs i \in \mathcal{I}_n^{(U)}}
    |w_{\bs i}|
    \right)+M_2$, Proposition~\ref{prop:bregman_converge} indicates $B_n=O_p(1)$. 
    Therefore the condition (iii) holds; \citet{newey1991uniform} Corollary 2.2 proves
    \begin{align}
    |\varepsilon_n^{(2)}(\hat{\bs \theta}_{\varphi,n})|
    \le
    \sup_{\bs \theta \in \bs \Theta}
    |\varepsilon_n^{(2)}(\bs \theta)|
    \overset{\text{\citet{newey1991uniform} Corollary 2.2}}{=}
    o_p(1),
    \label{eq:sup_evaluation}
    \end{align}
    indicating that $\varepsilon_n^{(2)}(\hat{\bs \theta}_{\varphi,n})=o_p(1)$. 
\qed


\subsection{Proof of Theorem~\ref{theo:sgd_convergence}}
\label{app:proof_of_theo:sgd_convergence}

Proof is two-folded. 
In the following, we first verify that 
(i) $\mathbb{E}_{\mathcal{M}^{(t)}}(\tilde{g}^{(t)}_{\eta}(\bs \theta))=\alpha \frac{\partial}{\partial \bs \theta}Q_{\eta}(\bs \theta)$, where 
\[
Q_{\eta}(\bs \theta):=D_{\varphi}(
\{\eta w_{\bs i}\}_{\bs i \in \mathcal{I}_n^{(U)}},
\{\mu_{\bs \theta}(\bs X_{\bs i})\}_{\bs i \in \mathcal{I}_n^{(U)}}),
\quad 
\alpha:=
\begin{cases}
|\mathcal{I}_n^{(U)}|/|\mathcal{K}_{\bs u}| & (v=1) \\
|\mathcal{I}_n^{(U)}| & (v=0) \\
\end{cases},
\] 
and we next prove 
(ii) $\mathbb{E}_{\tau}\left(\mathbb{E}_{\{\mathcal{M}^{(t)}\}_{t \in [\tau]}}(\|\frac{\partial}{\partial \bs \theta}Q_{\eta}(\tilde{\bs \theta}^{(\tau)})\|_2^2)\right)=O(1/\log T)$ by referring to (i) and \citet{ghadimi2013stochastic} Theorem 2.1~(a). 
Then, the assertion is proved.

\begin{itemize}
    \item[(i)] We first verify that $\mathbb{E}_{\mathcal{M}^{(t)}}(\tilde{g}^{(t)}_{\eta}(\bs \theta))=\alpha \frac{\partial}{\partial \bs \theta}Q_{\eta}(\bs \theta)$. 
    Here, we first consider the case $U \geq 2,v \geq 1$. A vector $\bs u=(u_1,u_2,\ldots,u_v)$ representing which of the entries in the index $\bs i=(i_1,i_2,\ldots,i_U)$ is fixed, is preliminary specified from the set $\{\bs u=(u_1,u_2,\ldots,u_v) \in [U]^v \mid u_1 < u_2 < \cdots < u_v\}$ by users. 
Then, considering a set $\mathcal{I}_{n,\bs u}^{(U)}(\bs j):=\{\bs i:=(i_1,i_2,\ldots,i_U) \mid \bs i \in \mathcal{I}_n^{(U)},i_{u_1}=j_1,\ldots,i_{u_v}=j_v\}$ for $\bs j \in [n]^v$, 
Algorithm~\ref{alg:minibatch} that defines $\mathcal{M}^{(t)}=(\tilde{\mathcal{P}}_{\text{mini}}^{(t)},\tilde{\mathcal{I}}_{\text{mini}}^{(t)},s_+^{(t)},s_-^{(t)})$ consists of the following two-steps. 
At iteration $t$, 
\begin{enumerate}
    \item[step 1.] $\bs j$ is randomly selected from a set $\mathcal{K}_{\bs u}:=\{\bs j \in [n]^v \mid \mathcal{I}^{(U)}_{n,\bs u}(\bs j) \neq \emptyset\}$ with the probability $p_{\bs j}$~(in Theorem~\ref{theo:sgd_convergence}, $p_{\bs j}$ is assumed to be $1/|\mathcal{K}_{\bs u}|)$, 
    \item[step 2.] $m_-,m_+$ entries are uniformly randomly selected from sets $\tilde{\mathcal{I}}_n^{(U)}=\mathcal{I}_{n,\bs u}^{(U)}(\bs j)$ and $\tilde{\mathcal{P}}_n^{(U)}=\mathcal{P}_{n,\bs u}^{(U)}(\bs j):=\{\bs i' \mid \bs i' \in \mathcal{I}_{n,\bs u}^{(U)}(\bs j),w_{\bs i'} \neq 0 \}$, and denote the sets as $\tilde{\mathcal{I}}_{\text{mini}}^{(t)},\tilde{\mathcal{P}}_{\text{mini}}^{(t)}$. 
    Coefficients $s_+^{(t)}:=|\tilde{\mathcal{P}}_n^{(U)}|/m_+$ and $s_-^{(t)}:=|\tilde{\mathcal{I}}_n^{(U)}|/m_-$ are also defined. 
\end{enumerate}

Therefore, the expectation of the stochastic gradient $\tilde{g}^{(t)}_{\eta}(\bs \theta)$ with respect to sampling the minibatch $\mathcal{M}^{(t)}$ is, 
\begin{align}
    &\mathbb{E}_{\mathcal{M}^{(t)}}(\tilde{g}^{(t)}_{\eta}(\bs \theta)) \nonumber \\
    &\hspace{0.5em}=
    \scalebox{0.88}{$\displaystyle 
    \mathbb{E}_{\mathcal{M}^{(t)}}\left(
    s_-^{(t)}
    \sum_{\bs i \in \tilde{\mathcal{I}}_{\text{mini}}^{(t)}}
    \mu_{\bs \theta}(\bs X_{\bs i})
    \varphi''(\mu_{\bs \theta}(\bs X_{\bs i}))
    \frac{\partial \mu_{\bs \theta}(\bs X_{\bs i})}{\partial \bs \theta} 
    -
    \eta
    \cdot 
    s^{(t)}_+
    \sum_{\bs i \in \tilde{\mathcal{P}}_{\text{mini}}^{(t)}}
    w_{\bs i}
    \varphi''(\mu_{\bs \theta}(\bs X_{\bs i}))
    \frac{\partial \mu_{\bs \theta}(\bs X_{\bs i})}{\partial \bs \theta}
    \right)$} \quad (\because \text{ the definition}~(\ref{eq:minibatch_gradient})) \nonumber \\
    &\hspace{0.5em}=
    \scalebox{0.88}{$\displaystyle
    \underbrace{
    \mathbb{E}_{\mathcal{M}^{(t)}}\left(
    s_-^{(t)}
    \sum_{\bs i \in \tilde{\mathcal{I}}_{\text{mini}}^{(t)}}
    \mu_{\bs \theta}(\bs X_{\bs i})
    \varphi''(\mu_{\bs \theta}(\bs X_{\bs i}))
    \frac{\partial \mu_{\bs \theta}(\bs X_{\bs i})}{\partial \bs \theta} 
    \right)
    }_{(\star 1)}
    -
    \eta \cdot 
    \underbrace{
    \mathbb{E}_{\mathcal{M}^{(t)}}\left(
    s_+^{(t)}
    \sum_{\bs i \in \tilde{\mathcal{P}}_{\text{mini}}^{(t)}}
    w_{\bs i}
    \varphi''(\mu_{\bs \theta}(\bs X_{\bs i}))
    \frac{\partial \mu_{\bs \theta}(\bs X_{\bs i})}{\partial \bs \theta}
    \right)
    }_{(\star 2)}
    $}, \label{eq:emg}
\end{align}
where the term $(\star 1)$ is evaluated by taking expectation with respect to the two steps in Algorithm~\ref{alg:minibatch} as
\begin{align}
    (\star 1)
    &=
    \underbrace{
    \mathbb{E}_{\bs j}\left(
        s_-^{(t)}
        \underbrace{
        \mathbb{E}_{\tilde{\mathcal{I}}_{\text{mini}}^{(t)}}
        \left(
        \sum_{\bs i \in \tilde{\mathcal{I}}_{\text{mini}}^{(t)}}
        \mu_{\bs \theta}(\bs X_{\bs i})
        \varphi''(\mu_{\bs \theta}(\bs X_{\bs i}))
        \frac{\partial \mu_{\bs \theta}(\bs X_{\bs i})}{\partial \bs \theta} 
        \mid \bs j
        \right)
        }_{(\text{expectation w.r.t. step 2})}
    \right)}_{(\text{expectation w.r.t. step 1})} \nonumber \\
    &=
    \mathbb{E}_{\bs j}\left(
        s_-^{(t)}
        \frac{m_-}{|\mathcal{I}_{n,\bs u}^{(U)}(\bs j)|}
        \sum_{\bs i \in \mathcal{I}_{n,\bs u}^{(U)}(\bs j)}
        \mu_{\bs \theta}(\bs X_{\bs i})
        \varphi''(\mu_{\bs \theta}(\bs X_{\bs i}))
        \frac{\partial \mu_{\bs \theta}(\bs X_{\bs i})}{\partial \bs \theta} 
    \right) \nonumber \\
    &=
    \mathbb{E}_{\bs j}\left(
        \sum_{\bs i \in \mathcal{I}_{n,\bs u}^{(U)}(\bs j)}
        \mu_{\bs \theta}(\bs X_{\bs i})
        \varphi''(\mu_{\bs \theta}(\bs X_{\bs i}))
        \frac{\partial \mu_{\bs \theta}(\bs X_{\bs i})}{\partial \bs \theta} 
    \right)    \quad 
    \left(\because s_-^{(t)}=\frac{|\mathcal{I}_{n,\bs u}^{(U)}(\bs j)|}{m_-}\right) \nonumber \\
    &=
    \frac{1}{|\mathcal{K}_{\bs u}|}
    \sum_{\bs j \in \mathcal{K}_{\bs u}}
    \sum_{\bs i \in \mathcal{I}_{n,\bs u}^{(U)}(\bs j)}
        \mu_{\bs \theta}(\bs X_{\bs i})
        \varphi''(\mu_{\bs \theta}(\bs X_{\bs i}))
        \frac{\partial \mu_{\bs \theta}(\bs X_{\bs i})}{\partial \bs \theta} 
    \quad 
    \left(
        \because p_{\bs j}=\frac{1}{|\mathcal{K}_{\bs u}|} \: (\forall \bs j \in \mathcal{K}_{\bs u})
    \right) \nonumber \\
    &=
    \frac{1}{|\mathcal{K}_{\bs u}|}
    \sum_{\bs i \in \mathcal{I}_n^{(U)}}
        \mu_{\bs \theta}(\bs X_{\bs i})
        \varphi''(\mu_{\bs \theta}(\bs X_{\bs i}))
        \frac{\partial \mu_{\bs \theta}(\bs X_{\bs i})}{\partial \bs \theta} 
    \quad 
    \left( \because
        \bigcup_{\bs j \in \mathcal{K}_{\bs u}}
        \mathcal{I}^{(U)}_{n,\bs u}(\bs j)
        =
        \mathcal{I}_n^{(U)} \right),
    \label{eq:star1}
\end{align}
and similarly, 
\begin{align}
    (\star 2)
    &=
    \frac{1}{|\mathcal{K}_{\bs u}|}
    \sum_{\bs i \in \mathcal{P}_n^{(U)}}
    w_{\bs i}
    \varphi''(\mu_{\bs \theta}(\bs X_{\bs i}))
    \frac{\partial \mu_{\bs \theta}(\bs X_{\bs i})}{\partial \bs \theta}. 
    \label{eq:star2}
\end{align}

Substituting (\ref{eq:star1}) and (\ref{eq:star2}) into (\ref{eq:emg}) leads to 
\begin{align*}
    \mathbb{E}_{\mathcal{M}^{(t)}}(\tilde{g}^{(t)}_{\eta}(\bs \theta)) 
    &=
    \frac{1}{|\mathcal{K}_{\bs u}|}
    \sum_{\bs i \in \mathcal{I}_n^{(U)}}
        \mu_{\bs \theta}(\bs X_{\bs i})
        \varphi''(\mu_{\bs \theta}(\bs X_{\bs i}))
        \frac{\partial \mu_{\bs \theta}(\bs X_{\bs i})}{\partial \bs \theta}
    -
    \eta \cdot 
    \frac{1}{|\mathcal{K}_{\bs u}|}
    \sum_{\bs i \in \mathcal{P}_n^{(U)}}
    w_{\bs i}
    \varphi''(\mu_{\bs \theta}(\bs X_{\bs i}))
    \frac{\partial \mu_{\bs \theta}(\bs X_{\bs i})}{\partial \bs \theta} \\
    &=
    \frac{|\mathcal{I}_{n}^{(U)}|}{|\mathcal{K}_{\bs u}|}
\underbrace{
    \frac{1}{|\mathcal{I}_{n}^{(U)}|}
    \left\{
    \sum_{\bs i \in \mathcal{I}_n^{(U)}}
        \mu_{\bs \theta}(\bs X_{\bs i})
        \varphi''(\mu_{\bs \theta}(\bs X_{\bs i}))
        \frac{\partial \mu_{\bs \theta}(\bs X_{\bs i})}{\partial \bs \theta}   
-
    \sum_{\bs i \in \mathcal{P}_n^{(U)}}
    \eta w_{\bs i}
    \varphi''(\mu_{\bs \theta}(\bs X_{\bs i}))
    \frac{\partial \mu_{\bs \theta}(\bs X_{\bs i})}{\partial \bs \theta}
\right\}
}_{=\frac{\partial}{\partial \bs \theta}D_{\varphi}( \{\eta w_{\bs i}\}_{\bs i \in \mathcal{I}_n^{(U)}} , \{\mu_{\bs \theta}(\bs X_{\bs i})\}_{\bs i \in \mathcal{I}_n^{(U)}}\left(=
\frac{\partial}{\partial \bs \theta}Q_{\eta}(\bs \theta) \right)} \\
&=
\alpha
\frac{\partial}{\partial \bs \theta}
Q_{\eta}(\bs \theta) \qquad \left(\because \alpha=\frac{|\mathcal{I}_n^{(U)}|}{|\mathcal{K}_{\bs u}|} \right).
\end{align*}

Thus (i) is proved for the case $U \geq 2,v \geq 1$. 
Here, we also consider the case $U \in \mathbb{N},v=0$. 
As $v=0$ indicates that there is no fixed entry in the index $\bs i$, meaning that 
the step~1 in the above explanation is skipped, 
Algorithm~\ref{alg:minibatch} consists of only the step~2. 
Thus, by noticing that $\tilde{\mathcal{P}}_n^{(U)}=\mathcal{P}_n^{(U)}, 
\tilde{\mathcal{I}}_n^{(U)}=\mathcal{I}_n^{(U)}$, 
following the same calculation leads to the equation $\mathbb{E}_{\mathcal{M}^{(t)}}(\tilde{g}^{(t)}_{\eta}(\bs \theta))=\alpha\frac{\partial}{\partial \bs \theta}Q_{\eta}(\bs \theta)$, which is the same as the case of $U \geq 2,v \geq 1$.

Since $v$ is limited to take value in $\{0,1,2,\ldots,U-1\}$, (i) is hereby proved for all the possible $(U,v)$.

\item[(ii)] We next prove that $\mathbb{E}_{\tau}\left(\mathbb{E}_{\{\mathcal{M}^{(t)}\}_{t \in [\tau]}}(\|\frac{\partial}{\partial \bs \theta}Q_{\eta}(\tilde{\bs \theta}^{(\tau)})\|_2^2)\right)=O(1/\log T)$ by referring to (i) and \citet{ghadimi2013stochastic} Theorem 2.1~(a). 
The following explanations are based on \citet{ghadimi2013stochastic}, with corresponding symbols $k \Leftrightarrow t$, $R \Leftrightarrow \tau$, 
$N \Leftrightarrow T$, $\gamma_k \Leftrightarrow \gamma^{(t)}$, $x_k \Leftrightarrow \tilde{\bs \theta}^{(t)}$, 
$f(x) \Leftrightarrow \alpha Q_{\eta}(\bs \theta)$, $G(\cdot,\xi_k) \Leftrightarrow \tilde{g}^{(t)}_{\eta}(\cdot)$, $L \Leftrightarrow H$, 
$D_f \Leftrightarrow D$, 
$\nabla \Leftrightarrow \frac{\partial}{\partial \bs \theta}$.

\citet{ghadimi2013stochastic} Theorem 2.1~(a) shows that, 
the iterative update
\begin{align}
    \tilde{\bs \theta}^{(t+1)}
    =
    \tilde{\bs \theta}^{(t)}
    -
    \gamma^{(t)}
    \tilde{g}^{(t)}_{\eta}(\tilde{\bs \theta}^{(t)})
    \label{eq:ghadmi_update}
\end{align}
satisfies 
\begin{align}
    \mathbb{E}_{\tau}\left(
    \mathbb{E}_{\{\mathcal{M}^{(t)}\}_{t \in [\tau]}}
    \left(
        \big\| 
        \alpha
        \frac{\partial}{\partial \bs \theta}Q_{\eta}(\tilde{\bs \theta}^{(\tau)}) \big\|_2^2
    \right)
    \right)
    &\leq 
    H \cdot 
    \frac{D^2+\sigma^2\sum_{t=1}^{T}\gamma^{(t)2}}{\sum_{t=1}^{T}(2\gamma^{(t)}-H \gamma^{(t)2})}
    , \label{eq:error_rate} 
\end{align}
where 
$D:=\sqrt{
    \frac{2}{H}
    \left(
        Q_{\eta}(\tilde{\bs \theta}^{(1)})-\inf_{\bs \theta \in \bs \Theta}Q_{\eta}(\bs \theta)
    \right)}$, 
    $H>0$ is the Lipschitz constant of $\alpha \frac{\partial}{\partial \bs \theta}Q_{\eta}(\bs \theta)$, 
    $\gamma^{(t)}$ represents the step size satisfying $\gamma^{(t)}<2/H$, and the number of iterations $\tau$ is chosen from $\{1,2,\ldots,T\}$ with the probability 
$\mathbb{P}(\tau=t)=\frac{2\gamma^{(t)}-H \gamma^{(t)2}}{\sum_{t=1}^{T}(2\gamma^{(t)}-H \gamma^{(t)2})}$, 
if assumptions (C-1) $\mathbb{E}_{\mathcal{M}^{(t)}}(\tilde{g}_{\eta}^{(t)}(\bs \theta))=\alpha \frac{\partial}{\partial \bs \theta} Q_{\eta}(\bs \theta)$ and (C-2) 
$\mathbb{E}_{\mathcal{M}^{(t)}}(\|\tilde{g}_{\eta}^{(t)}(\bs \theta)-\alpha \frac{\partial}{\partial \bs \theta} Q_{\eta}(\bs \theta)\|_2^2)<\sigma^2$ for some $\sigma \in (0,\infty)$, $(\forall \bs \theta \in \bs \Theta)$ hold.
These assumptions (C-1) and (C-2) correspond to eq.~(1.2) and eq.~(1.3) in \citet{ghadimi2013stochastic}, respectively.


In the case of Theorem~\ref{theo:sgd_convergence}, 
the minibatch SGD (\ref{eq:minibatch_sgd}) reduces to (\ref{eq:ghadmi_update}) due to the assumption $\bs \Theta=\mathbb{R}^q$, 
the step size satisfies $\gamma^{(t)}=\gamma t^{-1} \leq \gamma \overset{\text{(assumption)}}{<}2/H$, 
(C-1) is proved by the above calculation (i), and 
(C-2) is proved by 
\begin{align*}
    \mathbb{E}_{\mathcal{M}^{(t)}}
    \left( \big\|\tilde{g}_{\eta}^{(t)}(\bs \theta)-\alpha \frac{\partial}{\partial \bs \theta} Q_{\eta}(\bs \theta) \big \|_2^2
    \right)
    &=
    \mathbb{E}_{\mathcal{M}^{(t)}}\left(
    \sum_{\alpha=1}^{p}
    \left(\tilde{g}_{\eta}^{(t)}(\bs \theta)-\alpha\frac{\partial}{\partial \bs \theta} Q_{\eta}(\bs \theta)\right)_{\alpha}^2
    \right) \\
    &=
    \sum_{\alpha=1}^{p}
    \mathbb{E}_{\mathcal{M}^{(t)}}
    \left(
    \left(\tilde{g}_{\eta}^{(t)}(\bs \theta)-\alpha \frac{\partial}{\partial \bs \theta} Q_{\eta}(\bs \theta)\right)_{\alpha}^2
    \right) \\
    &=
    \text{tr}
    \mathbb{E}_{\mathcal{M}^{(t)}}
    \left(
    \left(\tilde{g}_{\eta}^{(t)}(\bs \theta)-\alpha \frac{\partial}{\partial \bs \theta} Q_{\eta}(\bs \theta)\right)^{\otimes 2}
    \right) \\
    &=
    \text{tr} \mathbb{V}_{\mathcal{M}^{(t)}}(\tilde{g}_{\eta}^{(t)}(\bs \theta)) \\
    &\leq 
    \sup_{\bs \theta \in \bs \Theta}\text{tr} \mathbb{V}_{\mathcal{M}^{(1)}}(\tilde{g}_{\eta}^{(1)}(\bs \theta))
    =:
    \sigma^2
    \overset{\text{(assumption)}}{<}
    \infty, 
\end{align*}
where $(\bs z)_{\alpha}$ represents the $\alpha$-th entry of the vector $\bs z=(z_1,z_2,\ldots,z_p)$, 
$\bs z^{\otimes 2}:=\bs z\bs z^{\top}$, and $\text{tr}\bs Z$ represents the trace of the matrix $\bs Z=(z_{ij})$, i.e., $\text{tr}\bs Z=\sum_{\alpha=1}^{p}z_{\alpha \alpha}$. 
Thus (\ref{eq:error_rate}) holds; we last evaluate the right hand side of (\ref{eq:error_rate}) in the following. 

Obviously, we have $H=O(1)$ and $\sigma^2=O(1)$ due to the assumptions, and $D=O(1)$ since the Lipschitz continuity of $\frac{\partial}{\partial \bs \theta}Q_{\eta}(\bs \theta)$ proves that $Q_{\eta}(\tilde{\bs \theta}^{(1)})$ is finite with any fixed $\tilde{\bs \theta}^
{(1)} \in \bs \Theta$. 
Then, it holds for $\gamma^{(t)}=\gamma t^{-1}$ that
\begin{align}
    H \cdot \frac{D^2+\sigma^2\sum_{t=1}^{T}\gamma^{(t)2} }{\sum_{t=1}^{T}(2\gamma^{(t)}-H \gamma^{(t)2})}
    &=
    H \cdot \frac{
        D^2+\sigma^2\gamma^{2}\sum_{t=1}^{T}t^{-2}
    }{
        2\gamma \sum_{t=1}^{T} t^{-1}
        -
        \gamma^{2}H\sum_{t=1}^{T} t^{-2}
    } \nonumber \\
    &\leq 
    H \cdot \frac{
        D^2+\sigma^2\gamma^{2} \pi^2/6
    }{
        2\gamma \log T
        -
        \gamma^{2}H \pi^2/6
    } 
    \quad 
    \bigg(\because \sum_{t=1}^{T}t^{-1} \geq \int_{t=1}^{T}t^{-1}\text{d}t=\log T \: (\geq 0), \nonumber \\ 
    &\hspace{4em} \text{ and } 
    \sum_{t=1}^{T}t^{-2} \leq \sum_{t=1}^{\infty}t^{-2}=\frac{\pi^2}{6}. \text{ See, e.g.,~\citet{hofbauer2002simple}.}
    \bigg) \nonumber \\
    &=
    O(1/\log T). \qquad \bigg(
        \because \: H=O(1),\sigma^2=O(1),D=O(1),\gamma=O(1)
    \bigg)
    \label{eq:evaluated_error_rate}
\end{align}
Thus, substituting $\alpha=O(1)$ and (\ref{eq:evaluated_error_rate}) into (\ref{eq:error_rate}) leads to 
\begin{align*}
\mathbb{E}_{\tau}\left(
\mathbb{E}_{\{\mathcal{M}^{(t)}\}_{t \in [\tau]}}\left(
\big\|
\frac{\partial}{\partial \bs \theta}
Q_{\eta}(\bs \theta)
\big\|_2^2
\bigg|_{\bs \theta=\tilde{\bs \theta}^{(\tau)}}
\right)
\right)
=
O(1/\log T)
\to
0, \quad 
(T \to \infty).
\end{align*} 
\end{itemize}

By noticing that $Q_{\eta}(\bs \theta)=D_{\varphi}(
\{\eta w_{\bs i}\}_{\bs i \in \mathcal{I}_n^{(U)}},
\{\mu_{\bs \theta}(\bs X_{\bs i})\}_{\bs i \in \mathcal{I}_n^{(U)}})$, Theorem~\ref{theo:sgd_convergence} is proved. 
\qed

\bibliographystyle{alpha}

\end{document}